\documentclass[a4paper, 11pt]{IWSCFF}		

\usepackage{bm}
\usepackage{amsmath}
\usepackage{gensymb}
\usepackage{amssymb}
\usepackage{multirow}
\usepackage{array}
\usepackage{enumitem}
\usepackage{lipsum}

\usepackage{caption}
\usepackage{subcaption}
\usepackage{verbatim}
\usepackage[colorlinks=true, pdfstartview=FitV, linkcolor=black, citecolor= black, urlcolor= black]{hyperref}
\usepackage{overcite}
\usepackage{footnpag}			      	
\usepackage[linesnumbered,ruled]{algorithm2e}

\PaperNumber{19-1964} 

\begin{document}

\title{Spacecraft Swarm Dynamics and Control about Asteroids}

\author{Corinne Lippe\thanks{Ph.D Candidate, Aeronautics and Astronautics, Stanford University, Durand Building, 496 Lomita Mall, CA 94305} \space
 and Simone D'Amico\thanks{Assistant Professor, Aeronautics and Astronautics, Stanford University, Durand Building, 496 Lomita Mall, CA 94305}
}

\maketitle{}

\begin{abstract}
This paper presents a novel methodology to control spacecraft swarms about single asteroids with arbitrary gravitational potential coefficients. This approach enables the use of small, autonomous swarm spacecraft in conjunction with a mothership, reducing the need for the Deep Space Network and increasing safety in future asteroid missions.  The methodology is informed by a semi-analytical model for the spacecraft absolute and relative motion that includes relevant gravitational effects without assuming J2-dominance as well as solar radiation pressure. The dynamics model is exploited in an Extended Kalman Filter (EKF) to produce an osculating-to-mean relative orbital element (ROE) conversion that is asteroid agnostic.  The resulting real-time relative mean state estimate is utilized in the formation-keeping control algorithm.  The control problem is cast in mean relative orbital elements to leverage the geometric insight of secular and long-period effects in the definition of control windows for swarm maintenance. Analytical constraints that ensure collision avoidance and enforce swarm geometry are derived and enforced in ROE space.   The proposed swarm-keeping algorithms are tested and validated in high-fidelity simulations for a reference asteroid mission.
\end{abstract}

\section{Introduction}
Missions to explore near-earth asteroids would provide valuable information about the solar system`s formation and potential in-situ resources for interplanetary missions.\cite{Coradini, Mazanek} As such, asteroid missions have become increasingly popular. However, information about the shape, gravity field, and rotational motion of the asteroid is required to ensure safe and effective satellite operations.  To obtain this information, previously flown missions about or to asteroids have utilized a single satellite tracked by the Deep Space Network (DSN). While the DSN enables ground-based control, it is a limited resource that introduces significant time delays in control implementation due to required human intervention.  In addition, the DSN has limited accuracy because of the distance to the object of interest. Recent work has attempted to reduce reliance on the DSN by proposing and analyzing the use of multiple satellite-probes equipped with cross-links for radiometric measurements and optical cameras for joint feature tracking, enabling autonomous navigation and estimation of gravity coefficients and the shape of asteroids.\cite{Leonard, Hesar} Preliminary studies indicate such measurements produce comparable accuracy to monolithic systems in shorter times without reliance on ground-based measurements. As a result, a swarm of small, autonomous spacecraft could be deployed by a larger mothership that also acts as a communication relay to obtain desired information about the asteroid's gravitational and rotational state.  

The benefits of this architecture are twofold.  First, the redundancy of the identical small swarm satellites decreases the risk associated with the mission.  Second, autonomous spacecraft control and navigation reduces reliance on the DSN and increases accuracy. This architecture and its corresponding benefits serve as the basis of the Autonomous Nanosatellite Swarming (ANS) mission concept. \cite{Stacey, ANS}  ANS poses three main requirements at a spacecraft dynamics and control level: 1) an accurate dynamics model including secular and long-period effects for swarm orbit propagation; 2) a conversion from osculating to mean orbital elements to filter out short-period oscillations from the estimated state; and 3) a formation-keeping guidance and control strategy which ensures safety, co-location, and retains swarm geometry to the prescribed level of accuracy. 

 State-of-the-art approaches are unable to meet these requirements.  Specifically, the literature lacks a computationally efficient dynamics model that captures the secular and long-period effects of solar radiation pressure (SRP) and gravitational potential on absolute and relative motion for arbitrary asteroids. While a model that captures SRP effects on orbits about asteroids has been derived by Guffanti, it has yet to be utilized in control development.\cite{Guffanti_SRP}  Conversely, the most advanced gravity perturbation models assume the central body has a dominant $J_2$ parameter, like Earth, \cite{Brouwer, Mahajan} and therefore do not achieve the desired level of accuracy for arbitrary asteroids as will be shown later in the paper.  Similar to current dynamic models, state-of-the-art conversions from an osculating to a mean state make assumptions that are incompatible with the most general asteroid missions.  These assumptions include $J_2$-dominance \cite{Mahajan, Gias} or a rotation rate of the primary attractor body that is equal to or slower than the orbit mean motion \cite{Ely}. Alternatively, Kalman filters have been proposed for earth-based systems, but this approach only considers the second-order gravity potential in reconstructing osculating measurements from mean states.\cite{Zhong}  This low-degree consideration is insufficient given how highly variable asteroid gravitational fields are.   For formation-keeping, the most advanced techniques have been developed for earth-orbiting swarms.  As a result, current approaches consider only atmospheric drag and $J_2$ effects on relative motion.\cite{Koenig_swarm, Morgan} Additionally, the approach by Morgan does not leverage fuel-efficient solutions \cite{Chernick} and requires frequent maneuvers.  Comparatively, Koenig's control approach uses fuel-efficient maneuvers and only requires maneuvers when satellite safety is at risk.  However, Koenig only requires in-plane control for Earth-based swarms.  

This paper addresses shortcomings in dynamics modeling, osculating-to-mean conversions, and formation keeping.  For dynamics modeling, this work leverages propagation models based on averaging theory because no assumptions are made about J2-dominance. \cite{Liu, Zhong}  The averaging-theory models are adapted and integrated with Guffanti's existing SRP model\cite{Guffanti_SRP} to produce a semi-analytical orbit propagation of absolute and relative motion for autonomous control purposes that is as asteroid-agnostic as possible. Considerations on an asteroid's accurate characterization \cite{Stacey} drive the requirements on absolute and relative orbit propagation and an osculating-to-mean conversion (see Table \ref{tab:ANS_req}).  Specifically, the absolute orbit propagation shall achieve a maximum error of 100 meters over 5 orbit periods for all errors represented as ROE with one exception. The error represented as the relative mean longitude shall achieve an error to within 500 m.  The absolute orbit requirement enables future control to within 1 km in position over five orbits. For relative motion, the propagation shall achieve a maximum error of 30 meters over 5 orbit periods in terms of ROE.  The relative motion requirement enables accurate orbit prediction during maneuvers and facilitates meter-level accuracy of the osculating-to-mean conversion described next.\cite{ANS}  The accurate and computationally efficient propagation enables proximity operation for imaging purposes of small satellites with limited resources.

Additionally, this paper presents a novel osculating-to-mean conversion for relative motion with meter-level precision.  The conversion is formulated as an Extended Kalman filter (EKF) with a mean relative orbital state.  In the EKF, the semi-analytical orbit propagation developed previously serves as the dynamics model. Measurements of the mean ROE state are approximated as the osculating relative orbital elements provided by the navigation. The EKF shall achieve specified $\sigma$ bounds, where $\sigma$ corresponds to the component-wise standard deviation of the EKF's state estimate from the ground truth. The EKF shall produce a $\sigma$ bound within 10 m on all ROE except $a\delta a$, which shall achieve a bound of 3 m. The requirement enables control within a few meters on $a\delta a$ and to within 100 meters on all other ROE.   A summary of requirements for the EKF and orbit propagation can be found in Table \ref{tab:ANS_req}.  Notably, all of the requirements have been defined with respect to a worst-case scenario gravity field, which is defined to be of a large mass with specified low-order zonal potentials as described later in the paper.

\begin{table}[htb]
    \caption{ANS Requirements for a worst-case scenario asteroid.}
    \label{tab:ANS_req}
    \centering
    \begin{tabular}{p{7cm} p{5cm} p{1cm} } 
    \noalign{\hrule height 2pt}
    \multirow{2}{*}{max absolute orbit propagation error (5 orbits)} & ($a\delta a, a\delta e_x, a\delta e_y, a\delta i_x, a\delta i_y$) & 100 m \\
    &  $a\delta \lambda$ & 500 m \\
    \hline
    max relative orbit propagation error (5 orbits) & ($a\delta a, a\delta \lambda, a\delta e_x, a\delta e_y, a\delta i_x, a\delta i_y$) & 30 m \\
    \hline
    \multirow{2}{*}{osculating to mean EKF ($\sigma$)} & $a\delta a$ & 3 m \\
    &  ($a\delta \lambda, a\delta e_x, a\delta e_y, a\delta i_x, a\delta i_y$) & 10 m \\
    
     \noalign{\hrule height 2pt}
\end{tabular}
\end{table}

The osculating-to-mean conversion provides a real-time estimate of the mean state for use in a guidance profile for formation-keeping.  The algorithm targets the over-constrained control problem of a limited number of thrusters present on small satellites. The novel profile developed in this paper aims to enable science objectives, as well as ensure swarm safety.  To promote uninterrupted gravity recovery and imaging, the profile ensures a user-specified time exists between maneuvers. Additionally, to facilitate stereoscopic imaging and communication, the guidance strategy prevents evaporation. All of this is achieved by casting the control problem in ROE space, where simple, analytical constraints are derived.

The rest of the paper is structured into six main parts.  First, relevant mathematical background material is reviewed.  Second, the development and validation of the dynamics model is presented.  Third, the EKF for the osculating-to-mean conversion is described. Fourth, the guidance profile and control logic is derived for formation-keeping.  Fifth, a validation of the dynamics model, filter, and guidance profile through a simulation of a reference asteroid mission is provided.  Finally, conclusions and next steps are presented.

\section{Background Material}
This section contains two review topics: the Gauss and Lagrange planetary equations and the quasi-nonsingular ROE.  The Gauss and Lagrange planetary equations are used in the semi-analytical orbit propagation based on averaging theory.  The quasi-nonsingular ROE are used for error description of absolute and relative motion propagation as well as in the formation-keeping control algorithm.

\subsection{Gauss and Lagrange Planetary Equations}
Primary attractor gravitational fields are represented using spherical harmonic coefficients, denoted $C$ and $S$, that arise from the Legendre polynomial expansion of the Newtonian potential. These harmonics describe the instantaneous gravity potential as a function of location, which can be written with dependence on Cartesian coordinates or orbital elements. In an orbital element representation, the short-period, long-period, and secular effects are distinguished by dependence on the mean anomaly, the argument of perigee, or neither parameter, respectively. For long-term orbit prediction and control implementation, a set of mean elements are adopted to capture the long-period and secular affects only.

The evolution of the mean elements can be expressed through a "mean" perturbing potential, which is computed by averaging the instantaneous perturbing potential over one orbital period. Note that the average is centered at the current time and therefore requires a half-orbit period of information in the future and the past.  The resulting average potential is used in the Gaussian variation of parameter equations as a disturbing function.  As a result, time derivatives of the orbital elements are created with long-term and secular contributions as defined in Vallado for the classical orbital elements.\cite{Vallado}  The expressions for the quasi-nonsingular orbital elements are given by
\begin{equation}
\begin{aligned}
	\label{eq:lagrange}
	\frac{da}{dt} = \frac{2}{na}\frac{\partial R}{\partial M_0} \\
	\frac{du}{dt} =  \bigg( \frac{\sqrt{1-e^2}}{na^2e} -\frac{1-e^2}{na^2e}\bigg) \frac{\partial R}{\partial e}- \frac{\cos{i}}{na^2\sqrt{1-e^2} \sin{i}} \frac{\partial R}{\partial i} -\frac{2}{na}\frac{\partial R}{\partial a}\\
	\frac{de_x}{dt} = \bigg( \frac{1-e^2}{na^2e}\frac{\partial R}{\partial M_0}-\frac{\sqrt{1-e^2}}{na^2e}\frac{\partial R}{\partial \omega} \bigg) \frac{e_x}{e}- \bigg(\frac{\sqrt{1-e^2}}{na^2e}\frac{\partial R}{\partial e}- \frac{\cos{i}}{na^2\sqrt{1-e^2} \sin{i}} \frac{\partial R}{\partial i}\bigg) e_y\\
	\frac{de_y}{dt} =  \bigg( \frac{1-e^2}{na^2e}\frac{\partial R}{\partial M_0}-\frac{\sqrt{1-e^2}}{na^2e}\frac{\partial R}{\partial \omega} \bigg) \frac{e_y}{e}+ \bigg(\frac{\sqrt{1-e^2}}{na^2e}\frac{\partial R}{\partial e}- \frac{\cos{i}}{na^2\sqrt{1-e^2} \sin{i}} \frac{\partial R}{\partial i}\bigg) e_x \\
	\frac{di}{dt}=\frac{2}{\sqrt{1-e^2}\sin{i}}\bigg(\frac{\partial R}{\partial \omega}\cos{i}-\frac{\partial R}{\partial \Omega}\bigg) \\
	\frac{d\Omega}{dt}= \frac{2}{\sqrt{1-e^2}\sin{i}}\frac{\partial R}{\partial i}\\
\end{aligned}
\end{equation}
where the disturbing function is represented by $R$\footnote{ A list of relevant symbols is found in Tables~\ref{tab:not1} and \ref{tab:not2} in the appendix Notation.}, the mean motion is denoted by $n$, and the eccentricity $e$ is defined as the magnitude of the 2-D vector described by components $e_x$ and $e_y$. The quasi-nonsingular set of orbital elements  are represented by $(a, u, e_{x}, e_{y}, i, \Omega)$ corresponding to the semimajor axis, mean argument of latitude, the x- and y-components of the eccentricity vector, inclination, and right ascension of the ascending node (RAAN), respectively.  Notably, the expressions  display a singularity due to the presence of the eccentricity and sine of the inclination in the denominator. As noted by Vallado, these singularities present issues for arbitrary disturbing functions, but they can be avoided by thresholding as discussed in the sequel.  

Quasi-nonsingular orbital elements are defined in an asteroid-centered inertial  (ACI) frame, which is illustrated in Figure \ref{fig:ROE_description}.  The ACI frame is defined to translate with the center of mass of the asteroid but maintain a fixed orientation.  The orientation of this frame is defined by a z-axis that aligns with the mean axis of rotation, an x-axis that is aligned with the intersection of the asteroid equatorial and the ecliptic planes, and a y-axis to complete the right-handed triad.  This frame is defined for principal axis rotators, which is an assumption of the ANS project.

\subsection{Quasi-Nonsingular Relative Orbital Elements}
A set of relative orbital elements provides a comparison between two sets of absolute orbital elements.  The comparison is useful for interpreting error between two propagations or for visualizing relative motion. In this work, the comparison is parameterized using a set of quasi-nonsingular ROE defined by D'Amico due to their simple relationship with relative motion for near-circular\cite{DAmico_thesis} and eccentric \cite{Sullivan_ecc} orbits.  Specifically, the ROE approximate the invariants of the Hill-Clohessey-Wiltshire and Yamanaka-Ankerssen equations for near-circular and eccentric orbits, respectively \cite{Damico_int, Sullivan_ecc}.  

The ROE $(\delta a, \delta\lambda, \delta\mathbf{e}, \delta\mathbf{i})$ include the relative semimajor axis, the relative mean longitude, the relative eccentricity vector, and the relative inclination vector, respectively.  The individual components are constructed from the absolute orbital elements of a chief and deputy spacecraft, denoted with the subscripts $c$ and $d$, according to 
\begin{equation}
\begin{aligned}
	\label{eq:ROE}
	\delta\alpha = \begin{pmatrix}
           \delta a \\
           \delta\lambda \\
           \delta \bf{e} \\
           \delta \bf{i}
         \end{pmatrix} = \begin{pmatrix}
           \delta a \\
           \delta\lambda \\
           \delta e_x \\
           \delta e_y \\
           \delta i_x \\
           \delta i_y
         \end{pmatrix} = \begin{pmatrix}
         (a_d - a_c)/a_c \\
         (u_d - u_c) + \cos{i_c}(\Omega_d - \Omega_c) \\
         e_{x,d}-e_{x,c} \\
         e_{y,d}-e_{y,c}\\
         i_d - i_c \\
         \sin{i_c}(\Omega_d - \Omega_c)
         \end{pmatrix}
\end{aligned}
\end{equation}

\begin{figure}[htb]
	\centering
	 \begin{subfigure}[htb]{0.3\textwidth}
    \centering
    \caption{}
    \includegraphics[height=1.3in]{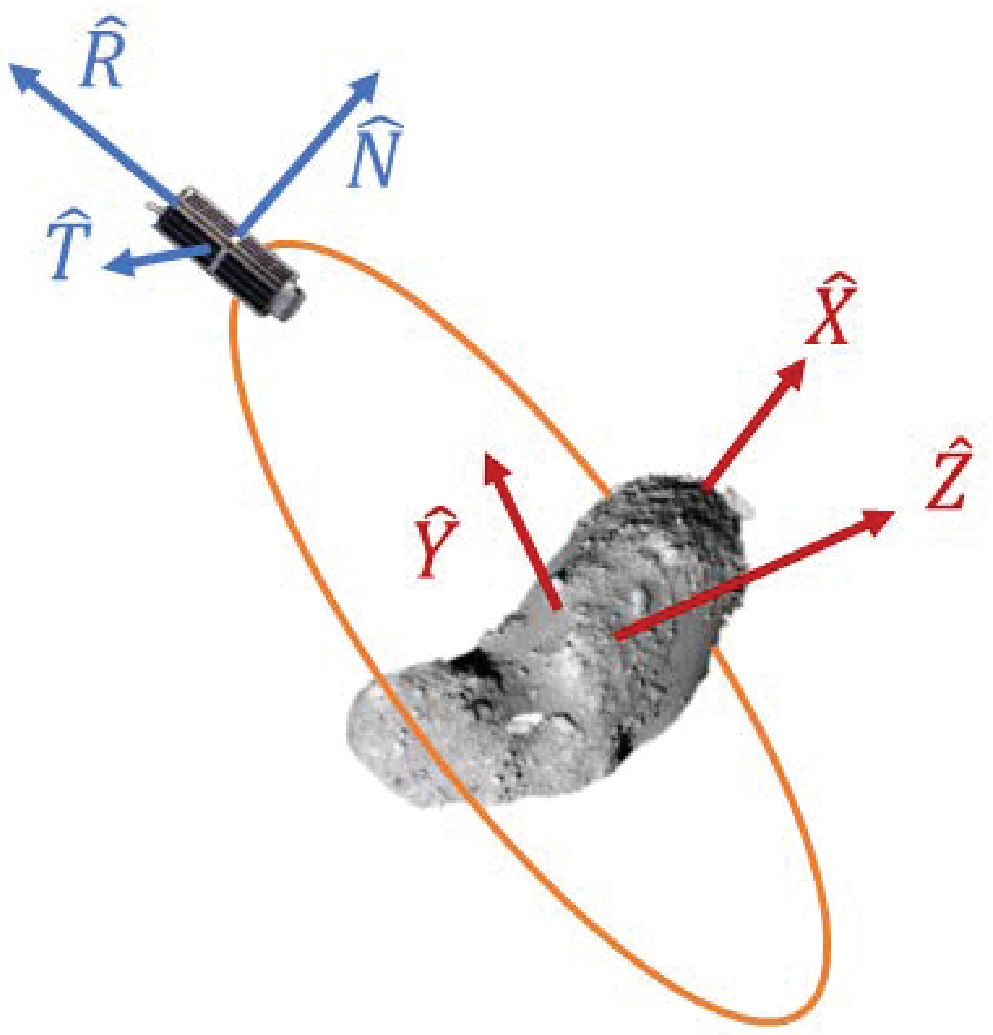}
    \end{subfigure}
	\begin{subfigure}[htb]{0.65\textwidth}
    \centering
    \caption{}
    \includegraphics[height=1.3in]{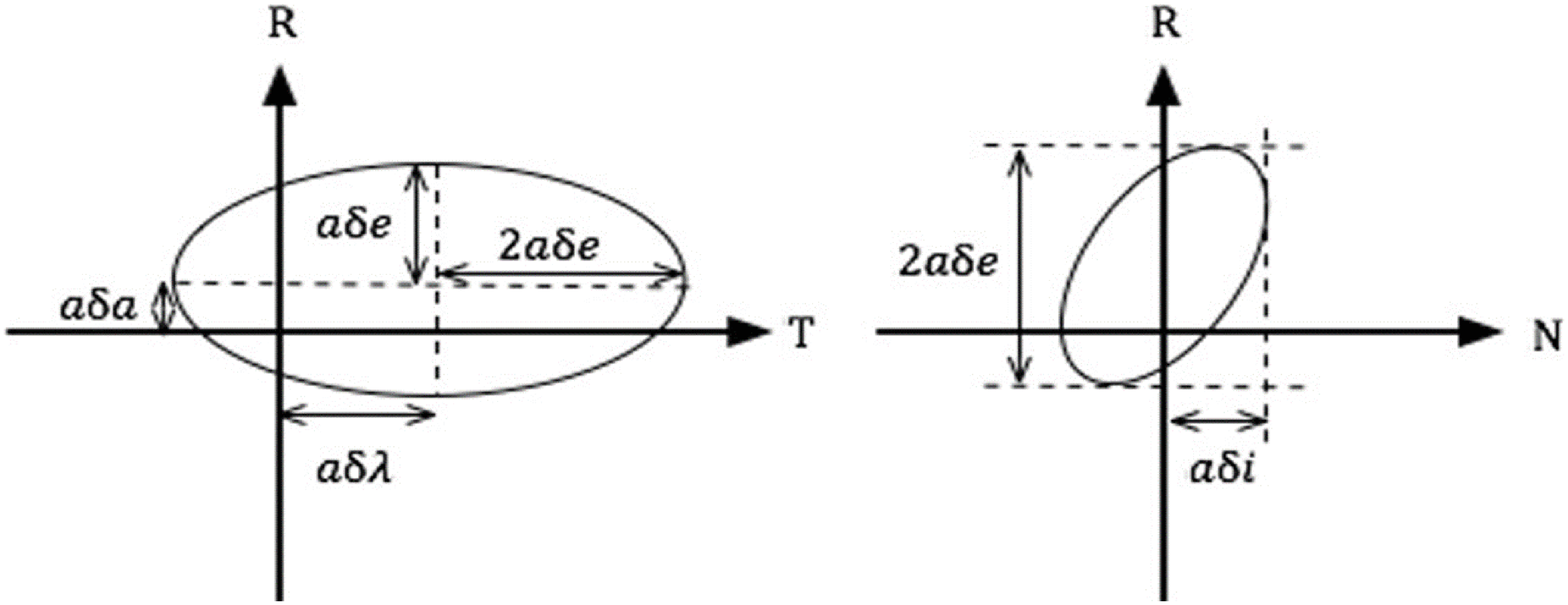}
    \end{subfigure}

	\caption{The Hill or RTN frame and the defined ACI frame is provided in subfigure a. Representation of relative motion in the Hill frame for near-circular orbits as described by quasi-nonsingular ROE is provided in subfigure b. Drift in the along-track direction due to $\delta a$ is omitted for simplicity.}
	\label{fig:ROE_description}
\end{figure}

The relationship between the ROE and the deputy's relative position and velocity in the Hill frame for near-circular orbits to the first order in separation is described by \cite{Damico_int}

\begin{equation}
\begin{aligned}
	\label{eq:ROE2cart}
	\begin{pmatrix}
           \delta r_R \\
           \delta r_T \\
           \delta r_N \\
           \delta v_R \\
           \delta v_T \\
           \delta v_N
         \end{pmatrix} = a \begin{bmatrix}
           1 & 0 & -\cos{u_c} & -\sin{u_c} & 0 & 0 \\
           0 & 1 & 2\sin{u_c} & -2\cos{u_c} & 0 & 0 \\
            0 & 0 & 0 & 0 & \sin{u_c} & -\cos{u_c}\\
            0 & 0 & n\sin{u_c} & -n\cos{u_c} & 0 & 0\\
            -\frac{3}{2}n & 0 & 2n\cos{u_c} & 2n\sin{u_c} & 0 & 0 \\
            0 & 0 & 0 & 0 & n\cos{u_c} & n\sin{u_c}
         \end{bmatrix}  \begin{pmatrix}
         \delta a \\
         \delta\lambda\\
         \delta e_x\\
         \delta e_y \\
         \delta i_x \\
        \delta i_y
         \end{pmatrix}
\end{aligned}
\end{equation}
where $\delta \mathbf{r}$ and $\delta \mathbf{v}$ represent the relative position and velocity in the Hill frame.  The Hill frame is defined by the radial, along-track, and out-of-plane directions denoted as R, T, and N respectively. The radial direction is defined as the vector from the asteroid's center of mass to the chief spacecraft's center of mass; the out-of-plane direction is defined by the angular momentum vector of the chief satellite; and the along-track direction completes the right-handed triad, which is illustrated in Figure \ref{fig:ROE_description} subfigure (a).  The relative motion provided by Equation \eqref{eq:ROE2cart} is illustrated in Figure~\ref{fig:ROE_description} subfigure (b) for near-circular orbits. 

The mapping in Equation~\eqref{eq:ROE2cart} is more accurate when $\delta r$ and $\delta v$ are defined in curvilinear as opposed to cartesian coordinates.  Equation~\eqref{eq:ROE2cart} represents a linearization of the relative motion for small inter-spacecraft separations in a cartesian frame. \cite{DAmico_thesis,deBruinj} Specifically, the ratio of inter-satellite distances in the radial, along-track, and out-of-plane directions to the orbit semimajor axis should not exceed 0.01. In this work, the rectilinear relative position approximation is used for convenience with no significant loss in accuracy for collision avoidance requirements, where inter-satellite distances are within specified limits. 

The relative motion and therefore ROE are affected by radial, along-track, and out-of-plane accelerations. The changes in the ROE are described by the control matrix $\mathbf{\Gamma}$ developed by D'Amico\cite{DAmico_thesis} for small spacecraft separations in near-circular orbits.  The matrix is defined by
\begin{equation}
\begin{aligned}
	\label{eq:gamma}
	\delta \mathbf{\dot{\mathbf{\alpha}}}(t) = \mathbf{\Gamma}(\mathbf{\alpha_c}(t))\delta \mathbf{p}(t) = \mathbf{\Gamma}(\mathbf{\alpha_c}(t)) \begin{pmatrix}
	\delta p_R(t) \\
	\delta p_T(t) \\
	\delta p_N(t) \\
	\end{pmatrix}, \\
	\mathbf{\Gamma}(\mathbf{\alpha_c(t)}) = \frac{1}{an} \begin{bmatrix}
           0 & 2 & 0 \\
           -2 & 0 & 0 \\
           \sin(u_c) & 2 \cos(u_c) & 0 \\
           -\cos(u_c) & 2 \sin(u_c) & 0 \\
           0 & 0 & \cos(u_c) \\
           0 & 0 & \sin(u_c)
         \end{bmatrix} 
\end{aligned}
\end{equation}
where $\delta\mathbf{p}$ represents the acceleration vector in the RTN frame of the chief, as demonstrated in Figure \ref{fig:ROE_description} subfigure a. While the  control matrix is rigorously defined for osculating ROE, changes in the osculating ROE corresponds approximately to the same change in mean ROE due to the near-identity mapping between the states.\cite{Roscoe}

\section{Development of Dynamics Model Using Averaging-Theory}
This section contains two parts.  First, the development of the dynamics model for the orbit propagation is explained, and example results for absolute and relative motion are provided.  Included in the results is the demonstration of achieving the previously defined requirements, which are listed in Table \ref{tab:ANS_req}.  Second, the passive relative motion trends are described for asteroid environments to improve understanding of the dynamic environment and provide insight into the drifts to be countered in the formation-keeping algorithm.

\subsection{Orbit Propagation}
As mentioned earlier, the state-of-the-art for Earth-based gravity models is not sufficient for capturing the mean absolute orbital elements evolution under asteroid gravity. To demonstrate this insufficiency, Mahajan's state-of-the-art, $J_2$-dominant mean orbital element propagation model\cite{Mahajan} combined with Guffanti's mean absolute orbit SRP model\cite{Guffanti} is compared to a ground truth simulation.  

A ground-truth simulation is developed with parameters found in Table~\ref{tab:dynamic_param} for an Eros-like asteroid. The simulated asteroid differs from Eros in its spin-axis and low-order zonal parameters. The spin axis was chosen to align with the z-axis of the J2000 reference frame for convenience.  This adjustment does not affect the gravity potential effects, which are described in the ACI frame.  Secondly, the normalized second, third, and fourth zonal parameters are all set to the conservative upper limit for $C_{30}$ and $C_{40}$ defined to be $\pm0.03$ based on available literature for asteroids. \cite{castalia,takahashi,small_body,toutatis,McMahon}  This adjustment exacerbates any modeling errors in the propagation and provides an estimate of the highest expected errors. A sweep over the quasi-stable orbits defined by inclination $i$ and argument of perigee $\omega$ was completed to provide a holistic representation of the errors. Notably, a sweep over the right ascension is not included because the parameter does not affect any of the gravity potential time derivatives as demonstrated in the formulas in the Geopotentials appendix.  The ground truth mean orbital elements are computed from a numerical average of the simulated osculating parameters centered over one orbit period, where angular values are unwrapped before averaging.  

The numerically averaged mean orbital elements are compared with the mean orbital elements produced by Mahajan's propagation.  The propagation is initialized to the ground truth mean state provided from numerical averaging centered at one orbit period in simulation time.  

\begin{table}[htb]
    \caption{Ground Truth Simulation Parameters.}
    \label{tab:dynamic_param}
    \centering
    \begin{tabular}{p{6cm} p{7cm} } 
    \noalign{\hrule height 2pt}
    Simulation Time & 5 Orbits \\
    \hline
    Integrator & Runge-Kutta (Dormand-Price) \\
    \hline
     Step size & Fixed: 10s \\ 
    \hline
    Gravity Model & $15^{th}$ degree and order \\
     \hline
    Third Body Gravity &  point masses, analytical ephemerides  \\
    \hline
    Third bodies included & Sun, Pluto and the eight planets \\
    \hline
    Solar Radiation Pressure & constant satellite cross-section normal to sun \\
    \hline
    Central Body & Eros variant \\
    \hline
    $a$ & 60 km \\
    \hline
    $e$ & 0.01  \\
    \hline
    $i$ & [100\degree, 135\degree, 170\degree] \\
    \hline
    $\Omega$ & 135\degree\\
    \hline
    $\omega$ & [46\degree, 136\degree, 91\degree, 216\degree, 271\degree, 316\degree]\\
    \hline
    Satellite Cross-sectional Area & 0.02 m$^2$\\
    \hline
    Satellite Coefficient of Reflectively & 1 \\
    \hline
    Satellite Mass &  5 kg\\
     \noalign{\hrule height 2pt}
\end{tabular}
\end{table}

 \begin{figure}[htb]
	\centering 
    \begin{subfigure}[htb]{0.3\textwidth}
    \centering
    \caption{}
    \includegraphics[width=1.8in]{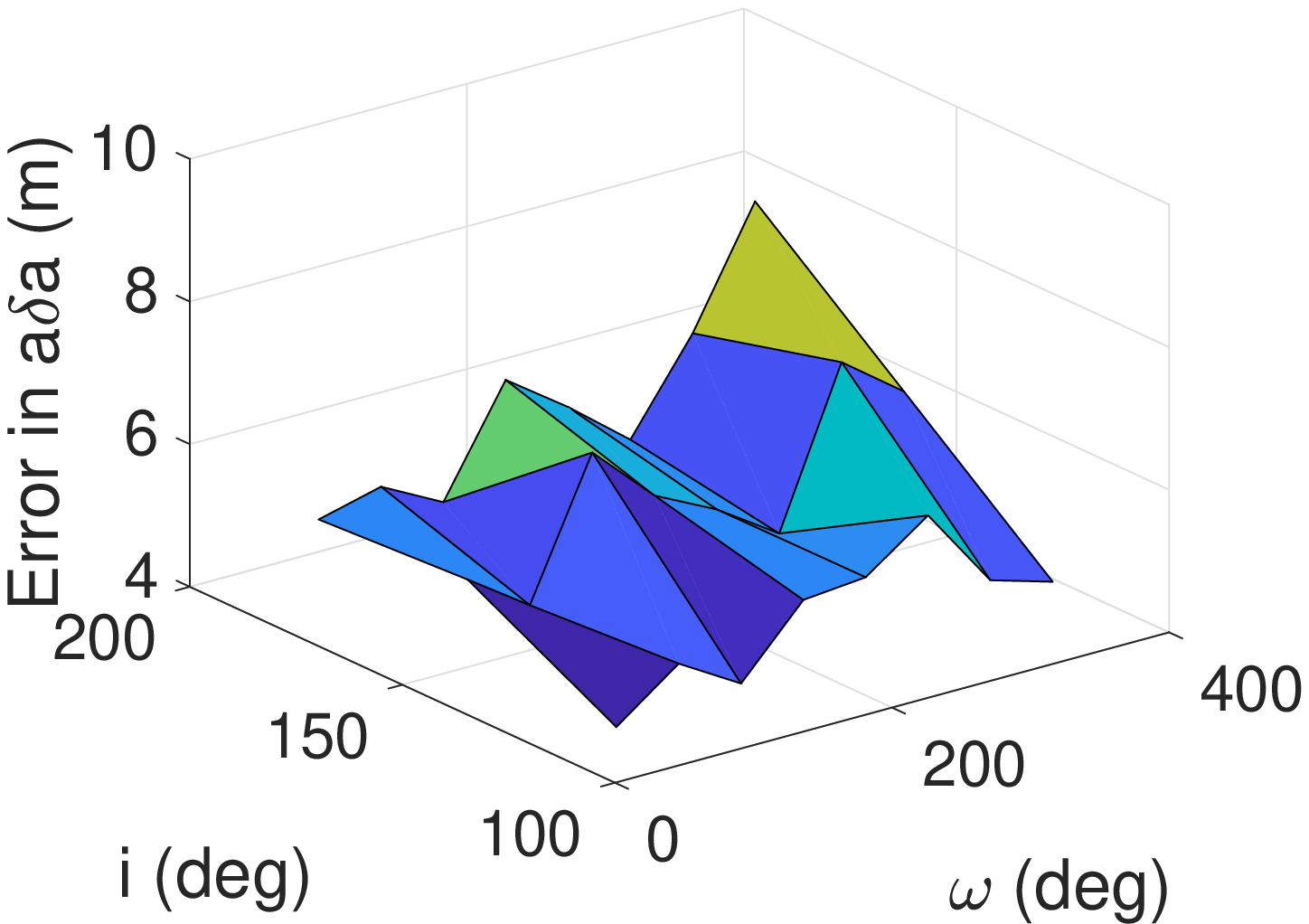}
    \end{subfigure}
    \begin{subfigure}[htb]{0.3\textwidth}
    \centering
    \caption{}
	\includegraphics[width=1.8in]{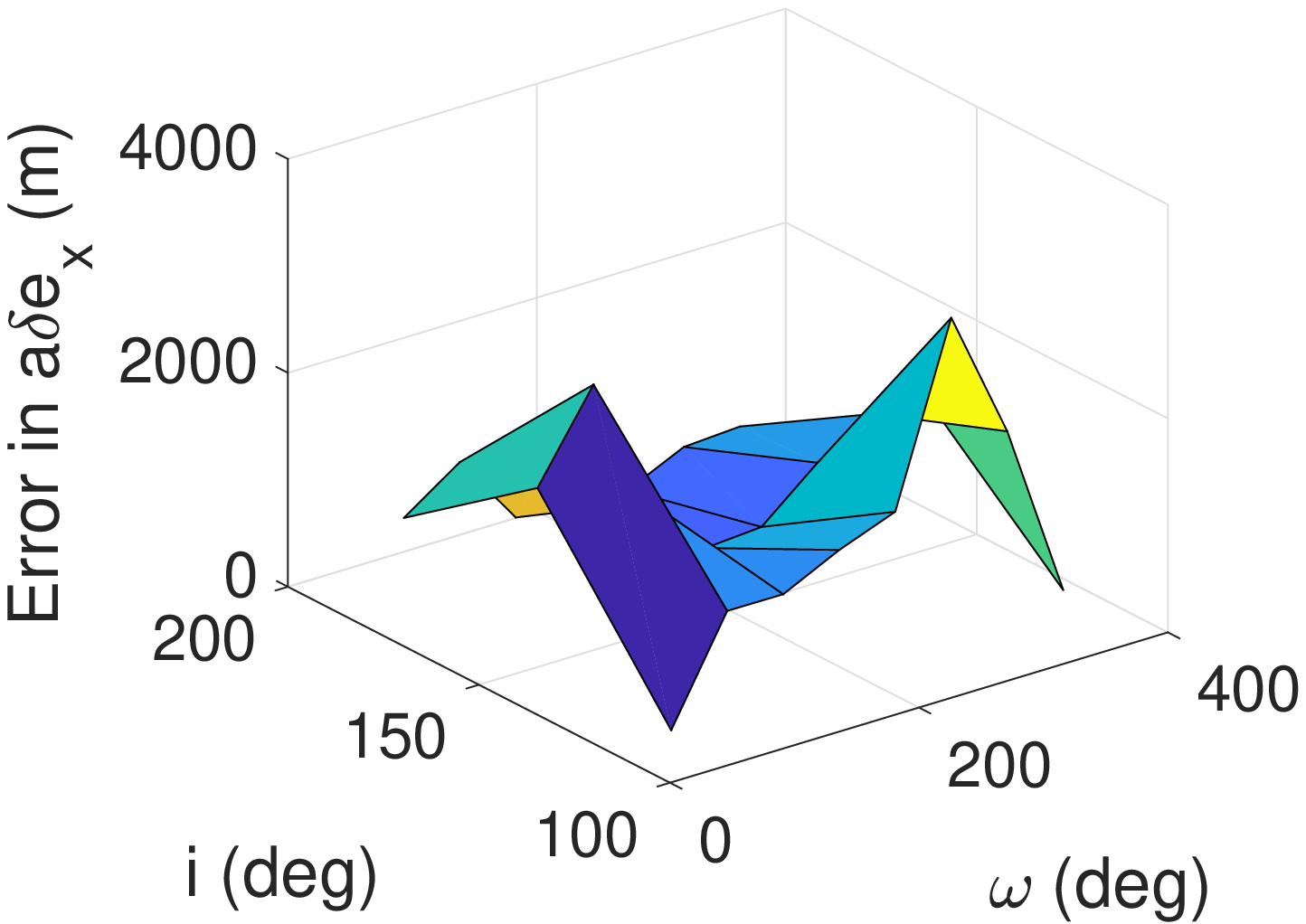}
    \end{subfigure}
	\begin{subfigure}[htb]{0.3\textwidth}
    \centering
    \caption{}
    \includegraphics[width=1.8in]{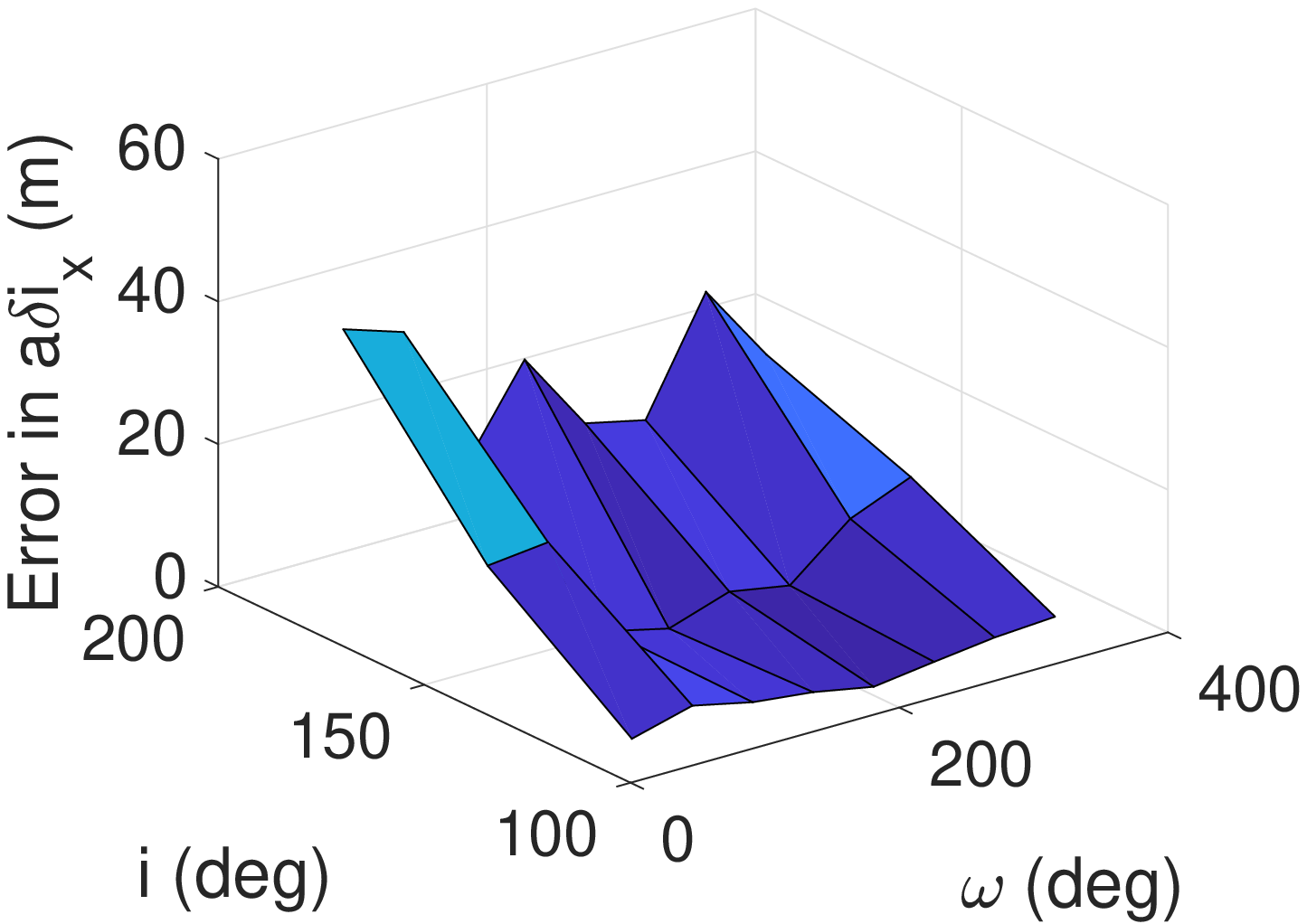}\\
    \end{subfigure}
    \begin{subfigure}[htb]{0.3\textwidth}
	\centering
	\caption{}
	\includegraphics[width=1.8in]{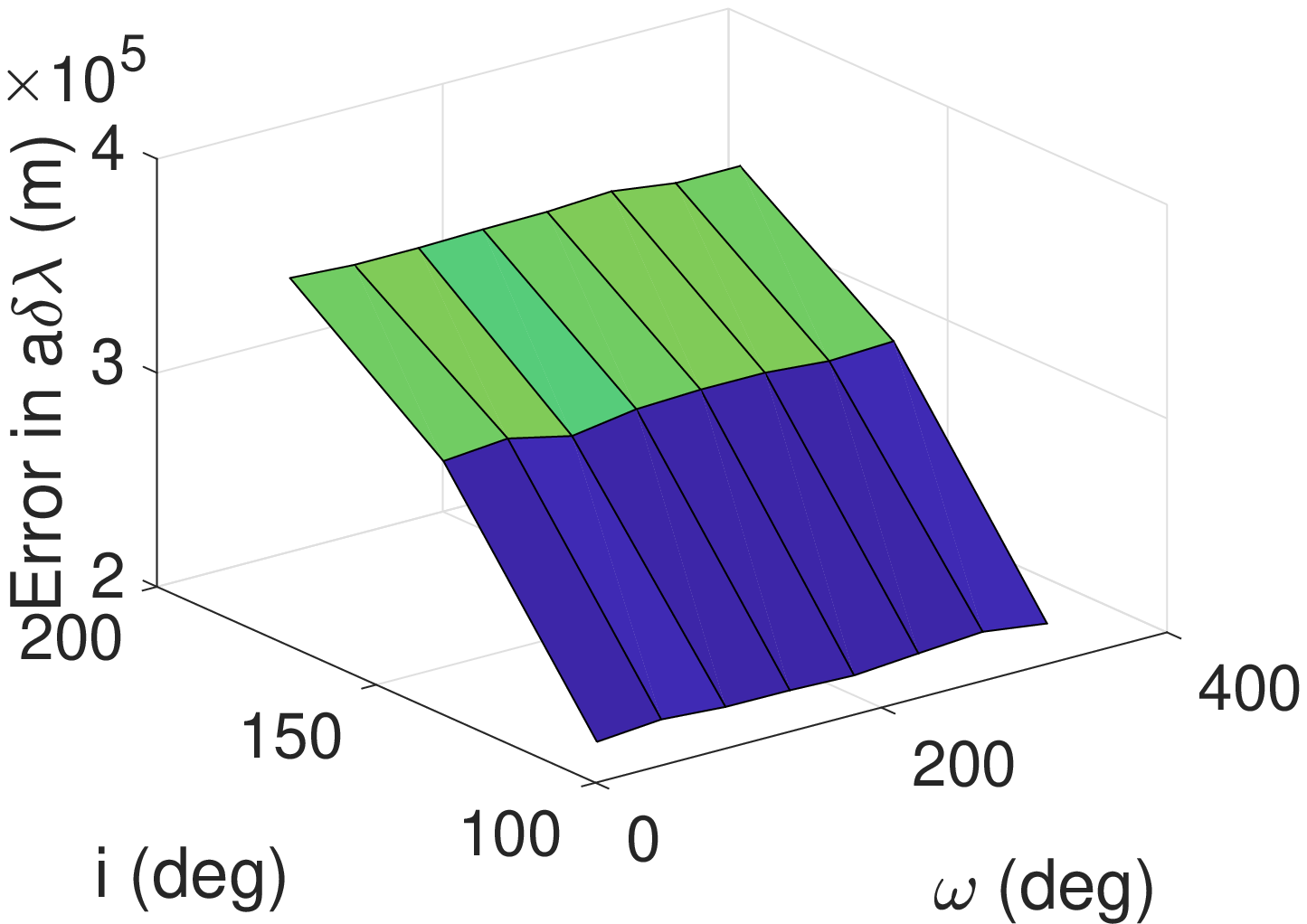}
    \end{subfigure}
    \begin{subfigure}[htb]{0.3\textwidth}
    \centering
    \caption{}
	\includegraphics[width=1.8in]{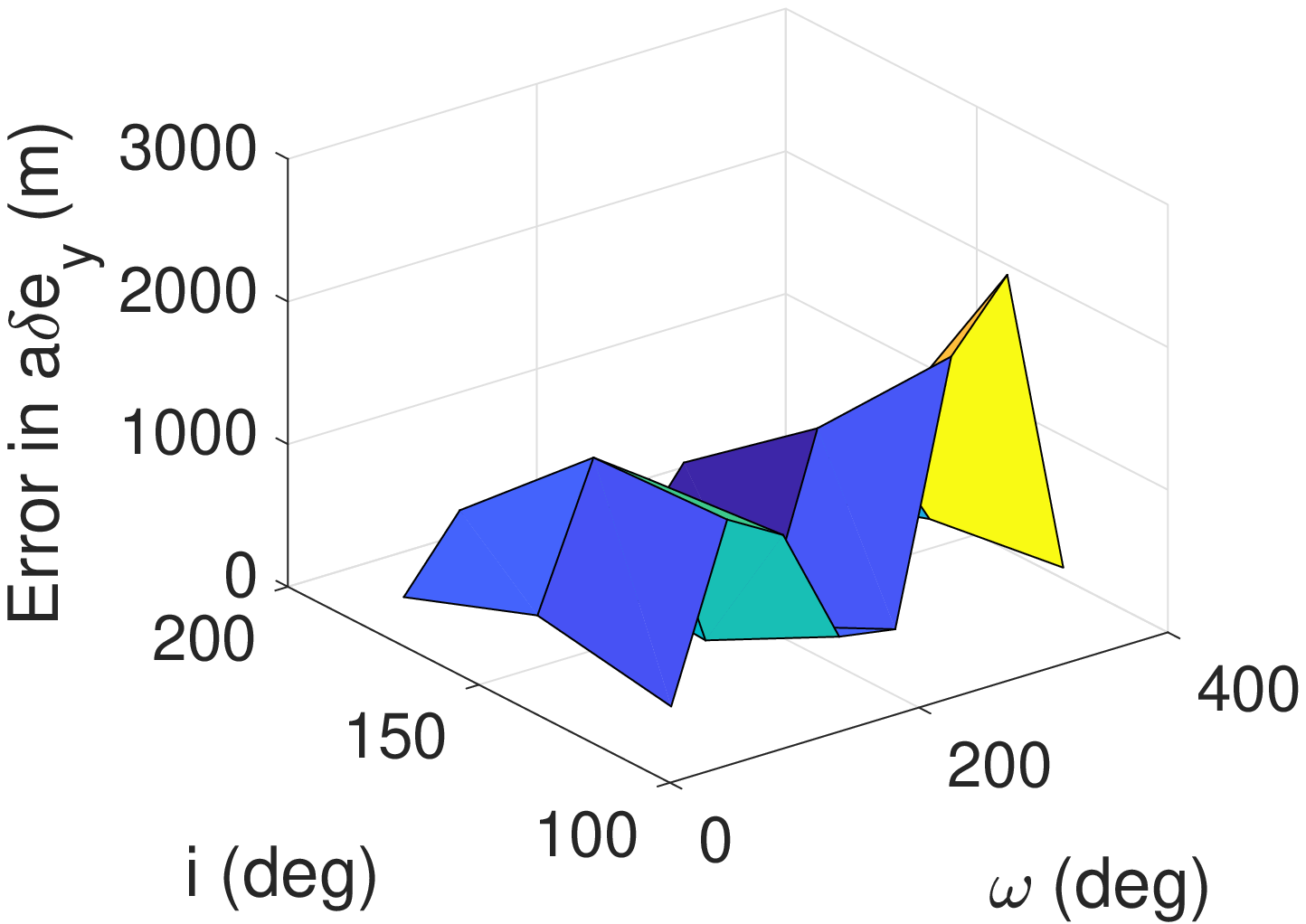}
    \end{subfigure}
    \begin{subfigure}[htb]{0.3\textwidth}
    \centering
    \caption{}
	\includegraphics[width=1.8in]{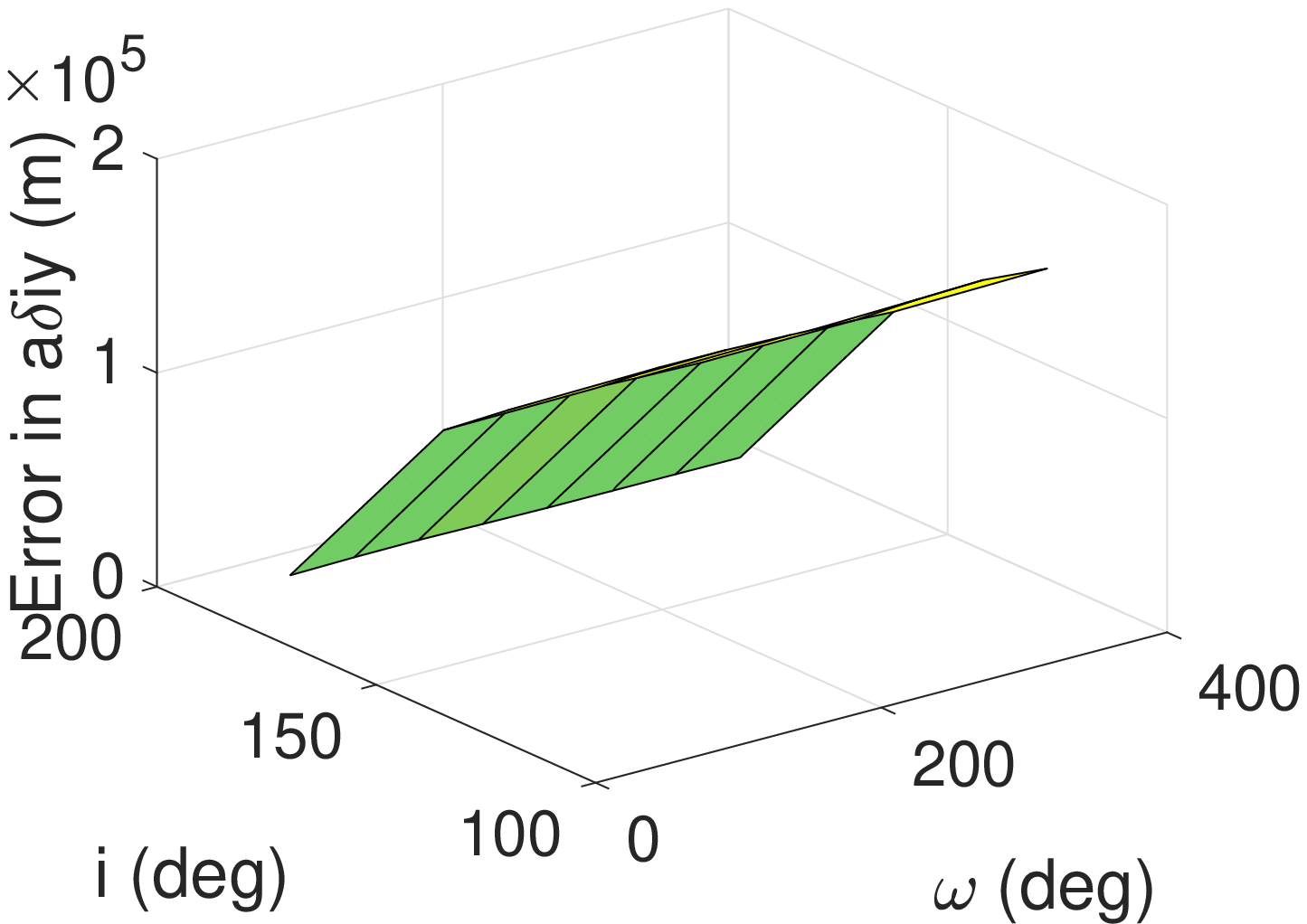}
    \end{subfigure}
	\caption{Errors in the mean absolute orbit propagation with Mahajan's\cite{Mahajan} 15th order zonal model in comparison to the ground truth represented as ROE for a near-circular orbit about an Eros-like asteroid over 5 orbits.}
	\label{fig:Mahajan}
\end{figure}

The propagation errors are provided in Figure~\ref{fig:Mahajan}.  The largest errors are those presented as $a\delta \lambda$ and $a\delta i_y$.  This is the result of the models failure to capture perturbations affecting $u$ and $\Omega$.  Notably, the error in the semimajor axis, represented as $\delta a$, is numerical because the gravity potential is a conservative force and the SRP does not produce any secular changes in the orbit's mean semimajor axis assuming a constant cross-sectional area is illuminated by the sun. Because the errors in the absolute orbit propagation are on the order of thousands of meters, the model is insufficient in predicting the absolute motion for the envisioned control applications.

Due to these limitations, the Lagrange Planetary equations given by Equation \eqref{eq:lagrange} are used here for asteroid applications with no assumption about $J_2$ dominance.   The gravity perturbations on the mean orbital elements for the gravitational potential zonal terms up to $J_4$ and including $J_2^2$ are provided in literature for the classical orbital elements\cite{Liu}.  Note that these expressions do not have singularities in eccentricity or inclination with the exception of terms associated with $J_3$.  As stated by Guffanti \cite{Guffanti}, the singularity for eccentricity exists in $\omega$ and $M$, and the singularity for equatorial orbits exists in $\Omega$ and $\omega$.  Therefore, the classical orbit derivatives are converted to quasi-nonsingular orbital element derivatives, where the singularity from the eccentricity in the $J_3$ expressions disappear due to cancellation of contributions from M and $\omega$\cite{Guffanti}.  Note that the singularity for inclination still exists but is only relevant for near-equatorial orbits, which are not ideal for the applications of ANS.  The converted expressions are found in the appendix, but a sample expression for the $e_y$ perturbation under $J_4$ is given by 
\begin{equation}
\begin{aligned}
	\label{eq:qns_sample}
   \frac{de_y}{dt} =  -\frac{15}{32} n J_4 \bigg(\frac{R_E}{a(1-(e_x^2+e_y^2))}\bigg)^4 \bigg [ \sin^2i (6-7 \sin^2i) (1-(e_x^2+e_y^2)) \frac{2e_y^2e_x}{e_x^2+e_y^2} \\
+ \bigg(16-62 \sin^2i+49 \sin^4i+ \frac{3}{4} (24-84 \sin^2(i) 
+63 \sin^4i) (e_x^2+e_y^2) \\ 
+(\sin^2i (6-7 \sin^2i)-\frac{1}{2} (12-70 \sin^2i
+63 \sin^4i) (e_x^2+e_y^2)) \frac{e_x^2-e_y^2}{e_x^2+e_y^2}\bigg)e_x \bigg ]
\end{aligned}
\end{equation}
where $R_e$ is the reference radius of the asteroid.  Remembering that $e_x$ and $e_y$ is proportional to the eccentricity, L'hopital's rule demonstrates that the terms in fractions (i.e. $\frac{2e_y^2e_x}{e_x^2+e_y^2}$) still approach 0 as eccentricity approaches 0.  The fraction approaches 0 as long as the numerator degree in eccentricity is larger than or equal to the denominator degree in eccentricity.  Therefore, for a near-circular orbit, the derivative is well defined and approaches 0 for all derivatives provided in the appendix.  

\begin{figure}[htb!]
	\centering
	\begin{subfigure}[htb]{0.3\textwidth}
	\centering
	\caption{}
	\includegraphics[width=1.8in]{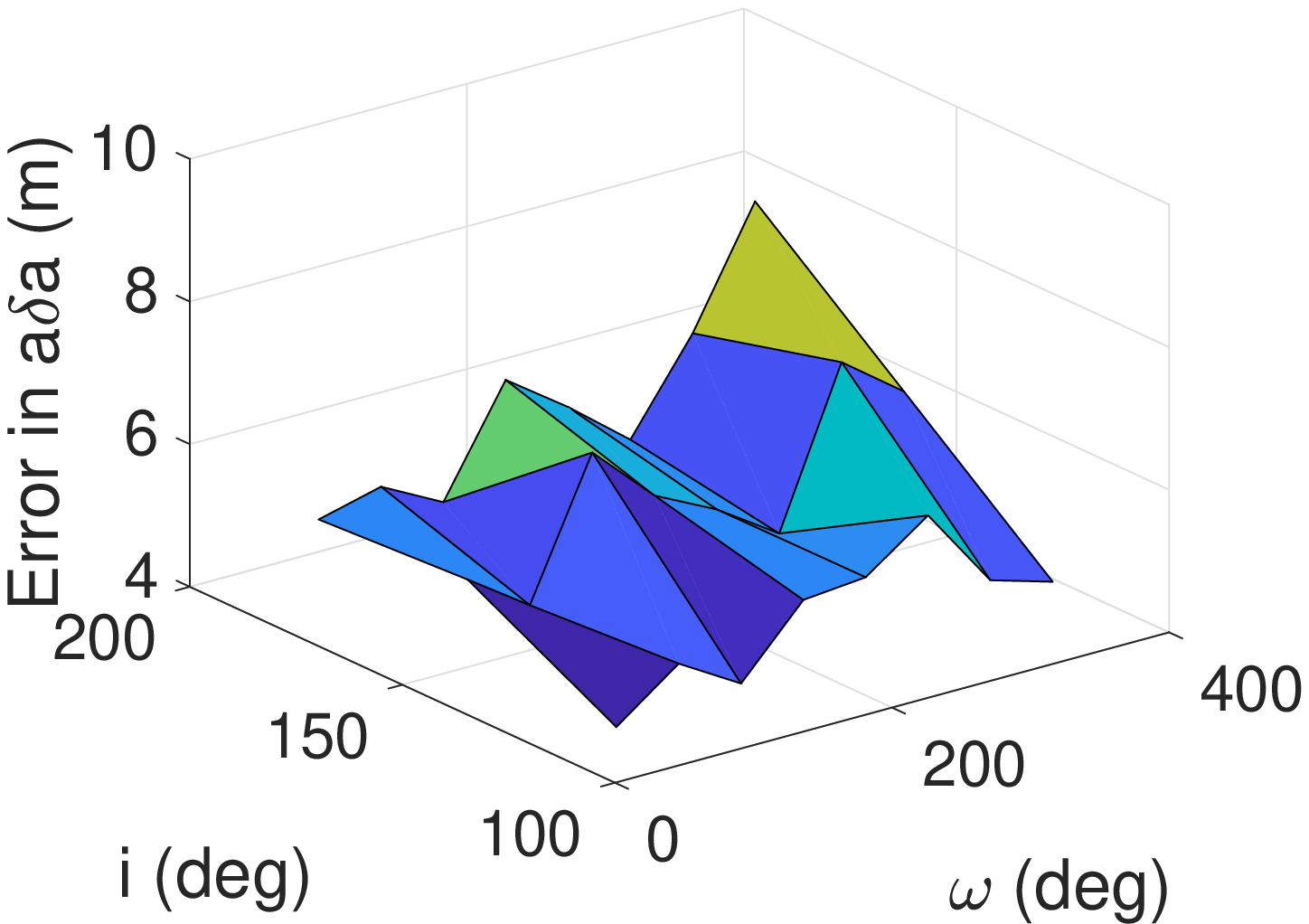}
	\end{subfigure}
	\begin{subfigure}[htb]{0.3\textwidth}
	\centering
	\caption{}
	\includegraphics[width=1.8in]{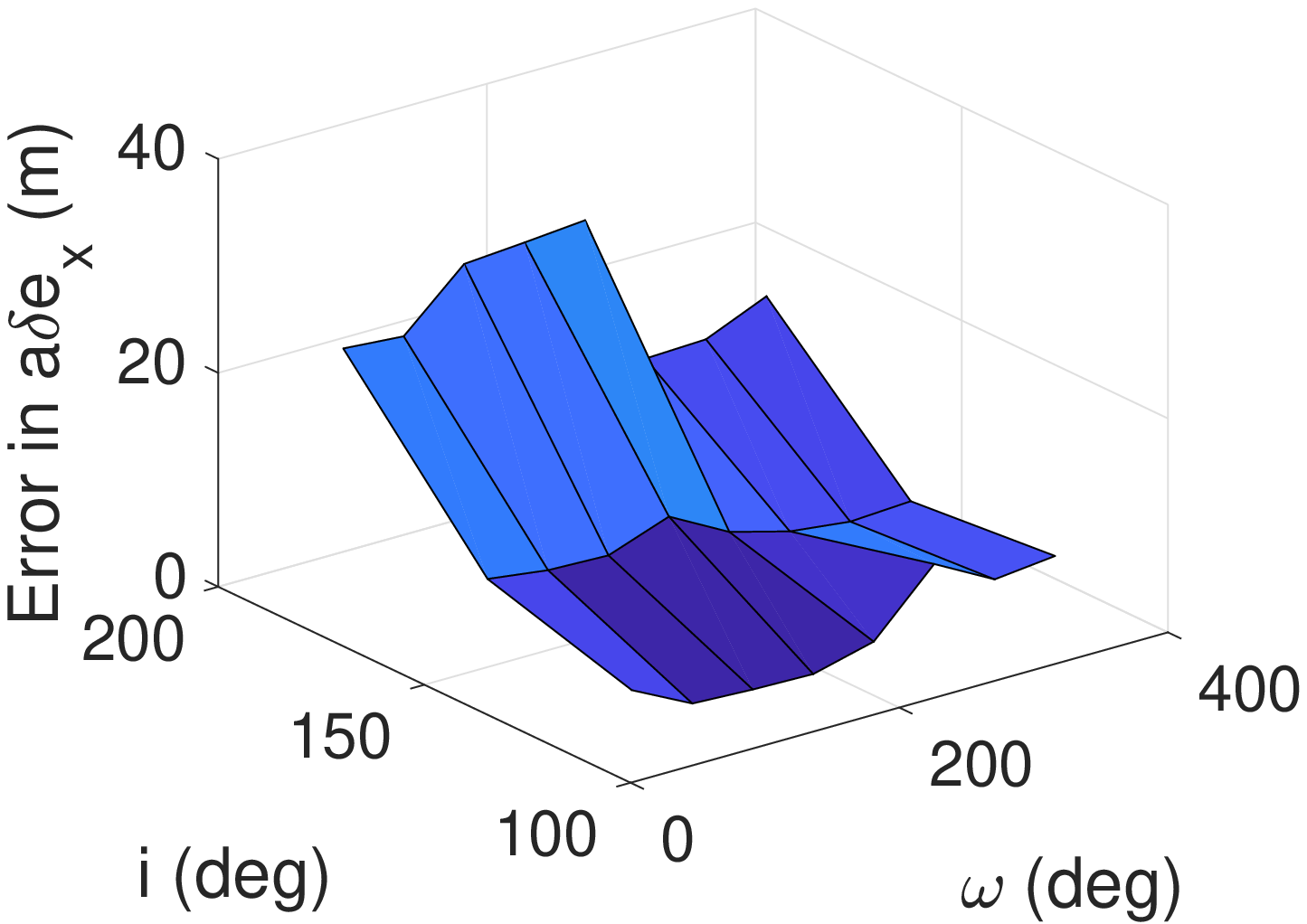}
	\end{subfigure}
	\begin{subfigure}[htb]{0.3\textwidth}
	\centering
	\caption{}
	\includegraphics[width=1.8in]{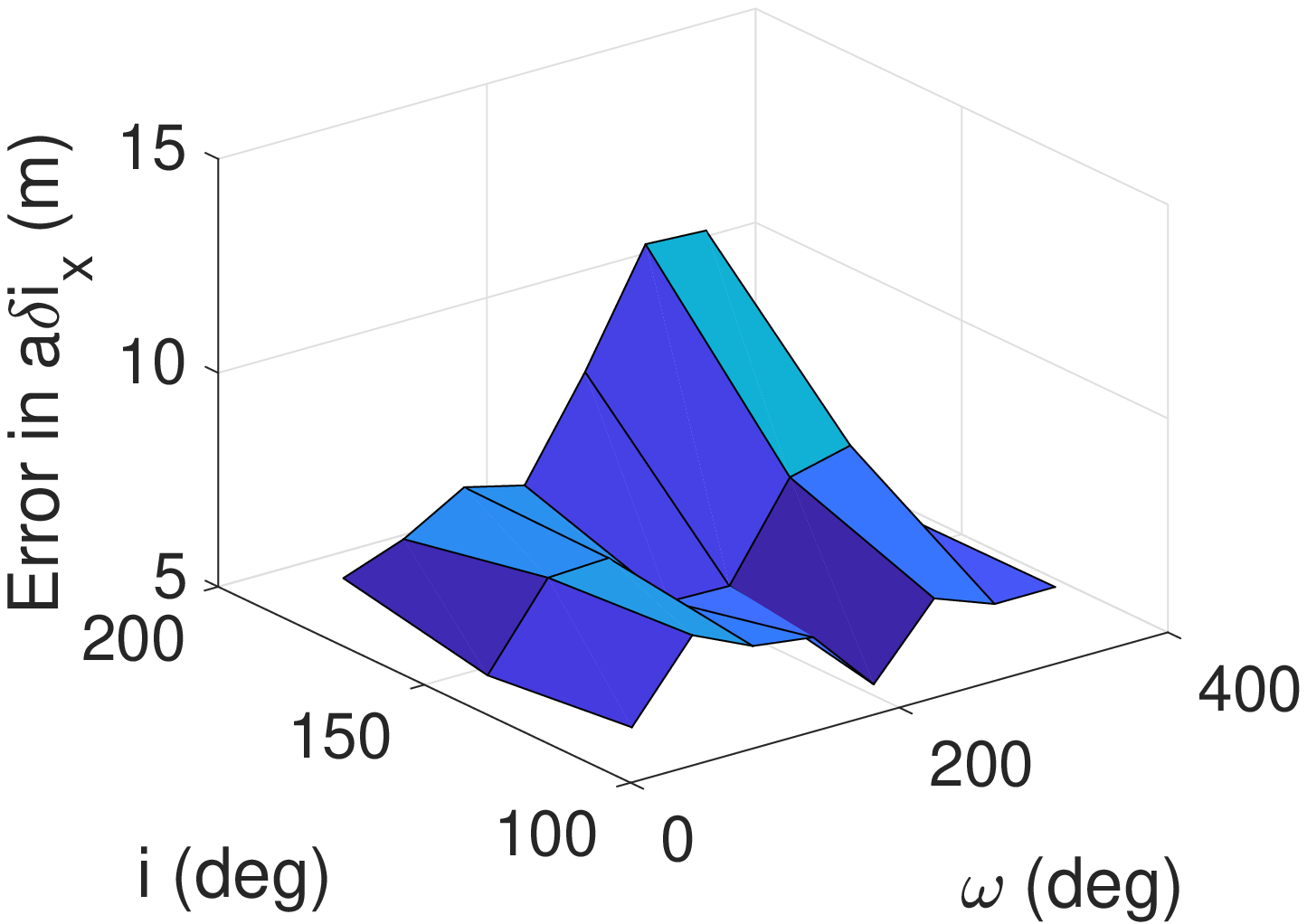}
	\end{subfigure}\\
	\centering
	\begin{subfigure}[htb]{0.3\textwidth}
	\centering
	\caption{}
	\includegraphics[width=1.8in]{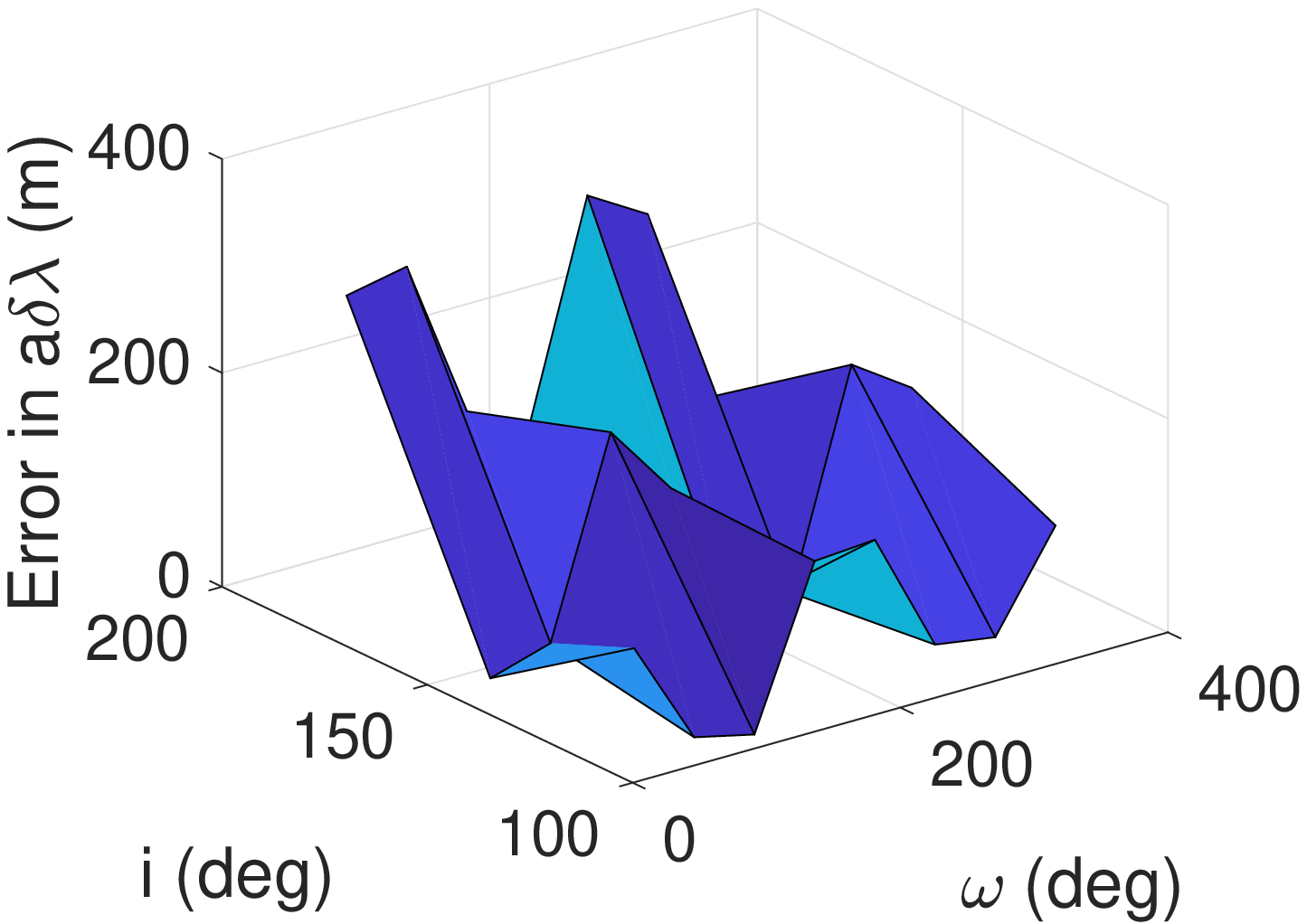}
	\end{subfigure}
	\begin{subfigure}[htb]{0.3\textwidth}
	\centering
	\caption{}
	\includegraphics[width=1.8in]{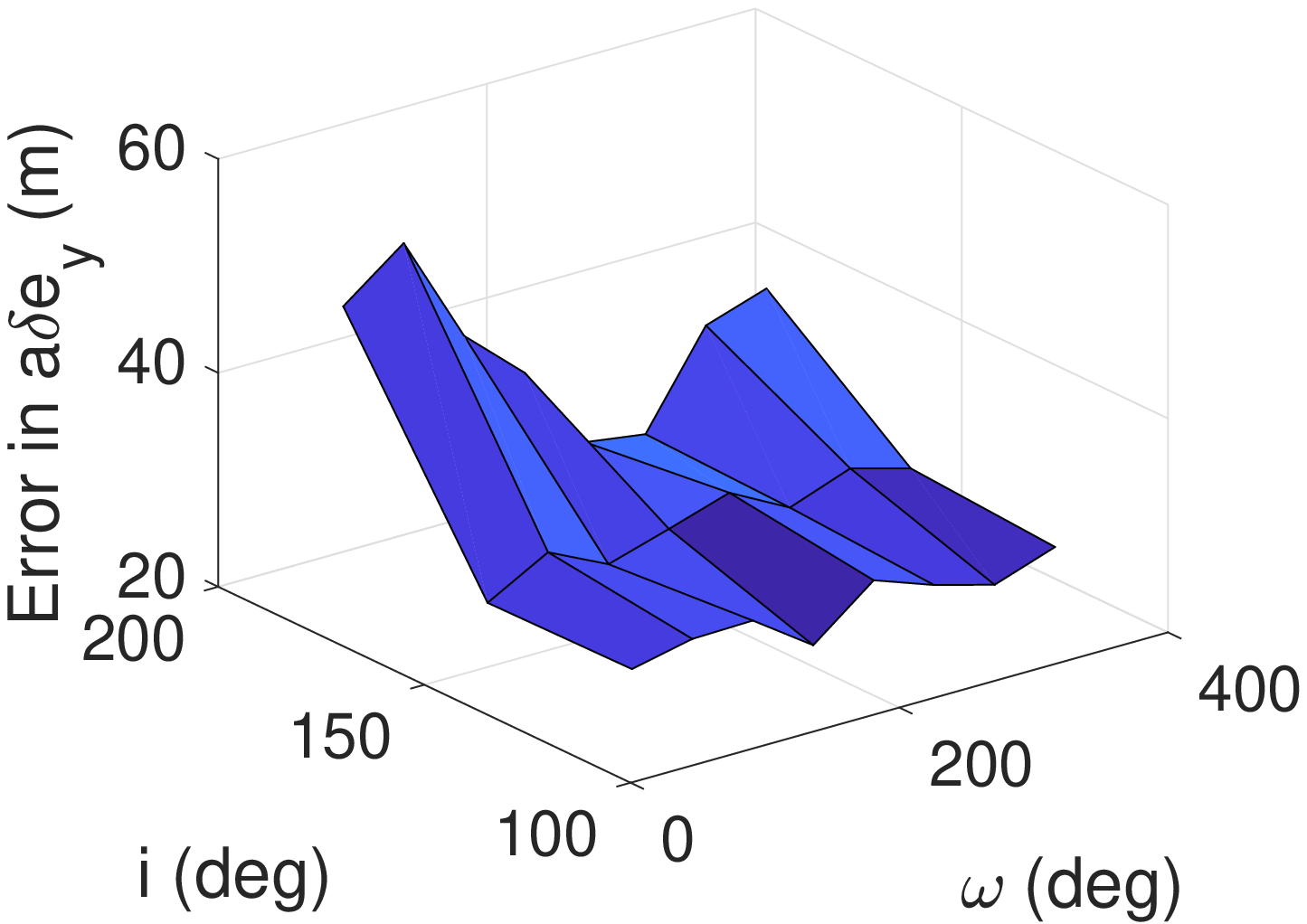}
	\end{subfigure}
	\begin{subfigure}[htb]{0.3\textwidth}
	\centering
	\caption{}
	\includegraphics[width=1.8in]{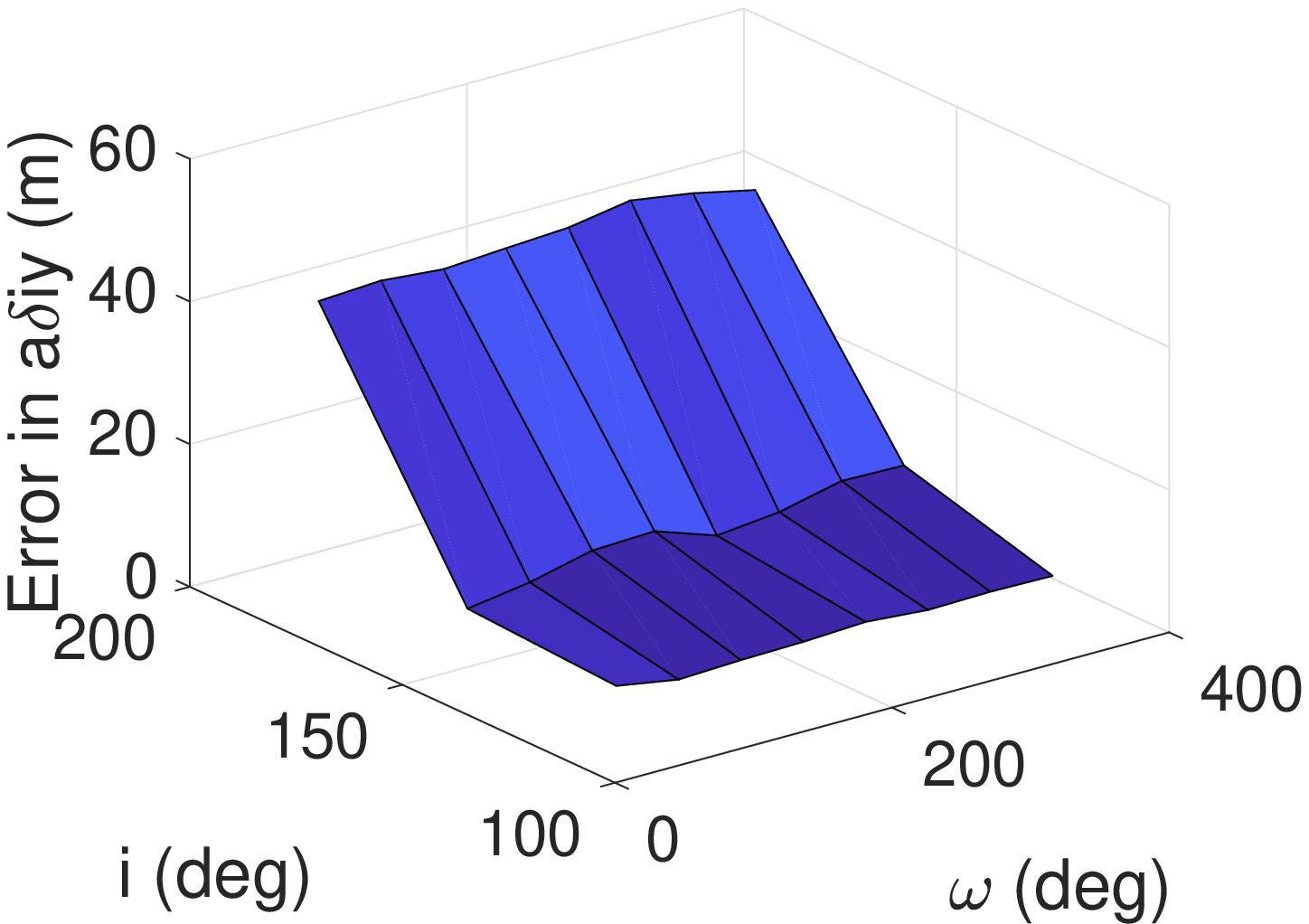}
	\end{subfigure}
	\caption{Errors in mean absolute orbit semi-analytical propagation in comparison to the ground truth represented as ROE for a near-circular orbit about an Eros-like asteroid over 5 orbits.}
	\label{fig:MEO_abs_orb}
\end{figure}

Because the resulting expressions are based on slow-varying mean parameters, a semi-analytical propagation is used for simplicity without loss in accuracy. Specifically, the semi-analytical propagator uses an Eulerian integration with a 100s time step with only 4 zonal and one SRP averaging-theory-based derivative expressions provided in the appendix for each quasi-nonsingular absolute orbital element.   To validate the propagation, the produced mean elements are compared with the same ground truth simulation described previously in Table \ref{tab:dynamic_param}.  The errors resulting from this propagation are provided in Figure \ref{fig:MEO_abs_orb}.

The error represented as $a\delta \lambda$ is the largest due to strong dynamic coupling with the error in the semimajor axis, represented as $a\delta a$.  Specifically, an error in the estimated semimajor axis produces a different orbital period.  The incorrect estimate of the orbital period results in an along-track separation, which appears as a difference in $a\delta \lambda$, over time with respect to the ground truth.  Even so, the maximum error described by $a\delta\lambda$ is less than 400 m for absolute motion prediction over 5 orbits. The remaining errors represented as ROE are all below 60 m. Thus, the developed orbit propagation meets specified requirements.  Additionally, errors in the propagation over 5 orbits decreased with increasing orbit altitude because the contributions from higher order, unmodeled gravitational potential terms decrease at a rate greater than the inverse quartic of orbit radius.

In comparison to the state-of-the-art, the semi-analytical propagator decreases errors by at least two orders of magnitude with two exceptions, $a\delta i_x$ and $a\delta a$.  Again, the non-zero $a\delta a$ error is a result of small variations in the averaged orbital elements.  However, the error presented as $a\delta i_x$  suggests Mahajan's model provides an accurate prediction of the mean inclination. Furthermore, the semi-analytical propagation requires less time to propagate the orbit because of the consideration of just 4 gravity potential zonal terms in comparison to 15 zonal terms.   The accuracy of the absolute propagation motivates the extension to a relative orbit propagation.

The relative motion can be computed by combining two absolute orbit propagations of independent spacecraft and constructing the ROE using Equation \eqref{eq:ROE}. The produced relative motion propagation must achieve a maximum error less than 30 meters in all ROE over 5 orbit periods.  This ensures that the model can be utilized over short-periods of time when navigation may not be available due to maneuvers and can be integrated into an EKF to produce a mean state estimate from osculating state measurements with meter level accuracy, as explained later. 

For the relative motion propagation, the deputy spacecraft was initialized to the osculating ROE state defined as $a\delta \alpha = (0, 0, 0, 400, 0, 400) m$, which defines an E-I vector separation formation. \cite{DAmico_thesis} Additionally, a 50\% differential ballistic coefficient was simulated.  This differential ballistic coefficient is defined as 
\begin{equation}
\begin{aligned}
	\label{eq:B_srp}
   \Delta B = \Delta B_d-\Delta B_c = \frac{C_{SRP,d}A_d}{m_d}-\frac{C_{SRP,c}A_c}{m_c}
\end{aligned}
\end{equation}
where $m$ represents the mass of the satellite, $A$ represents the cross-sectional area illuminated by the sun, and $C_{SRP}$ represents the SRP coefficient.  
\begin{figure}[htb]
	\centering
	\begin{subfigure}[htb]{0.3\textwidth}
	\centering
	\caption{}
	\includegraphics[width=1.8in]{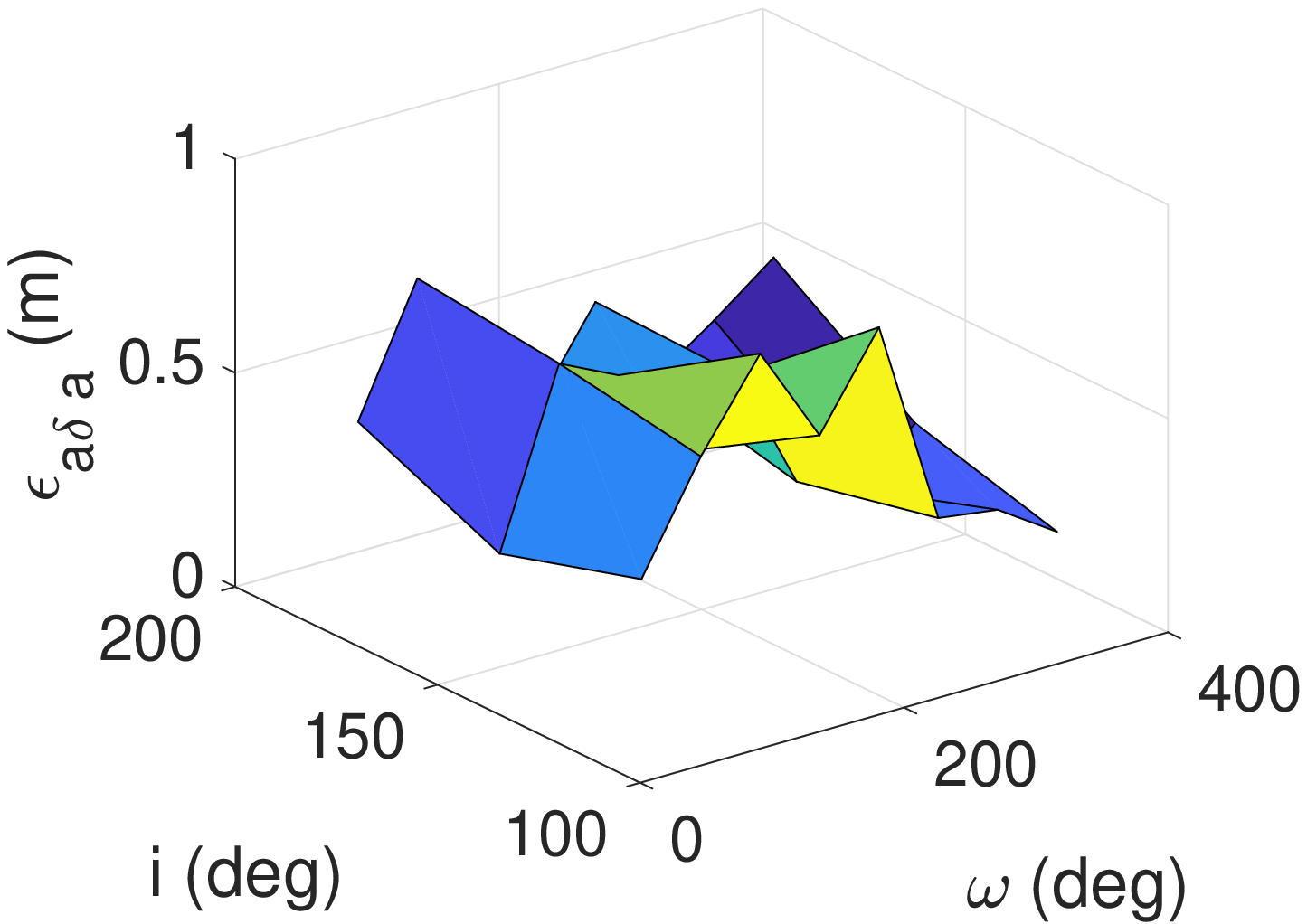}
	\end{subfigure}
	\begin{subfigure}[htb]{0.3\textwidth}
	\centering
	\caption{}
	\includegraphics[width=1.8in]{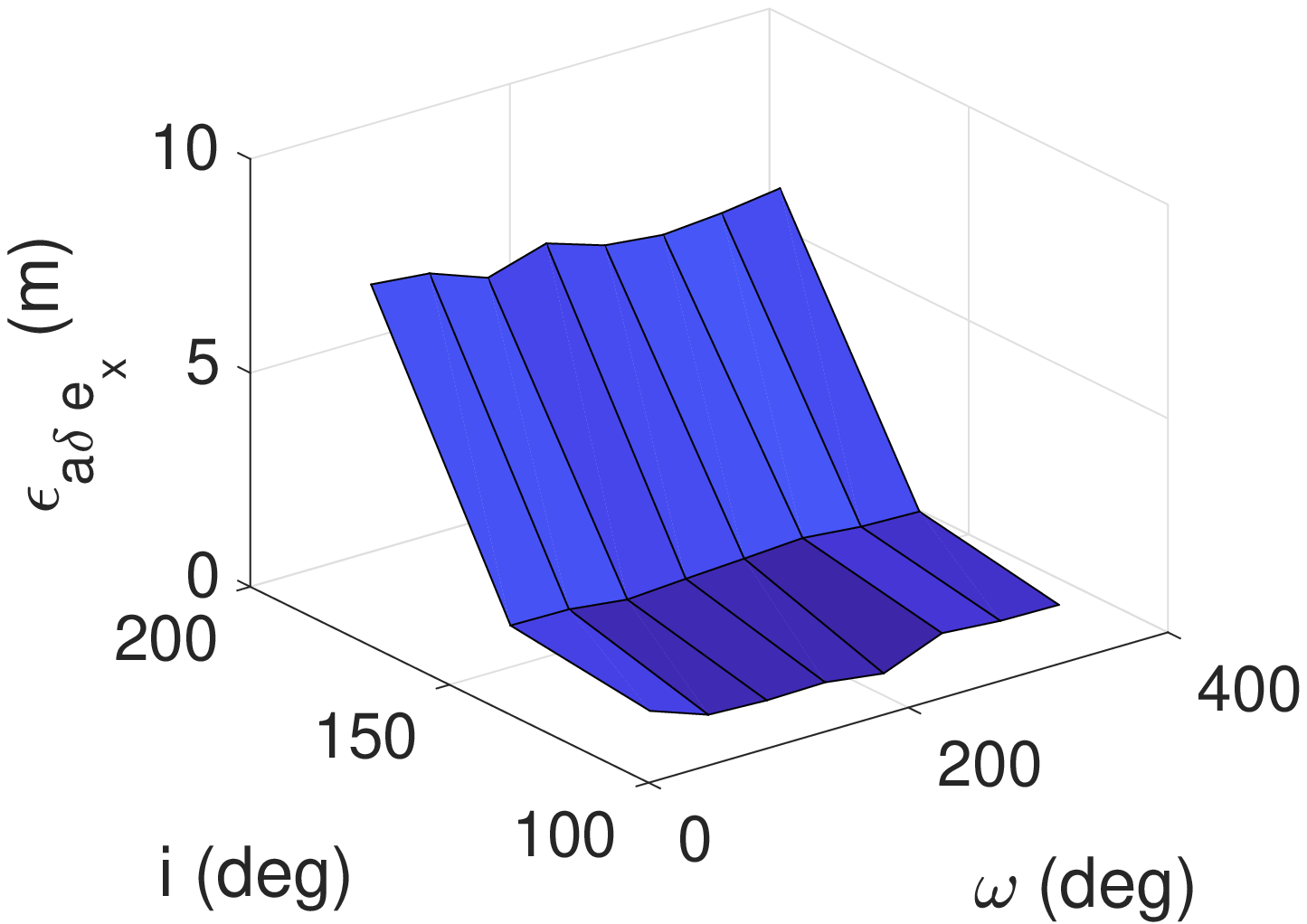}
	\end{subfigure}
	\begin{subfigure}[htb]{0.3\textwidth}
	\centering
	\caption{}
	\includegraphics[width=1.8in]{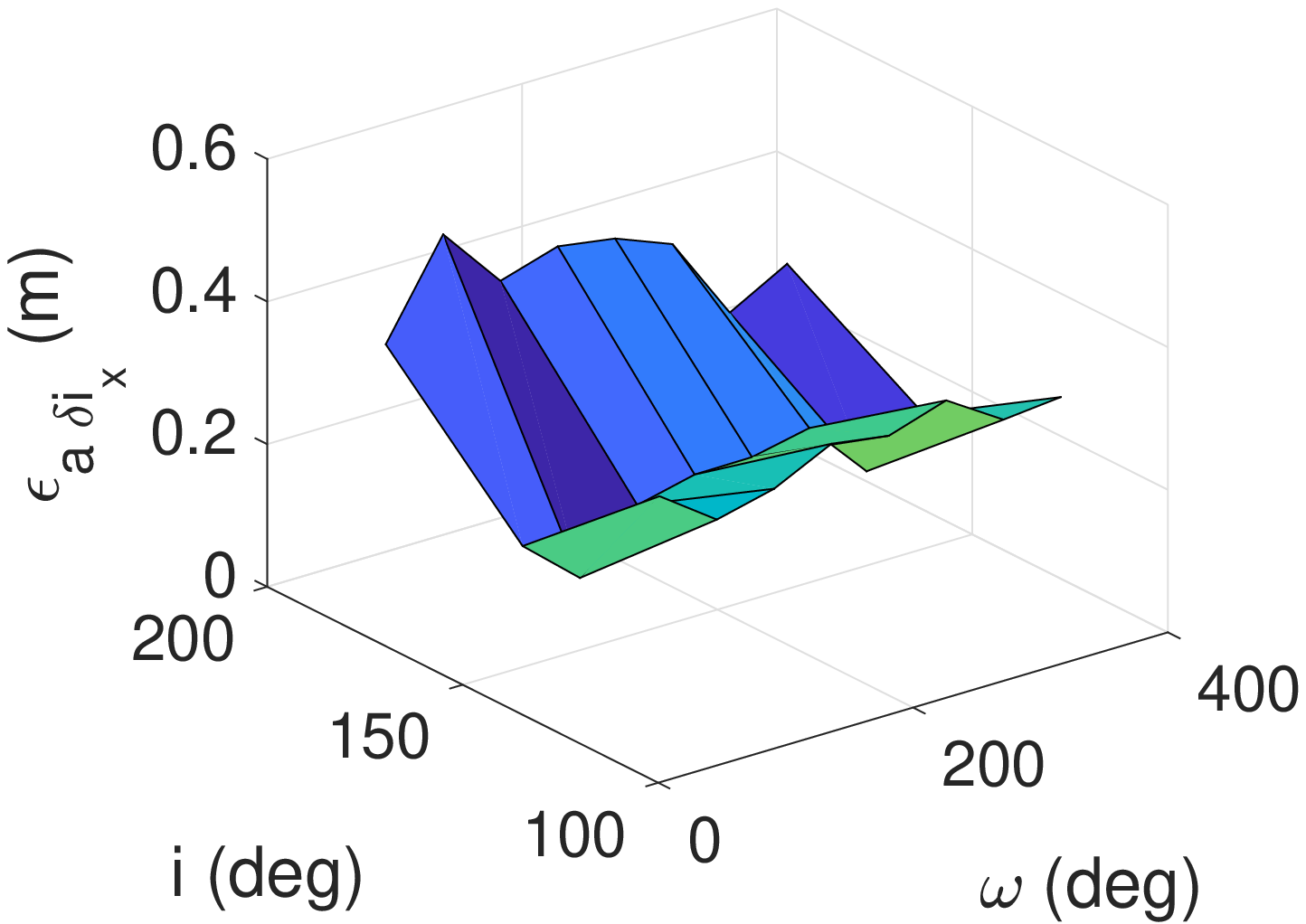}\\
	\end{subfigure}
	\centering
	\begin{subfigure}[htb]{0.3\textwidth}
	\centering
	\caption{}
	\includegraphics[width=1.8in]{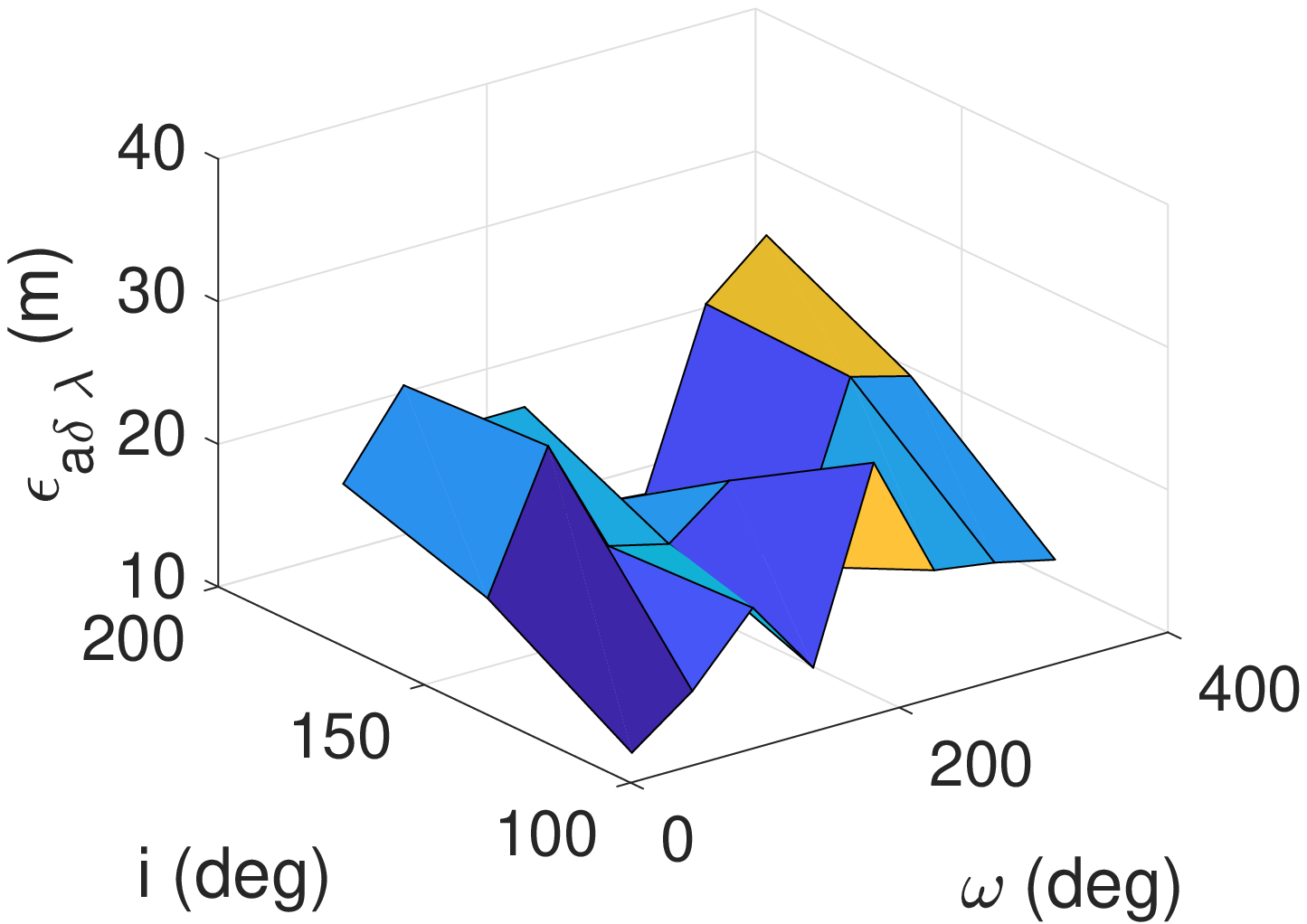}
	\end{subfigure}
	\begin{subfigure}[htb]{0.3\textwidth}
	\centering
	\caption{}
	\includegraphics[width=1.8in]{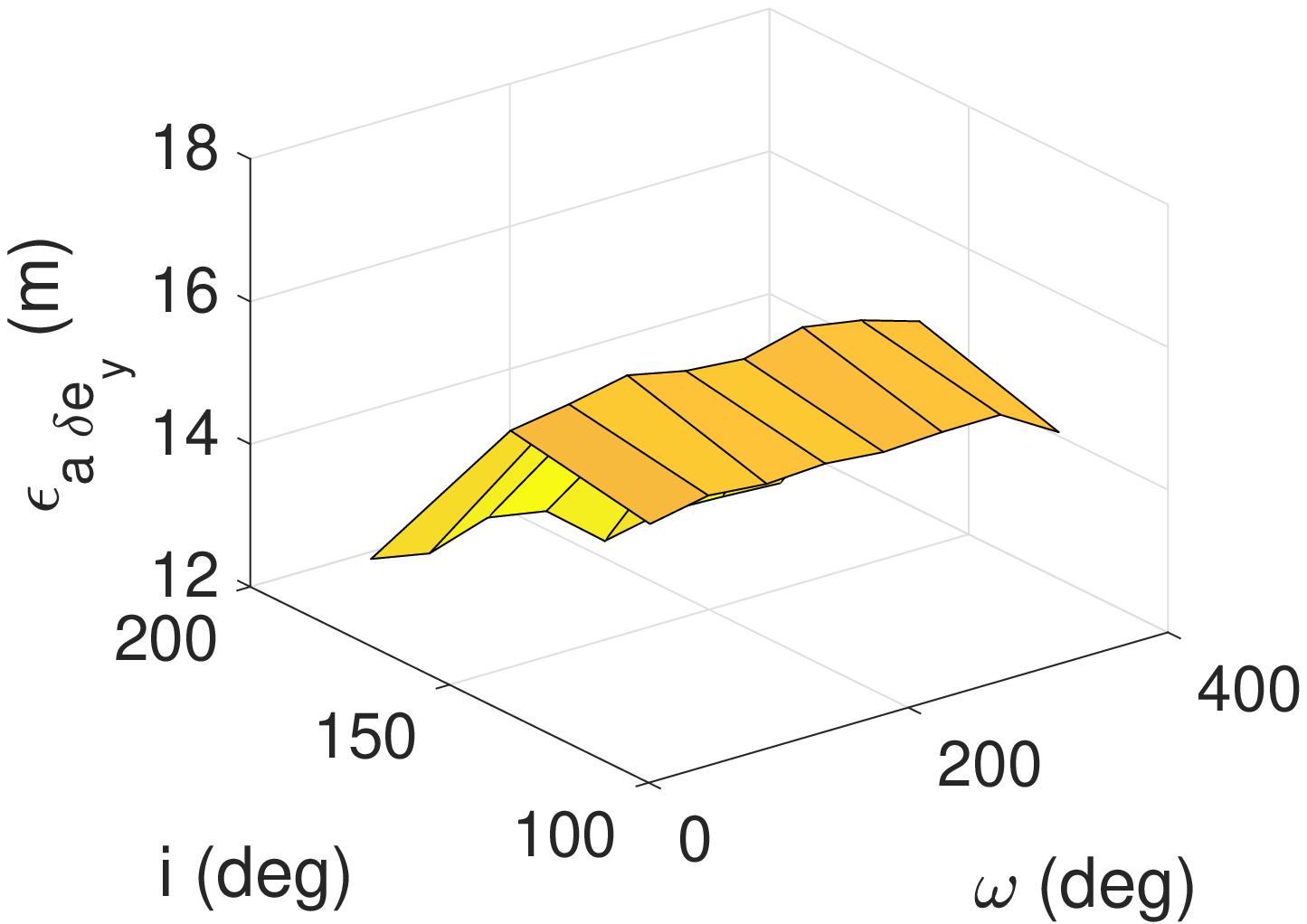}
	\end{subfigure}
	\begin{subfigure}[htb]{0.3\textwidth}
	\centering
	\caption{}
	\includegraphics[width=1.8in]{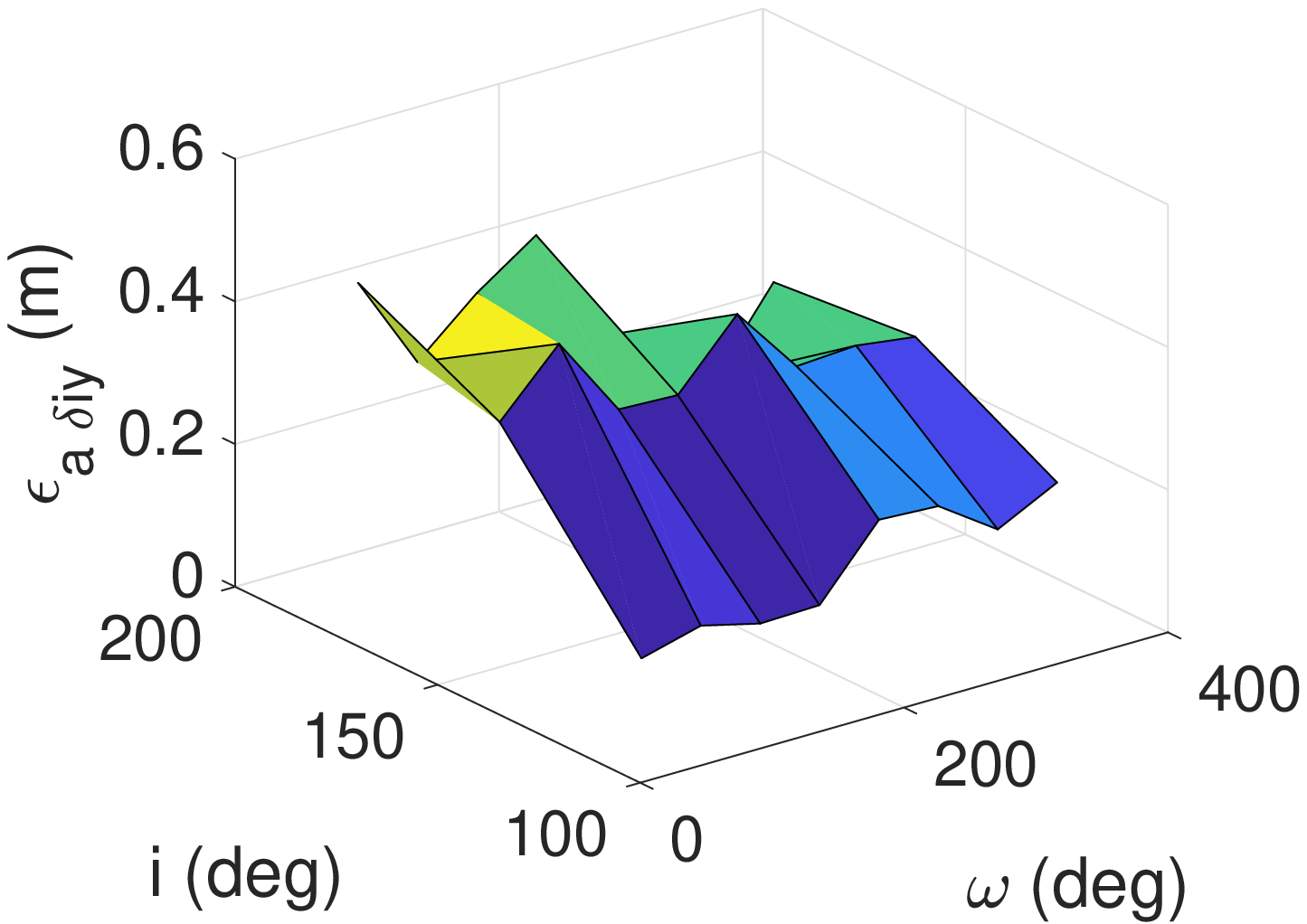}
	\end{subfigure}
	\caption{Errors in relative motion semi-analytical propagation in comparison to the ground truth for a near-circular orbit about an Eros-like asteroid over 5 orbit periods.  Errors are represented as ROE differences.}
	\label{fig:IAO_rel_orb}
\end{figure}

Figure \ref{fig:IAO_rel_orb} shows relative motion prediction errors in terms of ROE differences for the same absolute orbit previously simulated and assuming the differential ballistic coefficient is perfectly known.  For the error computation, the ground-truth ROE are constructed from the numerically averaged absolute orbital elements according to Equation \eqref{eq:ROE}. The errors in the relative motion prediction decrease by an order of magnitude in comparison to the absolute motion counterpart for all ROE except for the relative eccentricity vector.  The error in $\delta \mathbf{e}$ demonstrates a reduction by half in the x and y components. The error reduction follows from the close proximity of the two simulated spacecraft, which results in cancellations of unmodeled or mismodeled accelerations due to the location dependency of the gravity perturbation.  As expected, the relative mean longitude maintains the largest error of approximately 30 m due to coupling with the $a\delta a$ error.  This follows from a similar discussion for the absolute orbit propagation, where an error in $a\delta a$ corresponds to a mismodeling of the difference in the orbit size, and therefore the orbit period, between the two spacecraft. This causes Keplerian effects, which appear as $a\delta \lambda$ over time, to be mismodeled.  The remaining ROE have an accuracy to within 20 m over 5 orbits.  In summary, the presented dynamics model is both computationally efficient given the limited resources on small satellites by only considering $J_2-J_4$ and SRP and also accurate for autonomous control purposes even considering a worst-case scenario for the gravity potential.

\subsection{Relative Motion Trends}
The dynamics model presented in the previous section is used to isolate the individual perturbations on the relative motion. For this analysis, the instantaneous derivatives of the ROE are discussed.  These derivatives are constructed using the absolute orbit derivative expressions from the dynamics model presented in the previous section and provided in the appendix.   The time derivatives of the ROE are obtained by combining the time derivatives of the absolute orbital elements as described by\cite{Guffanti} 
\begin{equation}
\begin{aligned}
	\label{eq:ROE_deriv}
	\dot{\delta\alpha} = \begin{pmatrix}
           \dot{\delta a} \\
           \dot{\delta\lambda} \\
           \dot{\delta e_x} \\
           \dot{\delta e_y} \\
           \dot{\delta i_x} \\
           \dot{\delta i_y}
         \end{pmatrix} = \begin{pmatrix}
           (\dot{a}_d-\dot{a}_c)/a_c - \dot{a}\delta a/a_c \\
           (\dot{u}_d-\dot{u})+(\dot{\Omega}_d-\dot{\Omega}_c)\cos{i}_c-\dot{(i)}_c\delta i_y \\
           \dot{e}_{x,d}-\dot{e}_{x,c} \\
           \dot{e}_{y,d}-\dot{e}_{y,c} \\
           \dot{(i)}_{d}-\dot{(i)}_{c} \\
           (\dot{\Omega}_d-\dot{\Omega}_c)\sin{i}_c-\dot{(i)}_c\delta i_y/ \tan{i_c}
         \end{pmatrix} 
\end{aligned}
\end{equation} 

\subsubsection{Gravity Potential.} 
The relative semimajor axis and relative mean longitude experience little to no drift due to the gravity potential.  As stated previously, the gravity potential does not affect the orbit semimajor axis due to the conservative nature of gravity forces and is reaffirmed in the absolute orbital element derivatives in the appendix Geopotentials.  Therefore $\delta a$ remains unaffected according to the derivative in Equation \eqref{eq:ROE_deriv}. Furthermore, $\delta\lambda$ experiences drifts proportional to or less than $a_c^{-7/2}$ due to the gravity potential based on the derivatives of $u$, $\Omega$, and $i$ from $J_2$ through $J_4$ in the appendix Geopotentials.  The drift rate is small in comparison to Keplerian effects that are proportional to $a_c^{-3/2}$ for every 1 m difference in semimajor axis. \cite{Koenig}

\begin{figure}[htb]
	\centering
	\begin{subfigure}[htb]{0.3\textwidth}
	\centering
	\caption{}
	\includegraphics[width=1.9in]{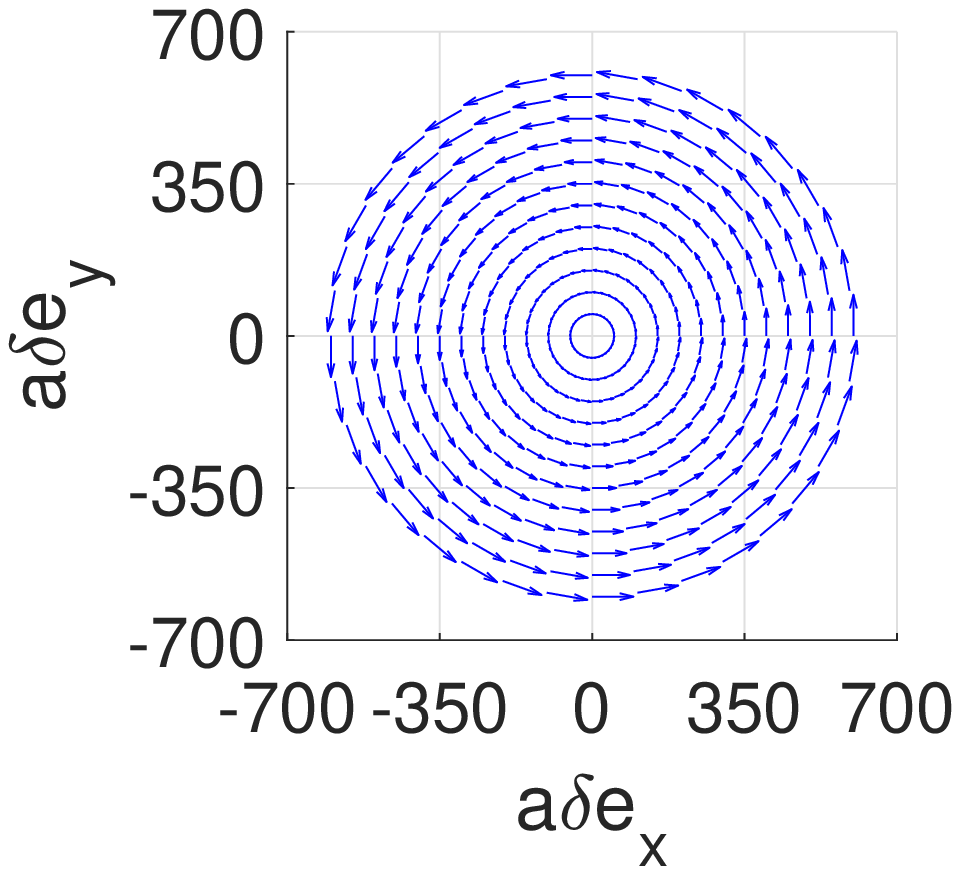}
	\end{subfigure}
	\begin{subfigure}[htb]{0.3\textwidth}
	\centering
	\caption{}
	\includegraphics[width=1.9in]{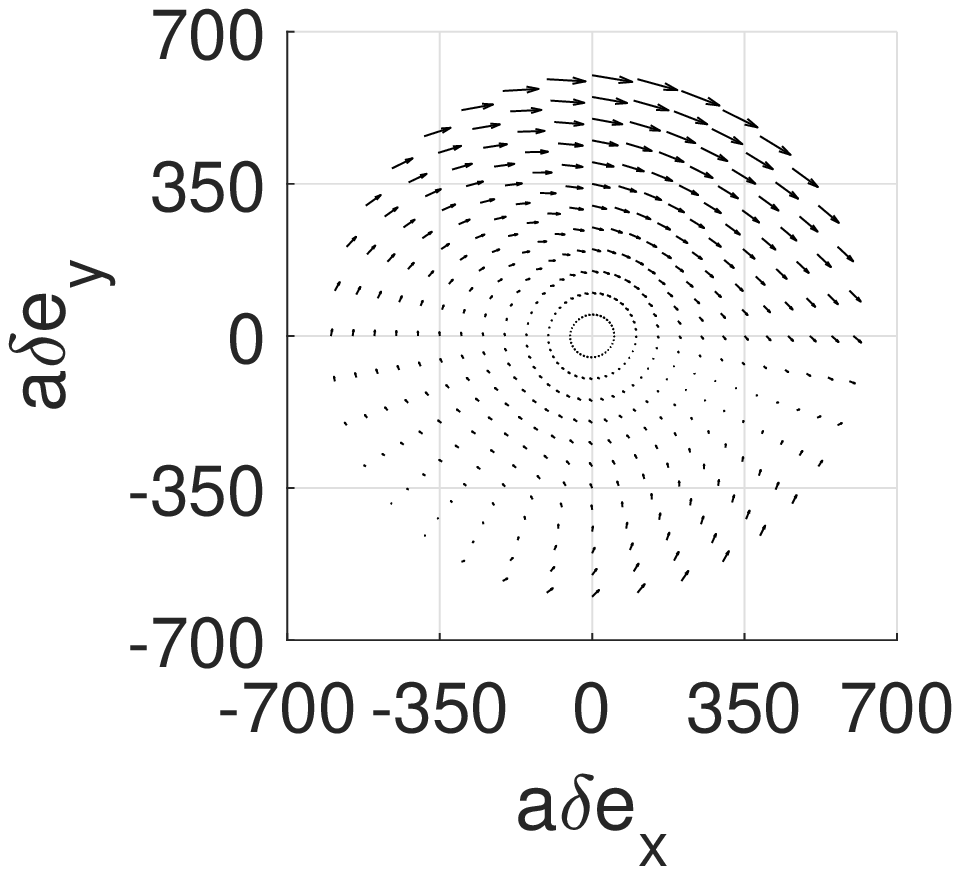}
	\end{subfigure}
	\begin{subfigure}[htb]{0.3\textwidth}
	\centering
	\caption{}
	\includegraphics[width=1.95in]{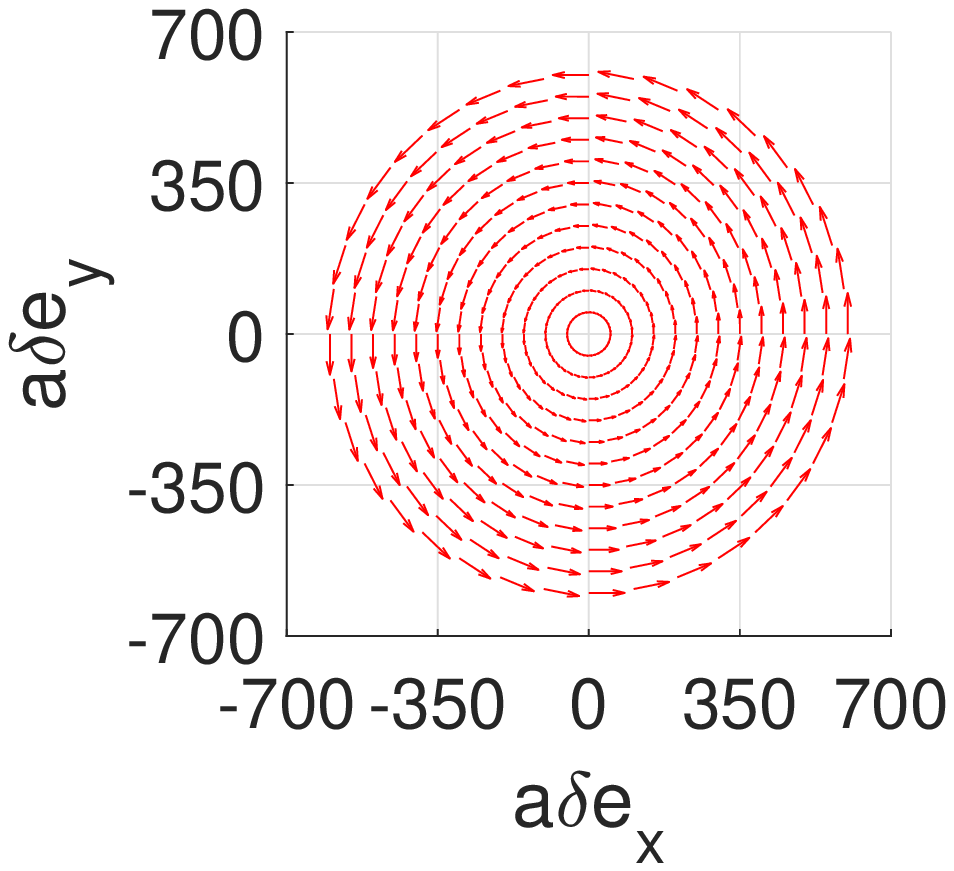}
	\end{subfigure}\\
	\centering
	\begin{subfigure}[htb]{0.3\textwidth}
	\centering
	\caption{}
	\includegraphics[width=1.9in]{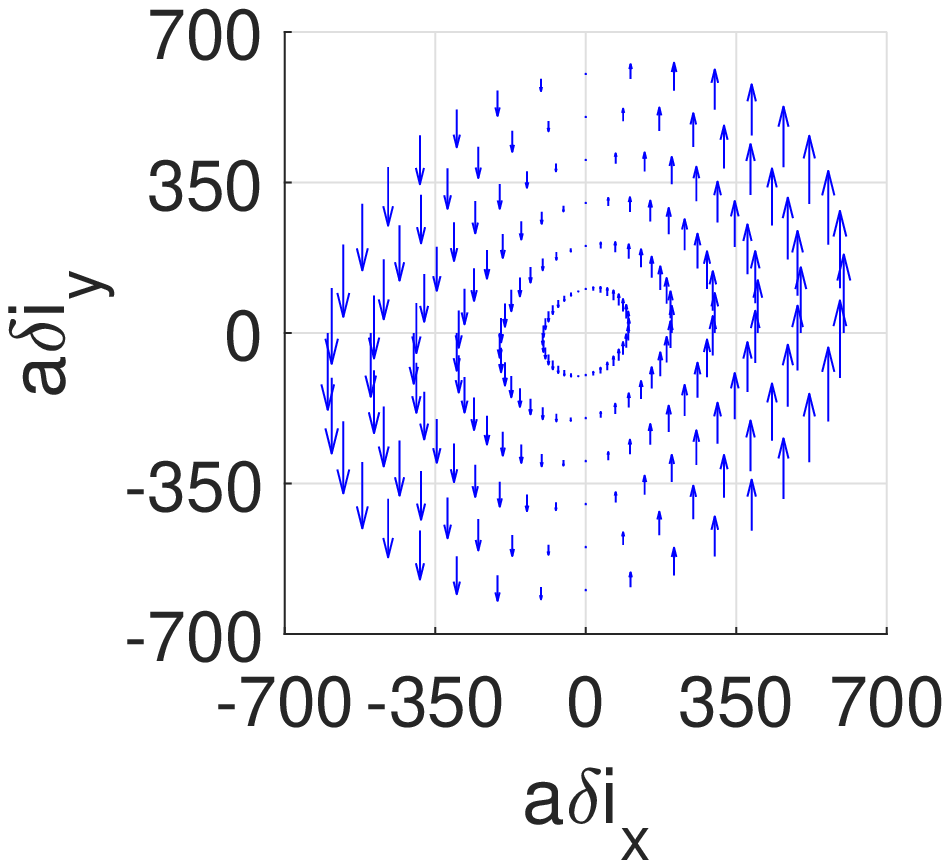}
	\end{subfigure}
	\begin{subfigure}[htb]{0.3\textwidth}
	\centering
	\caption{}
	\includegraphics[width=1.9in]{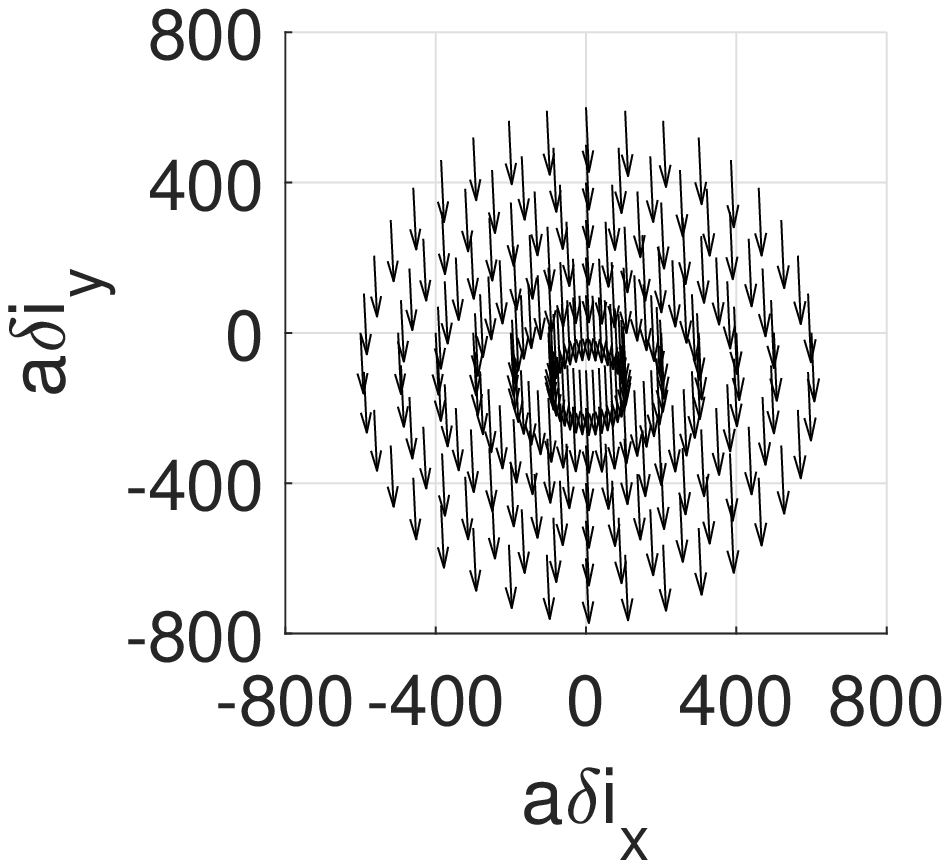}
	\end{subfigure}
	\begin{subfigure}[htb]{0.3\textwidth}
	\centering
	\caption{}
	\includegraphics[width=1.95in]{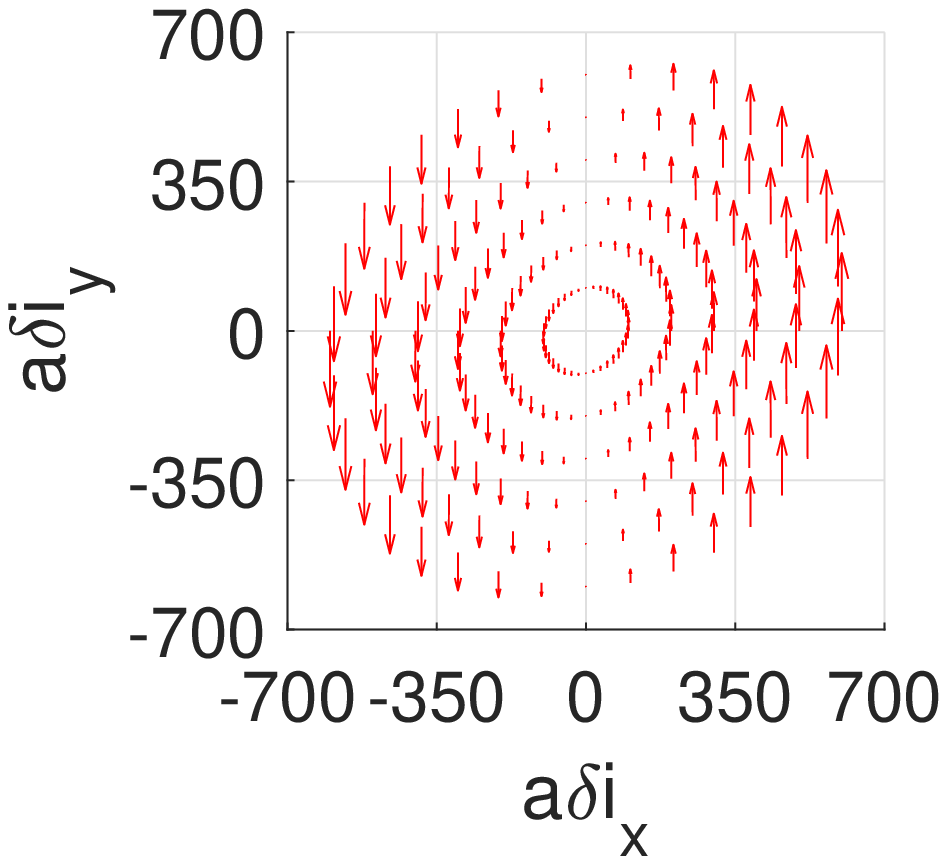}
	\end{subfigure} \\
	\caption{Secular and long period drifts caused by the $J_2$, $J_3$, and $J_4$ on the mean relative inclination and relative eccentricity vectors as defined by Equation \eqref{eq:ROE_deriv} and by the absolute orbital element time derivatives provided in the appendix Geopotentials.  The arrows represent instantaneous perturbations for a vector defined to the base of the arrow.}
	\label{fig:gp_drift}
\end{figure}

Comparatively, the drifts in the relative eccentricity and relative inclination vectors vary significantly depending on their orientation and the absolute orbit. Figure~\ref{fig:gp_drift} shows perturbing effects due to $J_2$, $J_3$, and $J_4$ on the relative motion for a near-circular orbit as obtained from Equation \eqref{eq:ROE_deriv} combined with the absolute orbital elment time derivatives in the appendices Geopotentials and SRP.  Specifically, the arrows in Figure~\ref{fig:gp_drift} are simply meant to indicate instantaneous directions and relative magnitude of drifts as affected by the orientation of the relative inclination and relative eccentricity vectors for each zonal term.  Therefore, this visualization is independent of specific asteroid properties.

A dominant trend is the rotation of the relative eccentricity vector caused by $J_2$ and $J_4$.  The contribution from each zonal perturbation is dependent on the sign of the zonal term and the absolute orbit inclination.  The rotational effect from the $J_2$ perturbation on the relative eccentricity vector has been derived previously \cite{Koenig, Guffanti,DAmico_thesis}, but the rotational effect of $J_4$ has not been discussed in literature. As noted by D'Amico \cite{DAmico_thesis}, the rotational drift rate of the relative eccentricity vector is equal to the drift rate of the argument of periapsis $\omega$.  This arises from the fact that the absolute orbit eccentricty vectors of both the chief and deputy rotate at the same rate to the first order, and therefore the difference vector rotates at the same rate.  As a result, the rotation rate of $\delta \mathbf{e}$ is equivalent to the secular drift of the argument of perigee according to Liu\cite{Liu} with the second-order terms in eccentricity ignored for a near-circular orbit.  The rate of change of the relative eccentricity vector phase $\phi$ is described as
\begin{equation}
\begin{aligned}
	\label{eq:de_drift}
    \dot{\phi}_{J_{4}} = \frac{-15}{32} J_4\frac{\sqrt{\mu}R_e^4}{a^{11/2}\eta^8}(16 - 62\sin{i_c}^2+49\cos{i_c}^4)\\
    \eta = \sqrt{1-e^2}
\end{aligned}
\end{equation}
where $\mu$ represents the asteroid gravitational parameter.

In relative inclination space, dominant trends are constant vertical drifts due to $J_2$ and $J_4$ that are proportional to the magnitude of $\delta i_x$. However, the directions of these drifts depend on the absolute orbit inclination and sign of the gravity potential term.  The expression for the drift due to $J_2$ has been developed by Guffanti\cite{Guffanti}. Guffanti's approach is reused here by taking a first-order Taylor expansion of the $\delta i_y$ perturbation induced by $J_4$ about zero separation.  This expansion is described by
\begin{equation}
\begin{aligned}
	\label{eq:diy_drift}
    \dot{\delta i_y} = \frac{\partial \dot{\delta i_y}}{\partial \delta i_x}\bigg\rvert_{\delta\mathbf{i} = 0}\delta i_x
\end{aligned}
\end{equation}

where $\dot{\delta i_y}$ refers to the secular perturbations of $J_4$ on the y-component of the relative inclination vector, which is constructed from the equations in the appendix Geopotentials and Equation \eqref{eq:ROE_deriv}.  The simplified expression is defined by 
\begin{equation}
\begin{aligned}
	\label{eq:diy_drift}
    \dot{\delta i_y} = \frac{15}{16} J_4 \frac{\sqrt{\mu}}{a^{11/2}} \bigg( \frac{R_e}{(1-(e_x^2+e_y^2))}\bigg)^4 (\sin{i_c}(4-7\sin{i_c}^2)+14(\sin{i_c}-\sin{i_c}^3)) \delta i_x
\end{aligned}
\end{equation}

In addition, the $J_3$ parameter induces a drift in the relative inclination vector that is proportional to eccentricity and therefore negligible for near-circular orbits.

\subsubsection{SRP.} 
As with the gravitational potential, the mean semimajor axis and therefore the mean $\delta a$ are unchanged under the assumptions of constant cross-section area and no eclipses for SRP. Also, the drift of $\delta \lambda$ from SRP constructed using the absolute orbital element derivatives in the appendix SRP is small compared to Keplerian effects, since the product of the solar flux and cross-sectional area is small in comparison to $\sqrt{\mu} a^{-3/2}$.
 
\begin{figure}[htb]
	\centering
	\begin{subfigure}[htb]{0.35\textwidth}
	\centering
	\caption{}
	\includegraphics[width=2.2in]{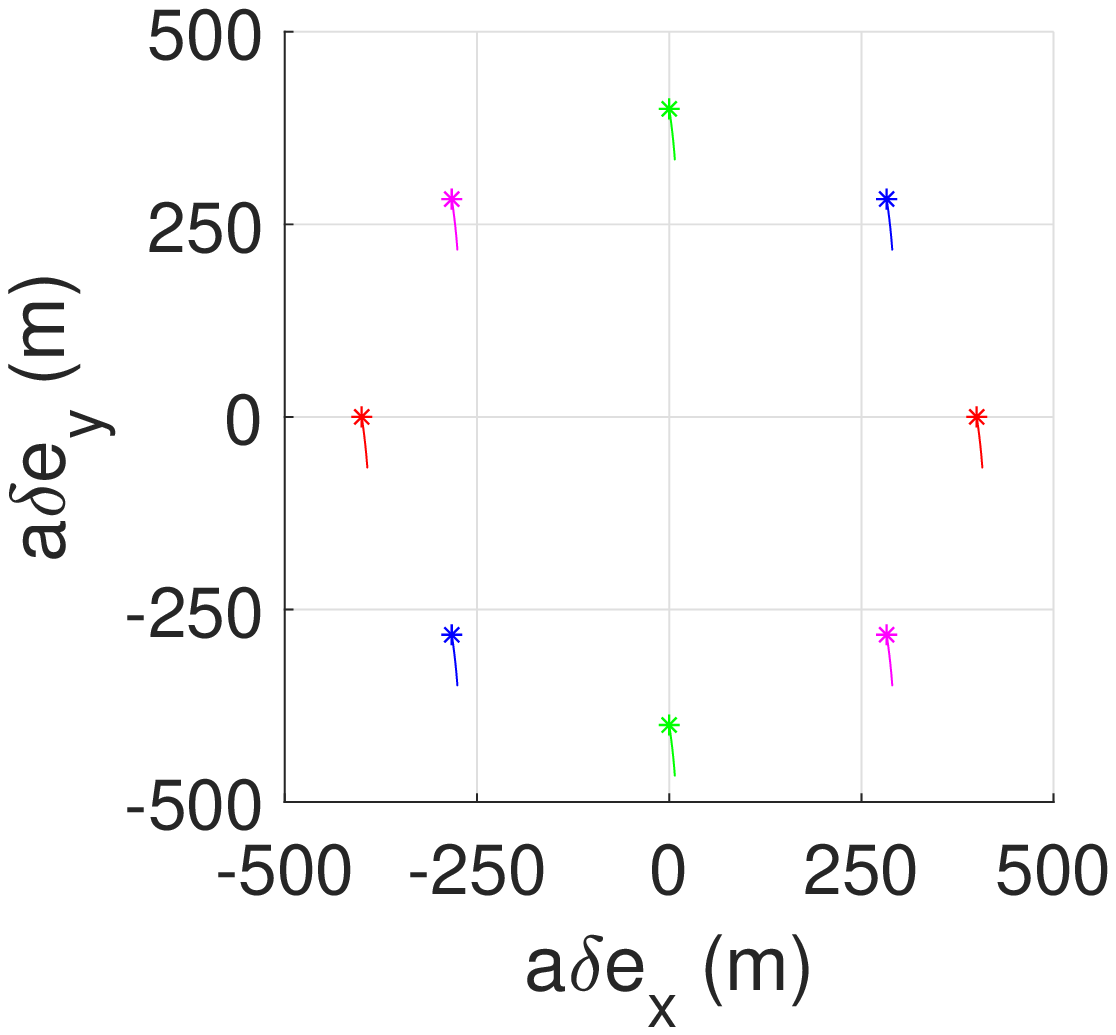}
	\end{subfigure}
	\begin{subfigure}[htb]{0.35\textwidth}
	\centering
	\caption{}
	\includegraphics[width=2.2in]{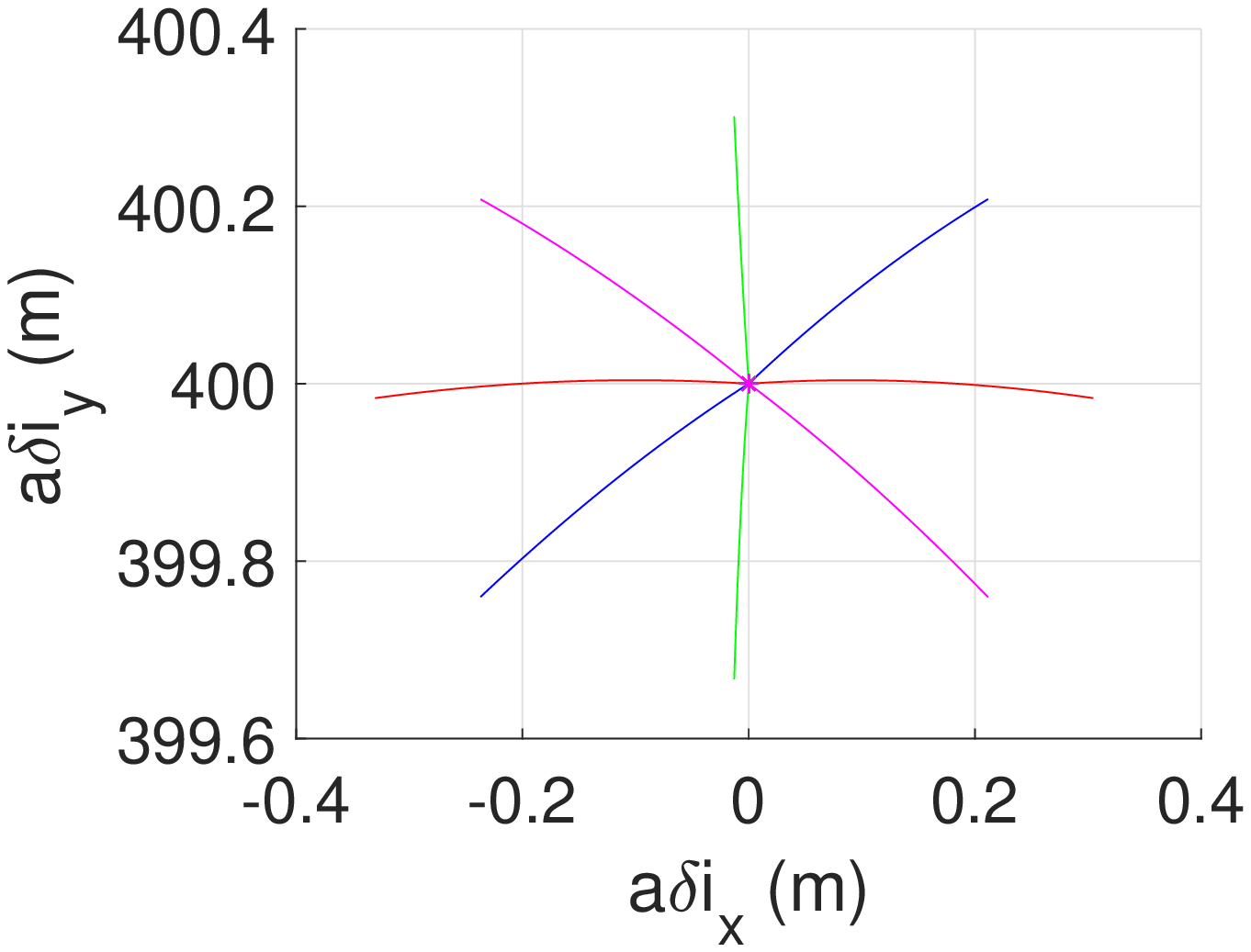}
	\end{subfigure}
	\caption{Secular drift in the mean relative eccentricity (subfigure a) and relative inclination (subfigure b) vectors  caused by SRP in a near-circular orbit about Eros for five orbit periods.  The asterisks indicate initial conditions.}
	\label{fig:SRP_de}
\end{figure}

Comparatively, the $\delta \mathbf{e}$ and $\delta \mathbf{i}$ drifts vary more significantly with the ROE and the absolute orbit. Therefore $\delta \mathbf{e}$ and $\delta \mathbf{i}$ are the focus of this section.  The perturbing effects of SRP repeat every orbit of the asteroid about the sun, which results in the $\delta\mathbf{e}$ and $\delta\mathbf{i}$ precessing in their respective planes over the asteroid's orbital period.  However, the drifts in $\delta\mathbf{e}$ and $\delta\mathbf{i}$ appear to be linear over fractions of the asteroid orbit period.  Figure~\ref{fig:SRP_de} shows the drift of $\delta\mathbf{e}$ and $\delta\mathbf{i}$ produced by a 0.002 $m^2/kg$ (50\%) differential ballistic coefficient over 5 orbits at an inclination of $135\degree$ produced by a semi-analytical propagation including only the SRP perturbation.  
 In Figure~\ref{fig:SRP_de}, the asterisk represents the initial conditions and the solid lines represent the drifts under SRP perturbations. Notably, the drift of $\delta\mathbf{e}$ is independent of the vector orientation and is approximately 70 m.  Comparatively, $\delta\mathbf{i}$ drifts in the direction of the relative eccentricity vector but is less than 1 m in magnitude.  The drift increases as the chief orbit eccentricity increases.  While the drift may be small, coupling with strong gravity effects can cause significant deviation.  These trends indicate the kinds of drifts that the proposed formation-keeping profile must counter, which are explored in a later section.

\section{Osculating-to-Mean Conversion}
To meet the second requirement of ANS, the osculating ROE state measurements, provided by the navigation systems, must be converted to a mean ROE state estimate.   The estimate must vary from the true mean state to within 10 meters for all ROE with the exception of $a\delta a$, which can vary up to 3 m.  This enables control to desired levels, as explained in the next section about guidance and control.  This section first presents results from the state-of-the-art models then moves to a description of the osculating-to-mean EKF developed in this work.

\begin{figure}[htb]
	\centering
	\begin{subfigure}[htb]{0.3\textwidth}
	\centering
	\caption{}
	\includegraphics[width=1.7in]{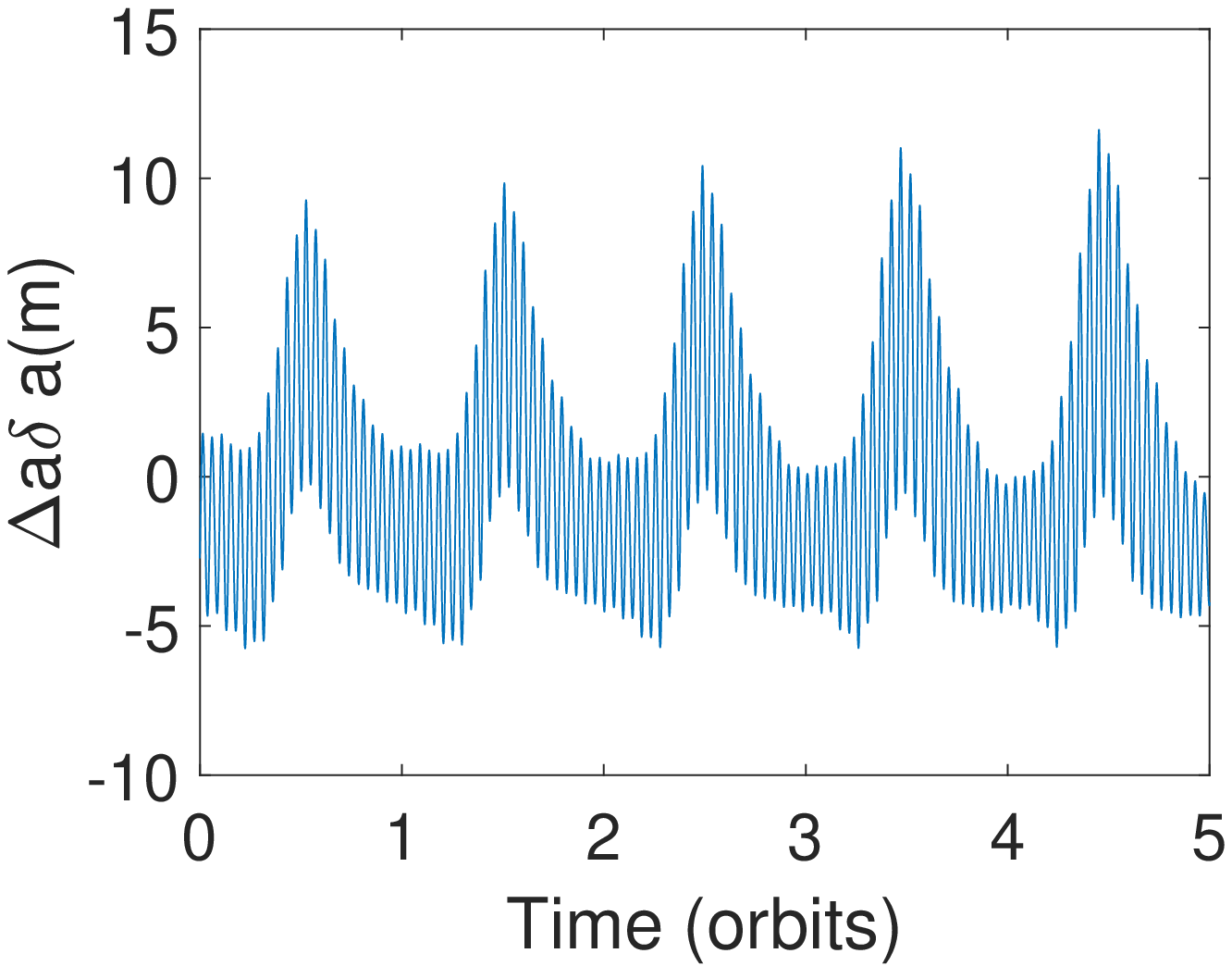}
	\end{subfigure}
	\begin{subfigure}[htb]{0.3\textwidth}
	\centering
	\caption{}
	\includegraphics[width=1.7in]{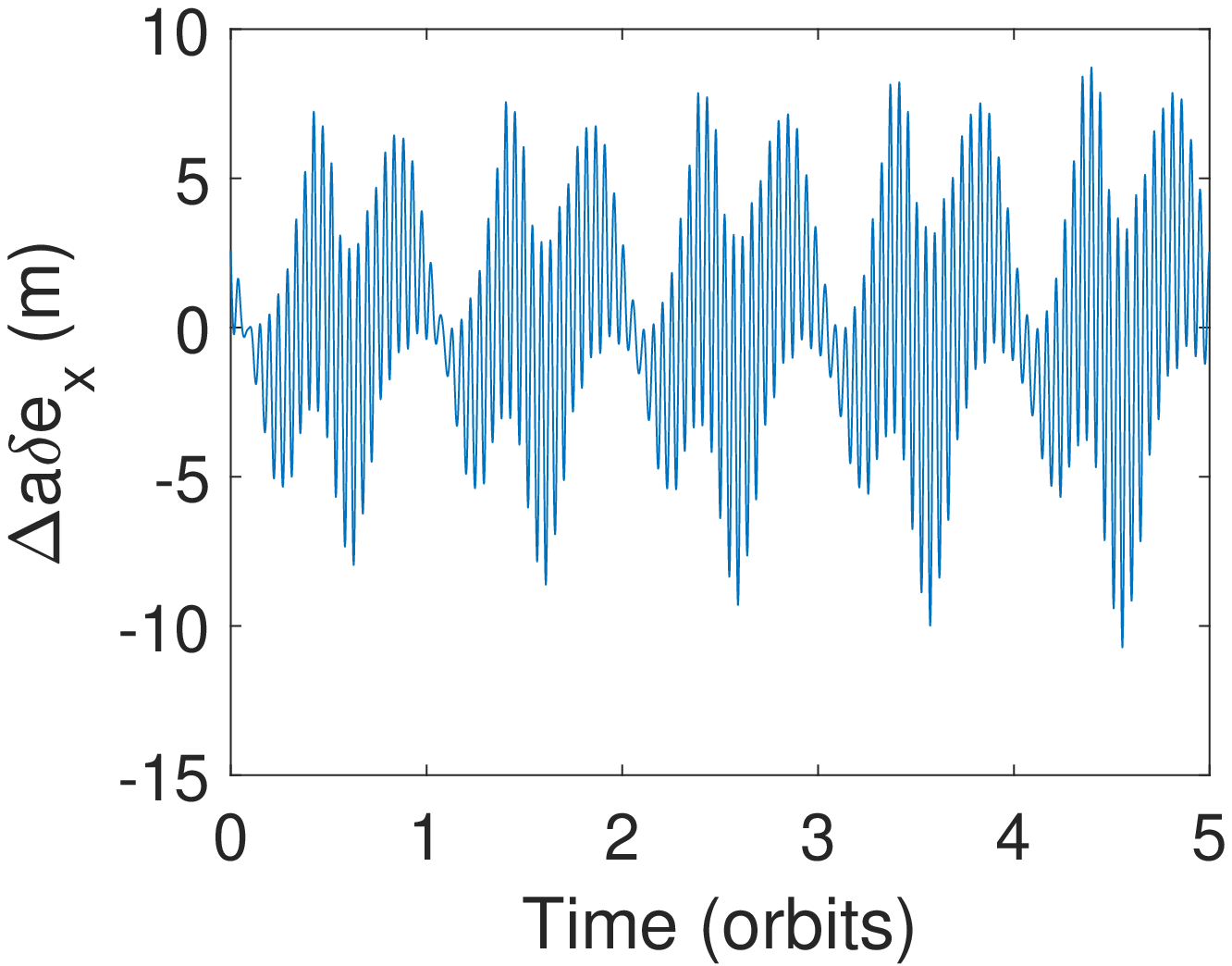}
	\end{subfigure}
	\begin{subfigure}[htb]{0.3\textwidth}
	\centering
	\caption{}
	\includegraphics[width=1.7in]{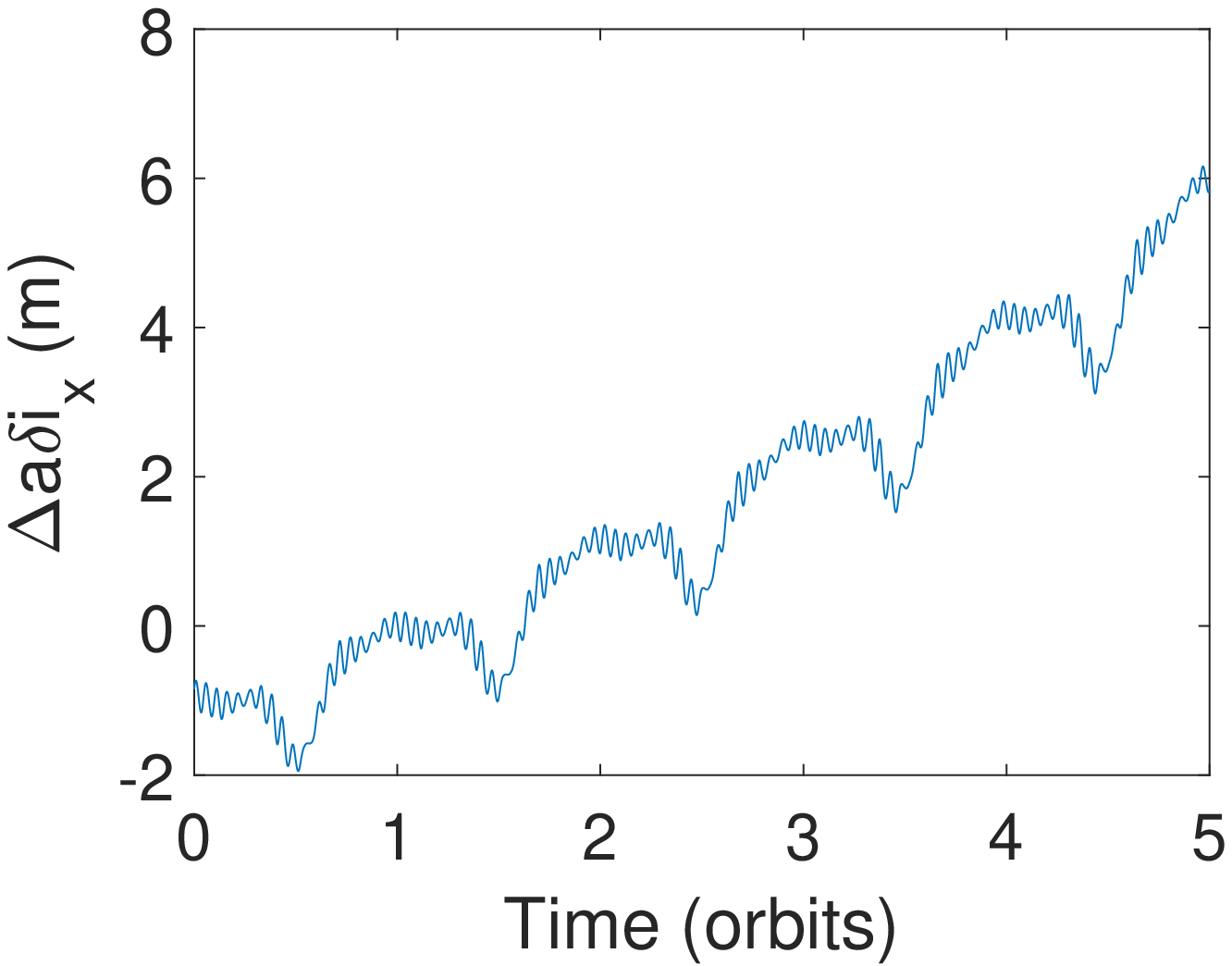}
	\end{subfigure}\\
	\centering
	\begin{subfigure}[htb]{0.3\textwidth}
	\centering
	\caption{}
	\includegraphics[width=1.7in]{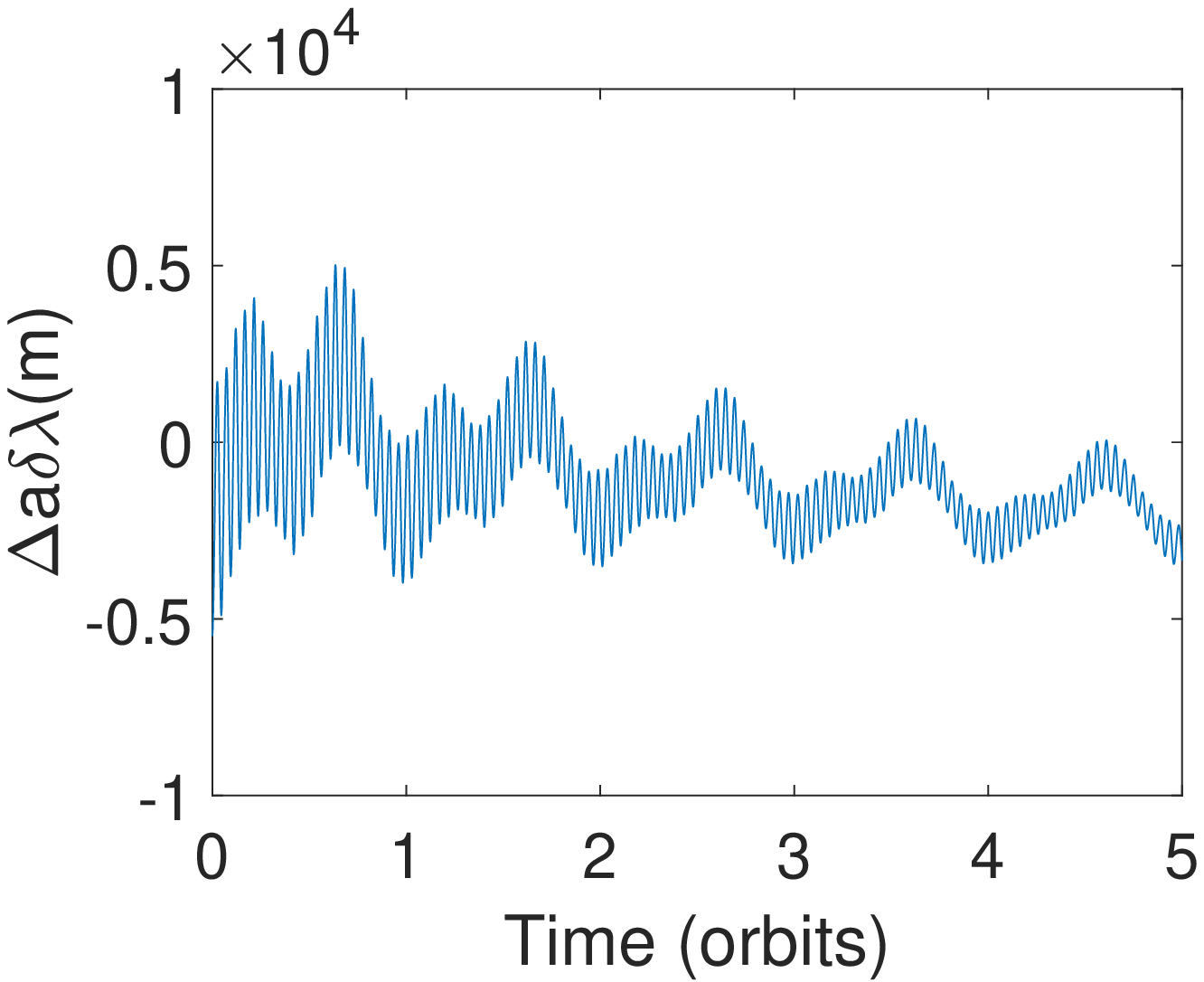}
	\end{subfigure}
	\begin{subfigure}[htb]{0.3\textwidth}
	\centering
	\caption{}
	\includegraphics[width=1.7in]{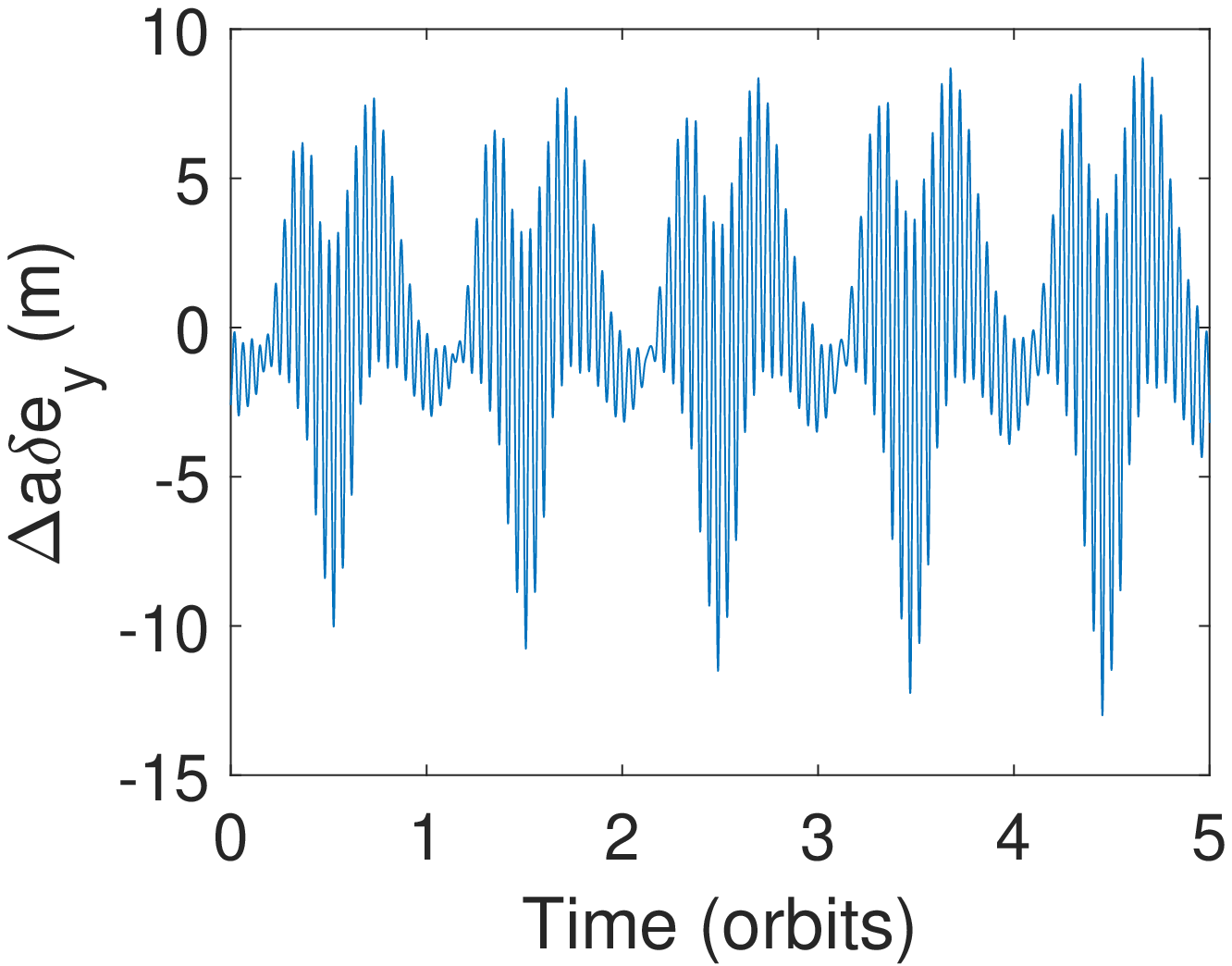}
	\end{subfigure}
	\begin{subfigure}[htb]{0.3\textwidth}
	\centering
	\caption{}
	\includegraphics[width=1.7in]{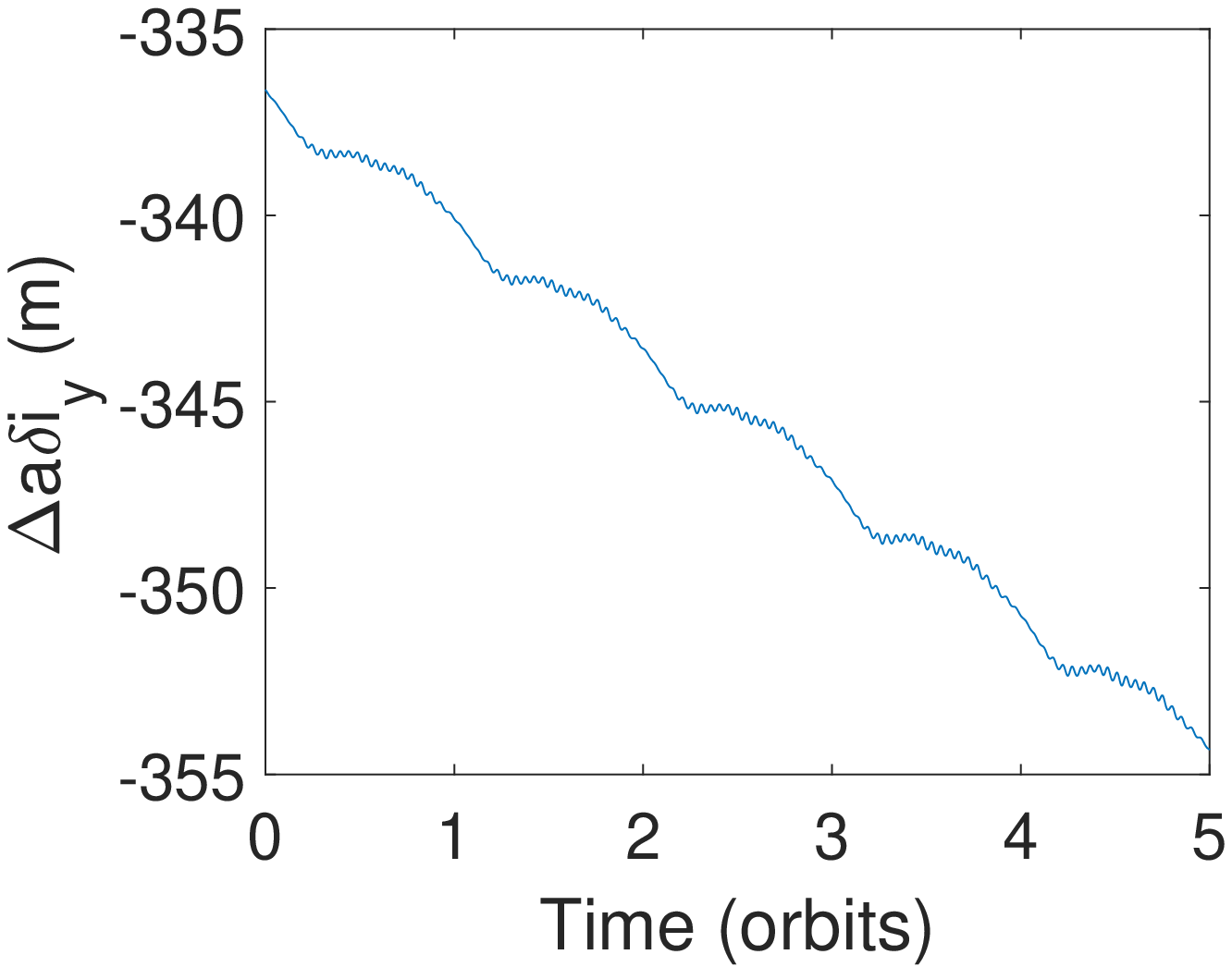}
	\end{subfigure} \\
	\caption{Error between the converted relative mean orbital elements using Schaub's iterative approach\cite{Schaub} and the ground-truth mean for a 60 km near-circular orbit around an Eros-like asteroid}
	\label{fig:osc2mean}
\end{figure}

Generally, osculating absolute orbital elements are converted to mean absolute orbital elements for both the chief and deputy satellites.  Then, the mean ROE are constructed from the mean orbital elements according to Equation \eqref{eq:ROE}.   The state-of-the-art for producing this absolute orbit conversion is demonstrated in Figure \ref{fig:osc2mean}.  This figure represents the error between the numerically averaged ground-truth relative orbital elements and the mean relative orbital elements produced by Schaub's iterative osculating-to-mean conversion\cite{Schaub} using ground-truth osculating orbital elements as the input.  The example results shown are for a quasi-stable 70 km near-circular orbit about the Eros-like asteroid presented in the Orbit Propagation section.  

Notably, the errors grow in time.  The error growth results from the absence of control that prevents the inter-satellite distance from increasing.  Therefore, oscillations in the osculating ROE increase, inhibiting the conversion.  While the errors of the relative eccentricity vector components are on reasonable scales, the remaining parameters are not. The $a\delta a$ and $a\delta \lambda$ parameters fail to meet requirements.  Also, the $a\delta \mathbf{i}$ components deviate from the truth significantly with the directionality of the trends suggesting the estimate only gets worse.  The large errors result from the fact that Schaub's approach assumes $J_2$ dominance. Therefore the approach fails to capture all relevant osculating effects from other spherical harmonic terms in order to remove them. 
 
To overcome these limitations, an EKF is proposed to provide mean ROE estimates.  The EKF leverages osculating ROE, which vary less in magnitude in comparison to the osculating absolute orbital elements due to cancellation of short-period effects from the close proximity of the chief and deputy spacecraft.  As such, fewer oscillations need to be removed to achieve requirements.  It follows that the osculating measurements serve as an approximation of the mean state measurement with noise.  Therefore, the proposed state $\mathbf{x} \in \mathbb{R}^{6}$ of the EKF is the mean ROE, and the proposed measurement $\mathbf{y} \in \mathbb{R}^{6}$ is the osculating ROE.

The specific mechanization of the EKF follows the notation and formulation of Montenbruck \cite{Montenbruck}.  For the time update, the previously presented semi-analytical propagator using the equations in the appendices Geopotentials and SRP is exploited with an added term that accounts for executed maneuvers. The dynamics update is therefore described as
\begin{equation}
\begin{aligned}
	\label{eq:dyn_update}
    \mathbf{x}_{k-} = \mathbf{x}_{k-1}+f(\mathbf{x}_{k-1}) +\mathbf{\Gamma}(\alpha_c(t))\mathbf{\Delta v}
\end{aligned}
\end{equation}
where $k$ indicates the current time step, '$-$' indicates the result from the time update step,  $f(\mathbf{x}_{k-1}) \in \mathbb{R}^{6}$ is the propagation provided by the dynamics model discussed previously, $\mathbf{\Gamma} \in \mathbb{R}^{6 \times 3}$ is the control matrix presented in Equation~\eqref{eq:gamma}, and $\Delta \mathbf{ v} \in  \mathbb{R}^{3}$ is the maneuver input in the RTN frame produced from the control logic.  

For the time update, the Jacobian matrix $\bm{\beta} \in  \mathbb{R}^{6 x 6}$ can be calculated numerically to produce an STM $\mathbf{\Phi} \in \mathbb{R}^{6 \times 6}$.  The Jacobian is calculated from the partial derivative of the dynamics update in Equation \eqref{eq:dyn_update} at each time step and must account for maneuver error.  Specifically, error in the estimation of the absolute orbit results in an misalignment of the estimated RTN directions from the truth inducing error in the maneuver direction.  Furthermore, a 5\% error in the maneuver execution magnitude to model realistic actuators is implemented.  For this work, it is assumed the accelerometers on the spacecraft can accurately measure the implemented thrust, but the error in the absolute orbit estimate and therefore the RTN directions must still be accounted for in the filter.  The Jacobian is therefore calculated as 

\begin{equation}
\begin{aligned}
	\label{eq:Jac}
	\mathbf{\Phi} = \mathbf{I}_{6x6}+ \bm{\beta} + \bm{\kappa} \\
	\beta_{kl} = \frac{\partial\delta \alpha_k}{\partial\delta \alpha_l} = \frac{\dot{\delta\alpha}_k(\delta\alpha + \dot{\delta\alpha}_l\Delta t)- \dot{\delta\alpha}_k(\delta\alpha - \dot{\delta\alpha_l} \Delta t)}{2\dot{\delta\alpha}_l\Delta t} \\
	\kappa_{kk} = \xi\Delta\delta\alpha_k
\end{aligned}
\end{equation}
where the $k$ and $l$ refer to components of the vector or matrix, the factor $\xi$ represents a measure of uncertainty in the executed maneuver, $\dot{\delta \alpha} \in \mathbb{R}^6$ is the time derivative of the ROE as a function of the ROE constructed by Equation~\eqref{eq:ROE_deriv}, and $\bm{\kappa} \in \mathbb{R}^{6 \times 6}$ is a diagonal matrix representing the maneuver contribution to uncertainties. Due to the slowly varying nature of the mean ROE and the approximately linear time derivatives, the Jacobian calculation can be completed with various user-selected time inputs.  This paper utilizes $\Delta t = 1000$s.

The covariance $\mathbf{P} \in \mathbb{R}^{6 \times 6 }$ is calculated by

\begin{equation}
\begin{aligned}
	\label{eq:cov_update}
	\mathbf{P}_{k^-}=\mathbf{\Phi} \mathbf{P}_k\mathbf{\Phi}^T+\mathbf{Q}
\end{aligned}
\end{equation}
where $\mathbf{Q} \in \mathbb{R}^{6 \times 6}$ represents the process noise matrix, provided by the accuracy of the dynamics model.  

Next, the measurement update is completed.  Since the measurements of the osculating elements is approximated as measurements of the mean elements, the sensitivity matrix $\mathbf{H}$ $\in  \mathbb{R}^{6 \times 6}$ is identity  and the model measurement vector $\mathbf{z} \in  \mathbb{R}^{6}$ is equivalent to the estimated mean ROE from the time update.  The covariance update follows the Joseph formulation, where $\mathbf{K} \in  \mathbb{R}^{6 \times 6}$ refers to the Kalman gain.  The measurement update is completed by 
\begin{equation}
\begin{aligned}
	\label{eq:meas_update}
	\mathbf{K}_k = \mathbf{P}_{k^-}\mathbf{H}_k^T(\mathbf{H}_k\mathbf{P}_{k^-}\mathbf{H}_k^T+\mathbf{S}_k)^{-1} \\
	\mathbf{P}_k = (\mathbf{I}-\mathbf{K}_k\mathbf{H}_k)\mathbf{P}_{k^-}(\mathbf{I}-\mathbf{K}_k\mathbf{H}_k)^T+\mathbf{K}_k\mathbf{S}_k\mathbf{K}_k^T \\
	\mathbf{x}_{k+1} = \mathbf{x}_{k-}+\mathbf{K}_k(\mathbf{y}-\mathbf{z}_k)
\end{aligned}
\end{equation}
where $\mathbf{S} \in \mathbb{R}^{6 \times 6}$ represents the measurement noise matrix.  

\begin{figure}[htb]
	\centering
	\begin{subfigure}[htb]{0.3\textwidth}
	\centering
	\caption{}
	\includegraphics[width=1.8in]{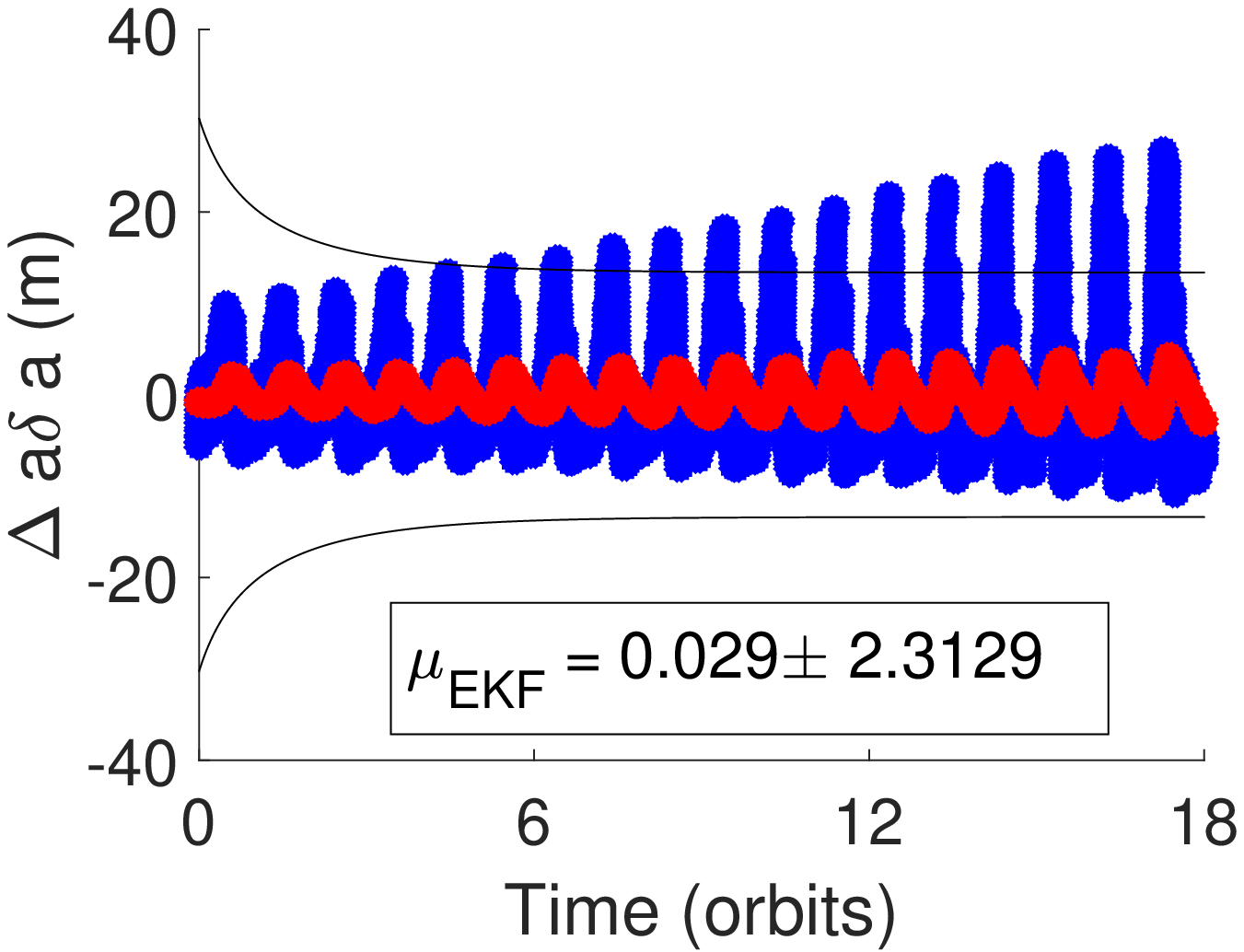}
	\end{subfigure}
	\begin{subfigure}[htb]{0.3\textwidth}
	\centering
	\caption{}
	\includegraphics[width=1.8in]{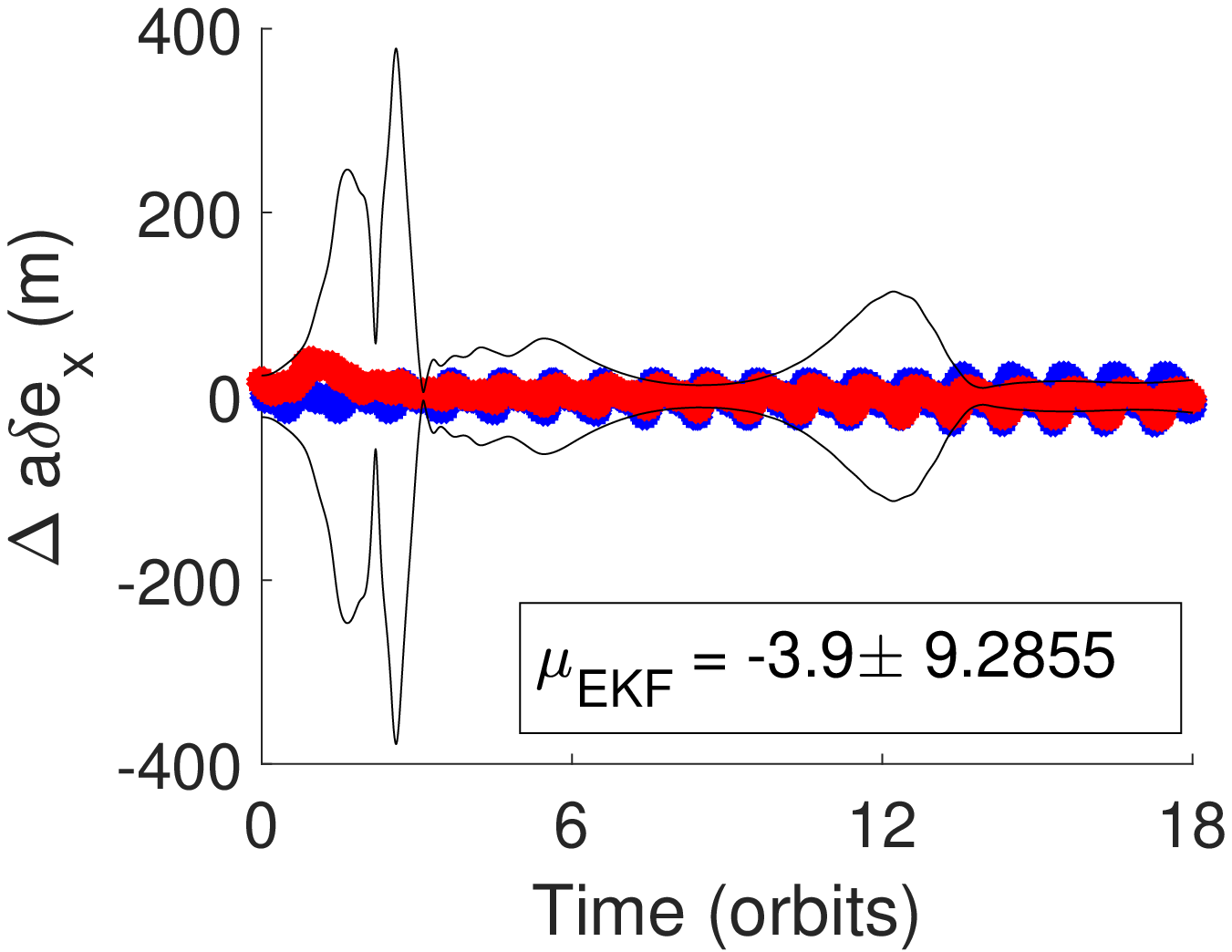}
	\end{subfigure}
	\begin{subfigure}[htb]{0.3\textwidth}
	\centering
	\caption{}
	\includegraphics[width=1.8in]{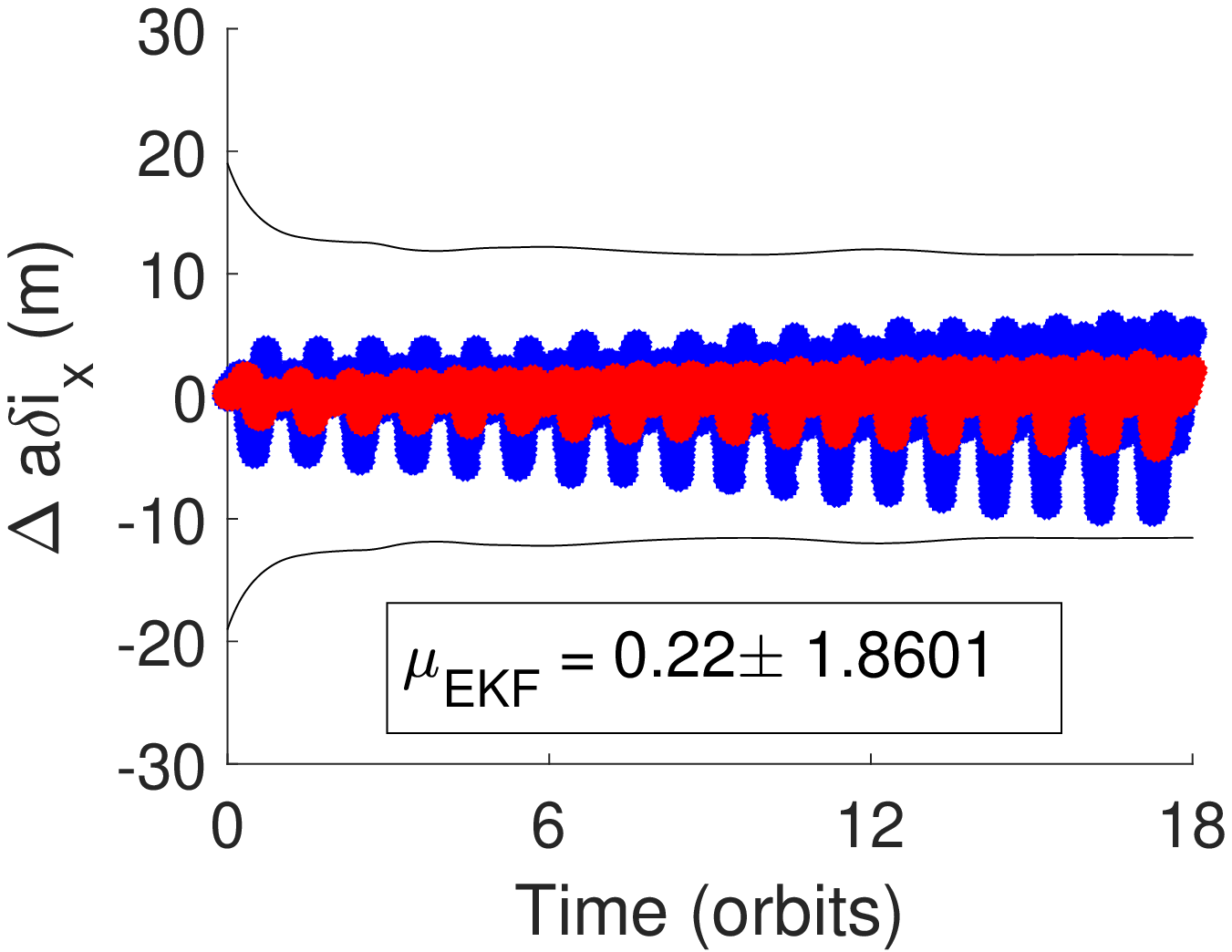}
	\end{subfigure}\\
	\centering
	\begin{subfigure}[htb]{0.3\textwidth}
	\centering
	\caption{}
	\includegraphics[width=1.8in]{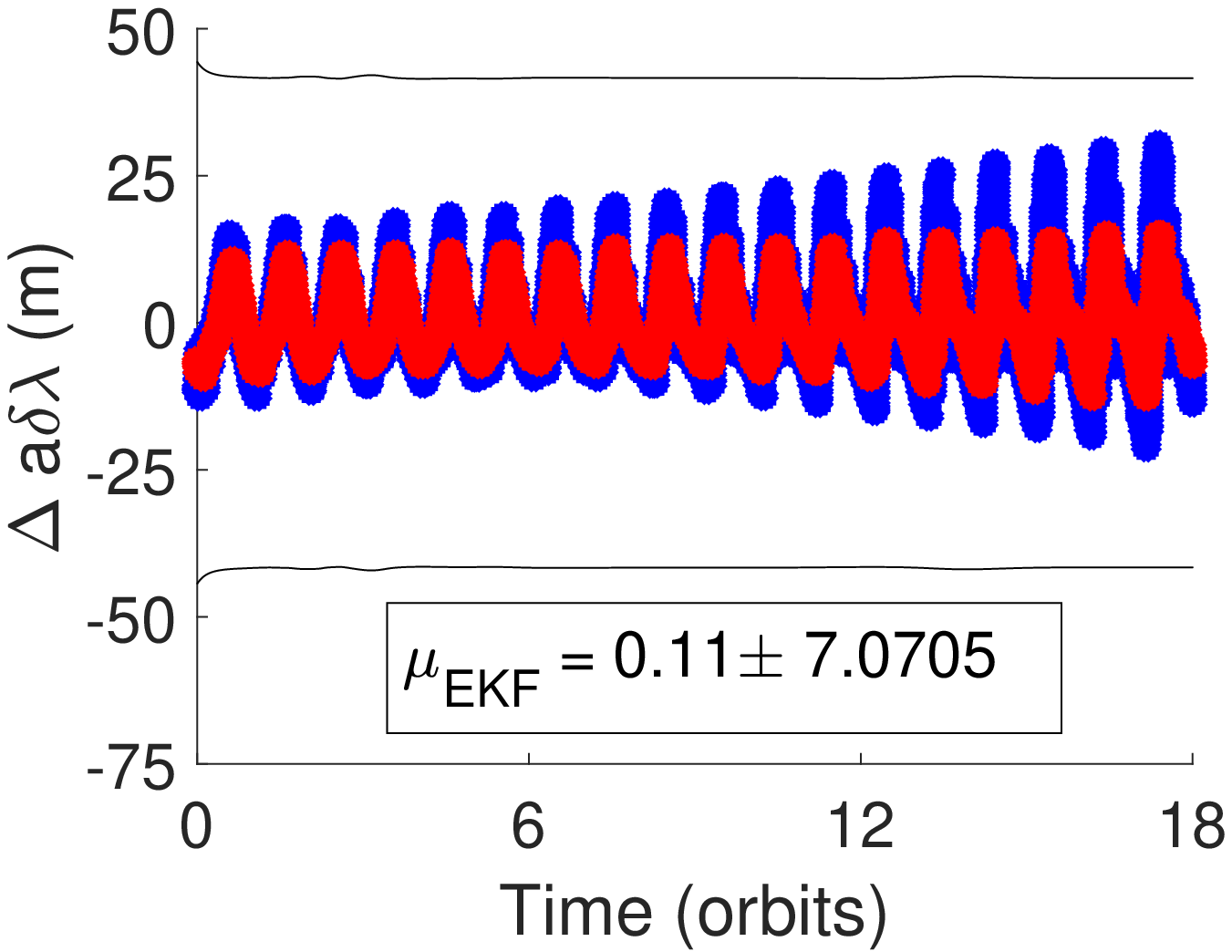}
	\end{subfigure}
	\begin{subfigure}[htb]{0.3\textwidth}
	\centering
	\caption{}
	\includegraphics[width=1.8in]{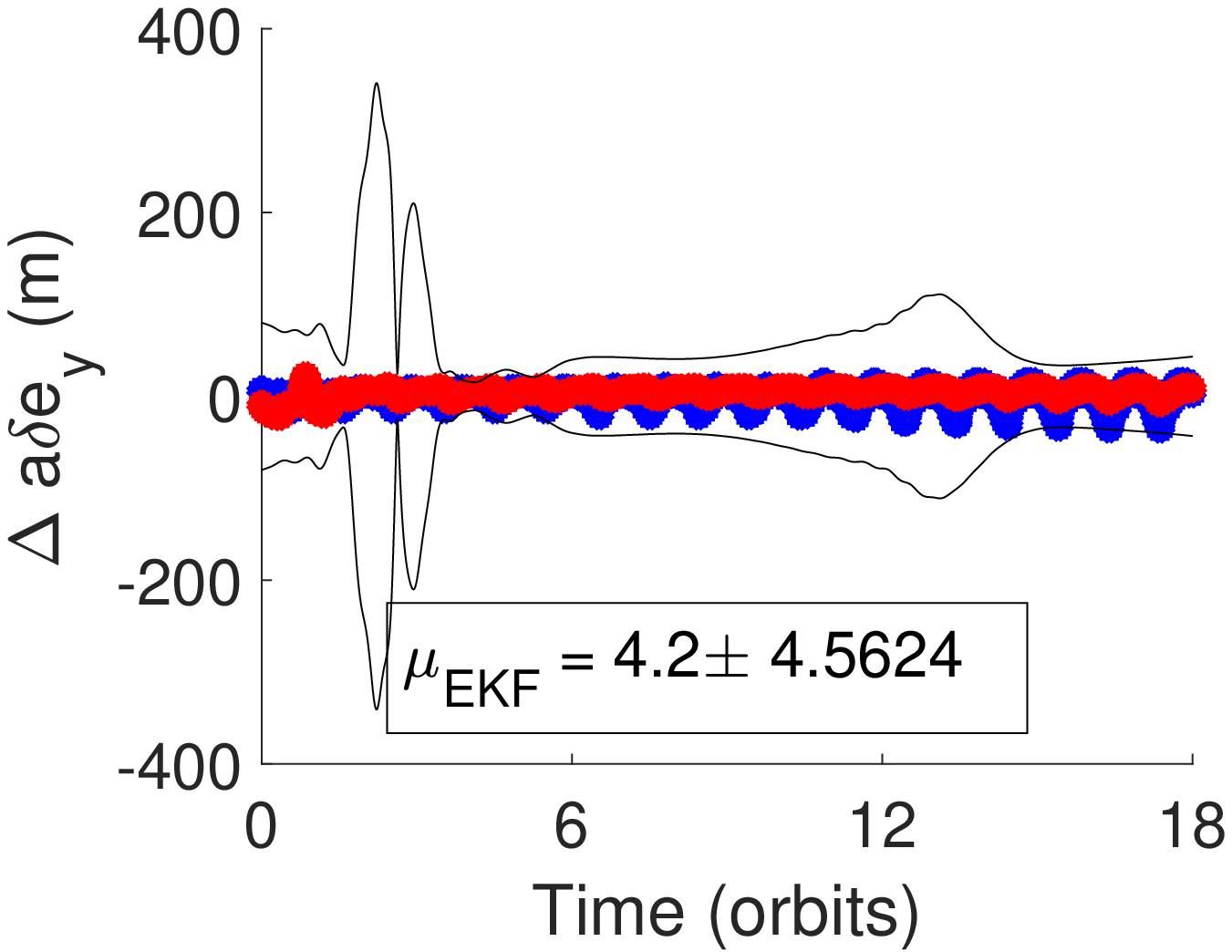}
	\end{subfigure}
	\begin{subfigure}[htb]{0.3\textwidth}
	\centering
	\caption{}
	\includegraphics[width=1.8in]{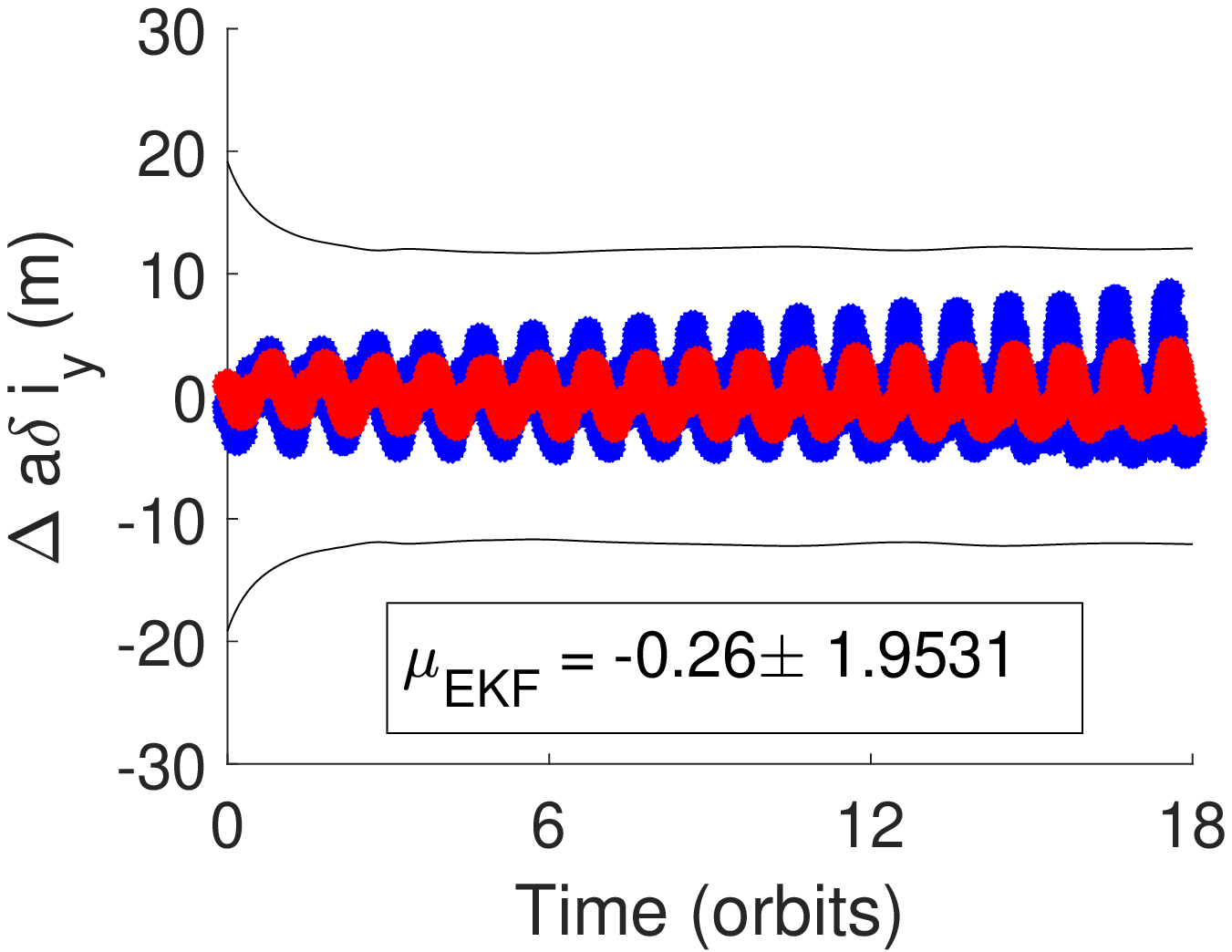}
	\end{subfigure}\\
	\caption{Comparison between ground truth mean ROE and osculating ROE measurements (blue) or mean estimates from the new filter (red).  The $3\sigma$ bound is shown in black.}
	\label{fig:osc2mean_res}
\end{figure}

The results from the filter are found in Figure \ref{fig:osc2mean_res} for a chief and deputy spacecraft in a near-circular orbit with a 70 km semimajor axis at an inclination of 160\degree \ about the Eros-variant asteroid presented in the Orbit Propagation section.  A Gaussian white noise with zero mean and $\sigma^2_{\delta \alpha} = 5 m^2$ is added to the true osculating ROE to model relative navigation measurements.  The variance is conservatively equivalent to 100 times the steady state mean absolute position error produced from previously published navigation results \cite{Stacey}. 

The figure displays the difference between the ground truth osculating ROE without noise and the ground truth mean ROE in blue.  The difference between the filter mean ROE estimate and the true mean ROE state is provided in red. The $3\sigma$ bound produced by the filter is shown in black. The mean error and the standard deviation $\sigma$ between the mean ROE estimate and ground truth mean state calculated over the last three orbit periods are included in each subfigure. Note again that the osculating measurements grow in time because the spacecraft drift apart due to lack of control input. 

The EKF demonstrates an accurate estimation and a reduction in the magnitude variation from the true osculating values for all parameters.  The most significant reduction in oscillations is for the relative semimajor axis, which is reduced by an order of magnitude from 30 m to 4 m. Compared to Schaub's iterative approach, the EKF produces a more accurate estimate of the mean $a\delta a$, $a\delta \lambda$, and $a\delta \mathbf{i}$.  The estimation of the relative eccentricity vector is comparable between the two approaches.  It should be noted that larger oscillations are maintained in the relative mean longitude and relative eccentricity vector due to the higher uncertainty in the dynamics model presented previously.  Some oscillations in all parameters remain due to two sources.  For one, measurements of the osculating orbit elements are not perfectly Gaussian with a mean equivalent to the mean ROE.  Secondly, variations appear in the ground truth mean ROE due to numerical averaging of the osculating orbital elements.  In all, the accuracy is on the meter level for all parameters.  Furthermore, the precision is within the 10-meter level for all ROE.  For $a\delta a$ a 3-meter level precision is reached.  Future work will implement adaptive noise processes to enable the filter to adjust to various satellite separations and improve precision even further in comparison to the state-of-the-art.

\section{Formation-Keeping Algorithm}
The third required advancement for ANS is a formation-keeping algorithm.  To accomplish mission objectives, the formation-keeping control law needs to guarantee safety, enable science objectives, and be robust to uncertainties in the gravity model.  First, a minimum separation is required to ensure spacecraft safety.  Second, the control law must ensure a specified time interval exists between maneuvers for uninterrupted science phase operations.  This follows from the fact that maneuvers require reorientation of spacecraft due to the limited thrusting capability of cubesats, thus preventing imaging.  Furthermore, thrusting produces navigation errors which require time for convergence. Therefore limiting separate maneuvering windows enables faster characterization of the asteroid. Additionally, the control algorithm must prevent evaporation, which inhibits stereoscopic imaging because the spacecraft are too far apart to see the same features of the asteroid surface.  Thirdly, the formation-keeping profile must be robust to uncertainties in the gravity model because little is known about gravitational potential of the asteroid before the swarm reaches the asteroid.  

The formation-keeping algorithm description is divided into two sections.  First, the specific swarm geometries are addressed with associated guidance profiles. Second, the state-space control law is presented.

\subsection{Guidance}
Two swarm geometries are discussed in this section: the E-I vector and condensed swarms. The swarms are defined in terms of ROE with respect to a reference, which may or may not contain a mothership.  If there is no mothership, the reference serves as an absolute orbit to be tracked.  This reference orbit may have reduced order dynamics in order for the swarm to reasonably track the reference without considering all of the perturbations that make the absolute orbit unstable in the long term. This is to be addressed in future work on absolute orbit control.  For this work, it is assumed the reference has an uncontrolled mothership that experiences the full dynamic environment.  Both swarms nominally have $\delta a = \delta \lambda = 0$.

\subsubsection{E-I vector.}
\begin{figure}[htb]
	\centering
	\begin{subfigure}[htb]{0.3\textwidth}
	\centering
	\caption{}
	\includegraphics[width=1.7in]{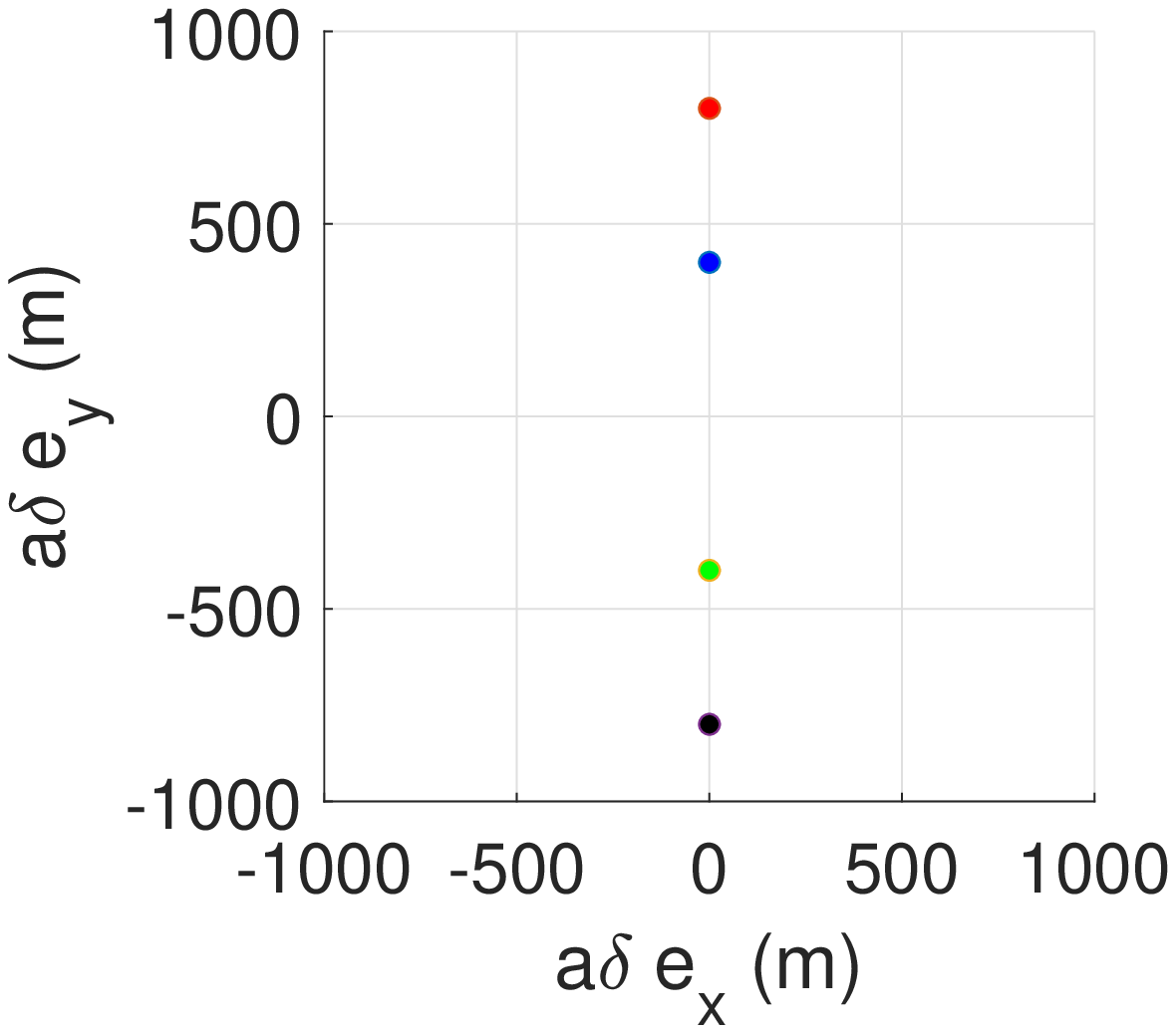}
	\end{subfigure}
	\begin{subfigure}[htb]{0.3\textwidth}
	\centering
	\caption{}
	\includegraphics[width=1.7in]{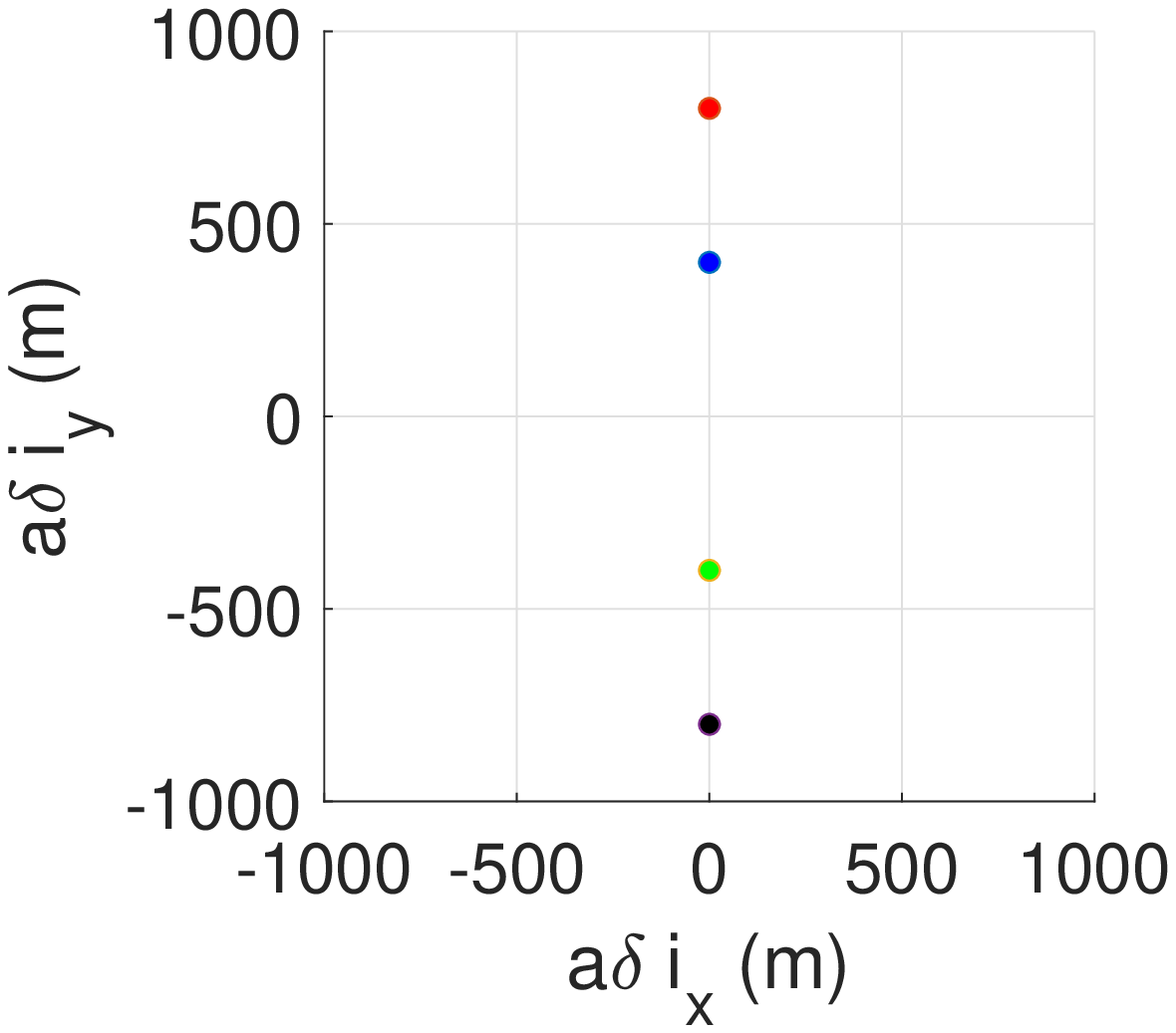}
	\end{subfigure}
	\centering
	\begin{subfigure}[htb]{0.3\textwidth}
	\centering
	\caption{}\includegraphics[width=1.7in]{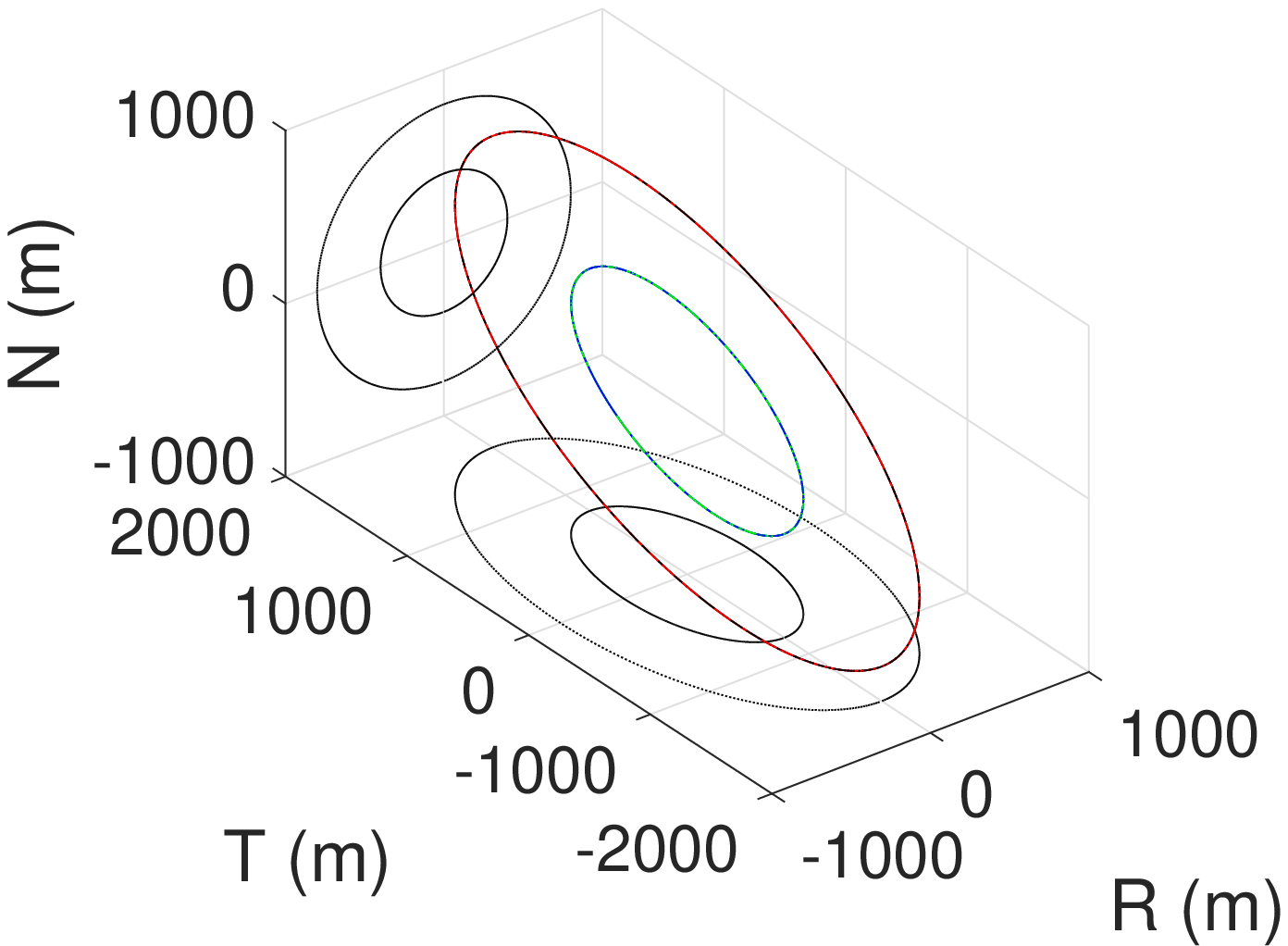}
	\end{subfigure}\\
	\caption{Visualization of the E-I vector swarm design. The parameters in mean relative eccentricity and inclination space are provided in subfigures a and b. The motion of the swarm in the Hill frame over one orbit period is provided in subfigure c with projections in the RT and RN plane provided.}
	\label{fig:ei_vec}
\end{figure}

The E-I vector separation geometry has been previously presented in literature \cite{Koenig_swarm} and is demonstrated in Figure~\ref{fig:ei_vec}.  This swarm geometry is defined by the fact that the $\delta \mathbf{e}$ and $\delta \mathbf{i}$ vectors between each pair of satellites are anti-parallel or parallel.  

Although defined for Earth applications, this swarm also has desirable traits for asteroid purposes.  One such trait is passive safety.  This results from the maximum out-of-plane separation at minimal radial separation and vice versa due to the parallel or anti-parallel orientation of the $\delta \mathbf{e}$ and $\delta \mathbf{i}$ vectors. A second desirable trait is the fact that the relative inclination vector can be oriented vertically to prevent secular drift from the gravity potential, as described in the Relative Motion Trends section.  

The E-I vector separation swarm has some drawbacks.  For one, the swarm satellites nominally rotate about the reference in 3D space over the orbit period but remain collinear.  Unfortunately, the collinear geometry limits the viewing angles of the asteroid surface and therefore stereoscopic vision. Additionally, to meet a minimum separation in the radial component when the out-of-plane component is minimal, $\delta \mathbf{e}$ must be restricted to a finite angular range from the vertical orientation \cite{Koenig}.  Therefore the rotation of $\delta\mathbf{e}$ caused by gravitational potential effects must be countered to ensure the vector remains within the desired range, which could become expensive.  In this work, the guidance will enforce that this vector remains vertical.

\begin{figure}[htb]
	\centering
	\begin{subfigure}[htb]{0.95\textwidth}
	\centering
	\caption{}
	\includegraphics[width=4.5in]{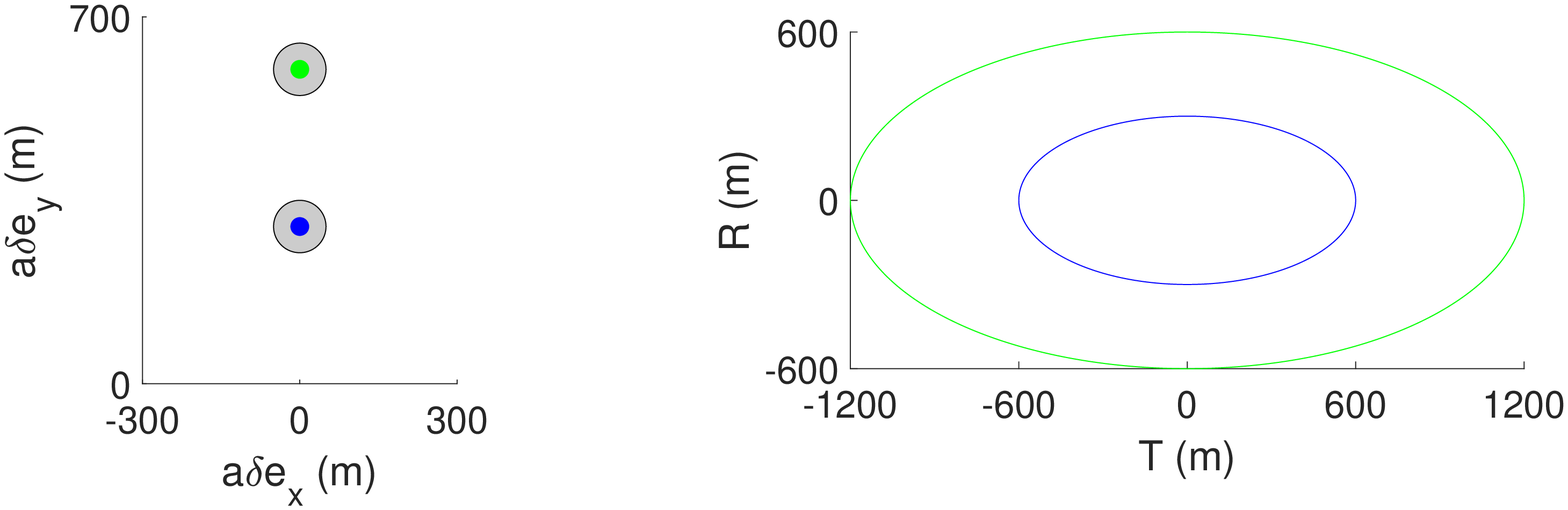}
	\end{subfigure}\\
	\centering
	\begin{subfigure}[htb]{0.95\textwidth}
	\centering
	\caption{}
	\includegraphics[width=4.5in]{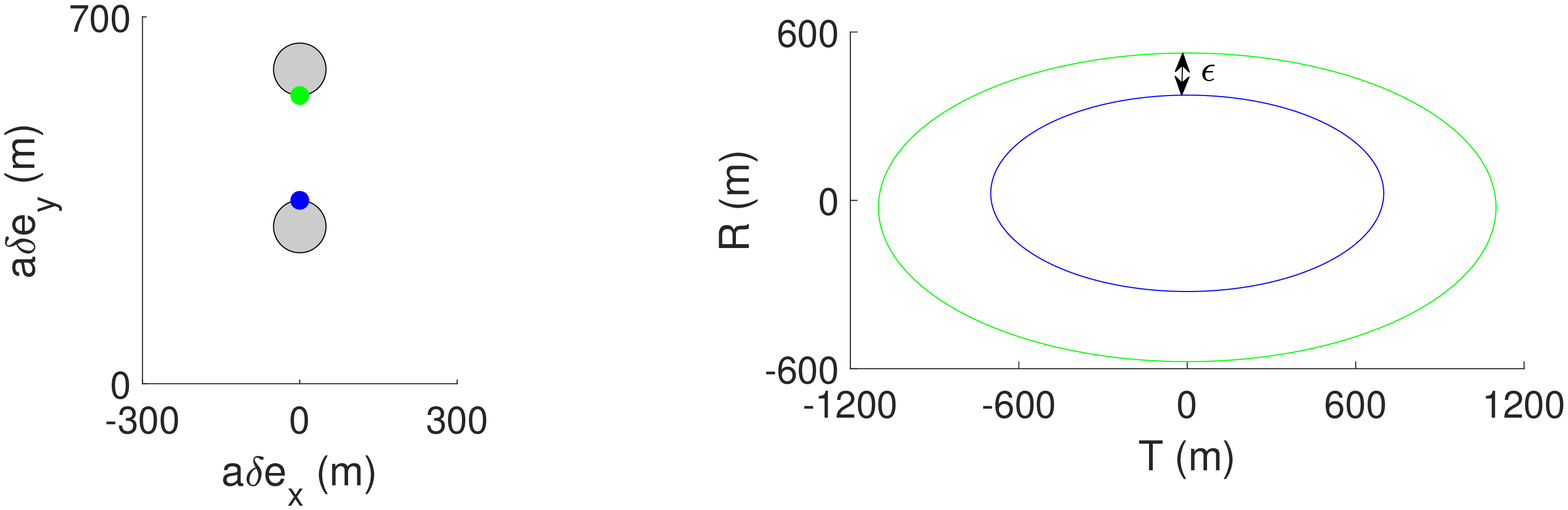}
	\end{subfigure}\\
	\caption{The nominal E-I vector swarm in the relative eccentricity space and the projected motion in the RT plane is shown in subfigure a.  The same swarm under worst-case variations in $\delta \mathbf{e}$ and $\delta a$ is shown in subfigure b with both the relative eccentricity space geometry and projected motion in the RT plane included.  The colors indicated the specific satellites, and gray indicates the deadbands on the relative eccentricity and relative inclination vectors. }
	\label{fig:ei_vec_safety}
\end{figure}

While the nominal geometry exists at $0  = \delta a = \delta \lambda$, spacing in $\delta \mathbf{e}$ allows for small deviations in $ \delta \lambda$ and $ \delta a$ while ensuring safety.  To guarantee a user-specified minimum safe separation $\epsilon$, the maximum allowable $\delta a$ must be contained to avoid collision in-plane.  This is demonstrated in Figure \ref{fig:ei_vec_safety}, where the nominal configuration is shown in subfigure a and the configuration under variations in $\delta \mathbf{e}$ and $\delta a$ is shown in subfigure b.  Note that  $\delta e_{db}$ is the allowable variation in the relative eccentricity vector described in detail in the control section.  The figure includes both $\delta \mathbf{e}$ and the motion projected onto the RT plane.  Notably, the satellites pass through this plane at the extremes of the radial position.  Therefore, this location is where collision could occur if the relative semimajor axis is allowed to vary significantly.

Based on the understanding provided by the diagram, the maximum allowed semimajor axis $\delta a_{s,ei}$ is defined as 
\begin{equation}
\begin{aligned}
	\label{eq:da_err_eivec}
	\delta a_{s,ei} \leq |\frac{\rho-\epsilon}{2}|	\\
    \rho = \min(||\delta \mathbf{e}_{i}-\delta \mathbf{e}_{j}|| - 2\delta e_{db}, ||\delta \mathbf{e}_{i}|| - \delta e_{db})
\end{aligned}
\end{equation}
where the indices $i,j$ refer to ROE describing the position of deputy $i$ and $j$.  The first expression in the definition of $\rho$ ensures no collision between the deputies.  The second expression ensures no collision with a spacecraft placed at the reference, if there is one.  Notably, the second expression always assumes the spacecraft exists exactly at the reference, hence the lack of a factor of 2 on $\delta e_{db}$.

\subsubsection{Condensed.}
\begin{figure}[htb]
	\centering
	\begin{subfigure}[htb]{0.3\textwidth}
	\centering
	\caption{}
	\includegraphics[width=1.7in]{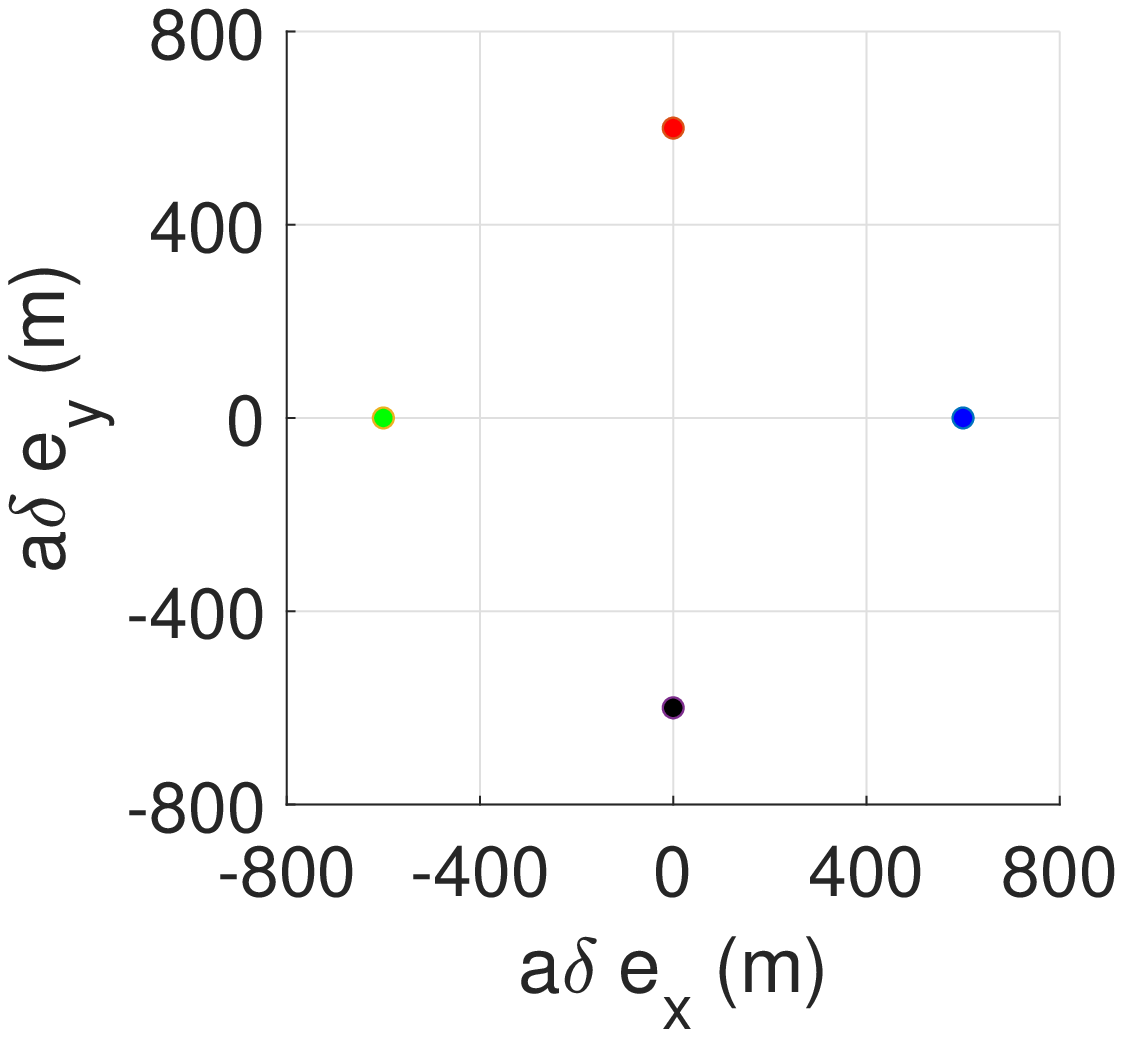}
	\end{subfigure}
	\begin{subfigure}[htb]{0.3\textwidth}
	\centering
	\caption{}
	\includegraphics[width=1.7in]{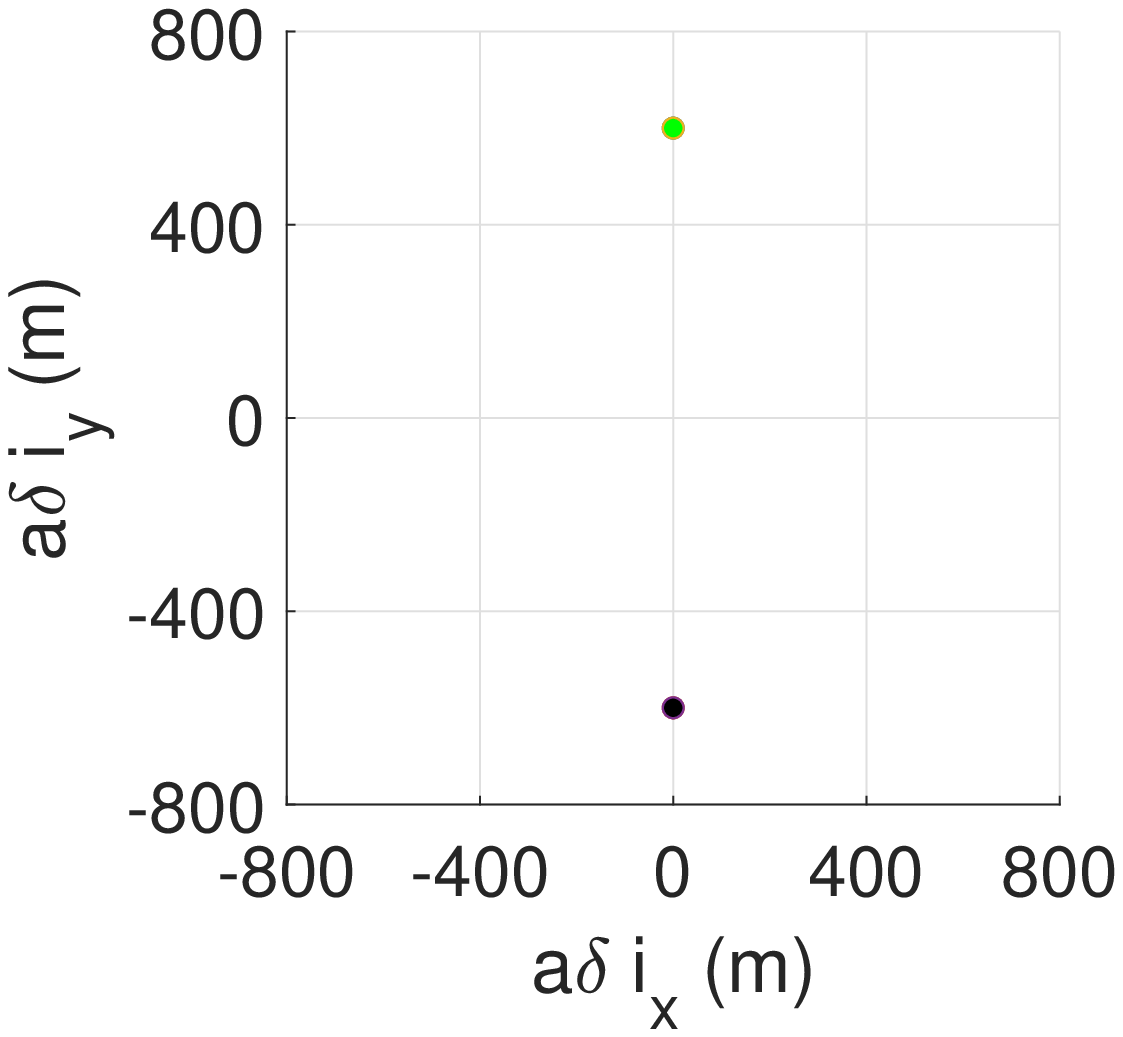}
	\end{subfigure}
	\centering
	\begin{subfigure}[htb]{0.3\textwidth}
	\centering
	\caption{}
	\includegraphics[width=1.7in]{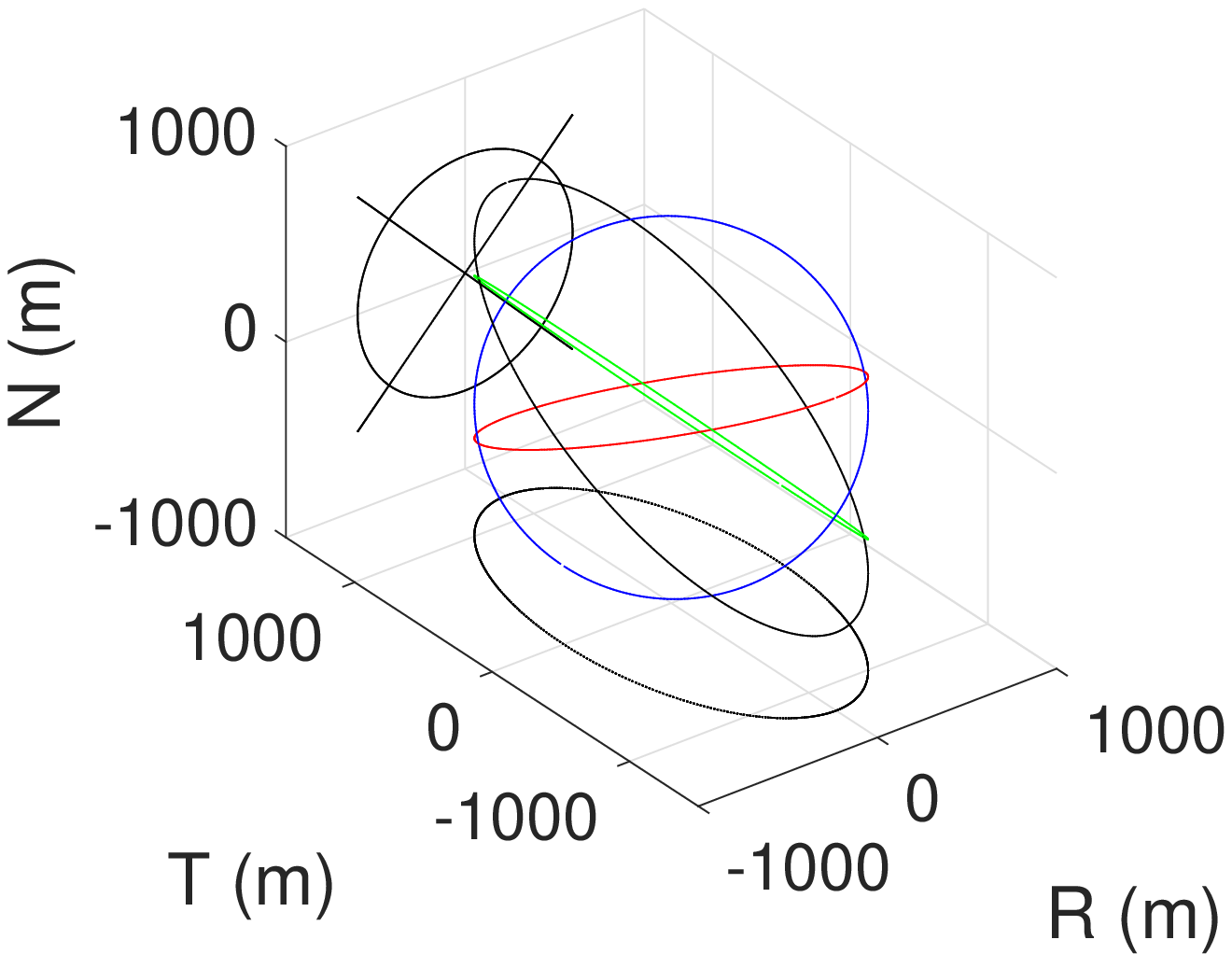}
	\end{subfigure}\\
	\caption{Visualization of the condensed swarm design. The parameters in mean relative eccentricity and inclination space are provided in subfigures a and b. The motion of the swarm in the Hill frame over one orbit period is provided in the subfigure c with projections in the RN and RT plane included.}
	\label{fig:cond}
\end{figure}

The condensed swarm geometry, plotted in Figure~\ref{fig:cond}, is a variation on the high-density swarm presented in literature \cite{Koenig_swarm}. The main difference between the high-density and the condensed swarm is the relative inclination vector of each satellite.  Because the high-density swarm relies on in-plane separation defined by $\delta \lambda$ and $\delta\mathbf{e}$, the out-of-plane parameter $\delta \mathbf{i}$ serves as a degree of freedom that can be exploited to produce a favorable geometry.  Therefore, a vertical relative inclination vector is used to not only provide motion of the swarm out-of-plane and prevent satellites from obstructing each other's view of the asteroid but also to avoid secular drifts induced by the gravitational perturbations. 

The condensed swarm geometry has a few benefits.  For one, the condensed swarm geometry does not require compensation for the rotation of $\delta \mathbf{e}$ because the formation geometry is driven by the relative phase between $\mathbf{\delta e}$ vectors. The secular rotation rate caused by $J_2$ and $J_4$ is independent of orientation, so all of the spacecraft rotate at the same rate and therefore maintain the same relative orientation.  As such, the guidance dictates that the $\delta \mathbf{e}$ vectors rotate at this secular rate.  Furthermore, allowing the absolute orientation of the relative eccentricity vectors to vary changes the position of the swarm satellites with respect to the reference at the same mean argument of latitude in subsequent orbits.  As a result, imaging angles between the spacecraft vary, aiding in reconstructing a 3D model of the asteroid features. 

The drawback of the geometry is the need for persistent safety measures to guarantee minimum in-plane separations.  The maximum separation in the relative mean longitude to ensure a user-specified safety $\epsilon$ for spacecraft encircling one another is defined in literature as \cite{Koenig_swarm}
\begin{equation}
\begin{aligned}
	\label{eq:safety}
	\Delta a\delta\lambda_{ij} \leq \sqrt{(3(a_c^2(||\delta \mathbf{e}_{i}-\delta \mathbf{e}_{j}||)^2-\epsilon^2))}
\end{aligned}
\end{equation}

where $\Delta a\delta\lambda_{ij}$ is the difference in $a\delta\lambda$ between two spacecraft. The expression in~\eqref{eq:safety} must hold for every pair of satellites $i,j$. As a result, to ensure passive safety, the upper limit on $\delta \lambda$, denoted $\delta \lambda_{s,cond}$ is then taken as the equality constraint in Equation~\eqref{eq:safety} based on the smallest allowable relative eccentricity vector between spacecraft.  This is described as
\begin{equation}
\begin{aligned}
    \label{eq:dl_safety}
    a\delta\lambda_{s,cond} = \min (\sqrt{(3(a_c^2\tau^2-\epsilon^2)}/2, \sqrt{3(a_c^2\xi^2-\epsilon^2)}) \\
    \tau = \min(||\delta \mathbf{e}_{i}- \delta \mathbf{e}_{j}|| - 2\delta e_{db}) \\
    \xi = ||\delta \mathbf{e}_{i}|| - \delta e_{db}
\end{aligned}
\end{equation}
where $\tau$ describes the separation between deputy spacecraft and $\xi$ pertains to the separation with the chief or reference. Note that the factor of 2 in the expression for $\tau$ ensures both spacecraft are equally responsible for limiting the along-track separation.  The factor of 2 is not in the $\xi$ expression because the reference is nominally placed at $a\delta\lambda = 0$.

Because of the reliance on in-plane separation, maneuvers may be required earlier than desired to ensure swarm safety.  Furthermore, a missed maneuver creates a risk of collision.  Such continuities will be addressed in future work.  A summary of the benefits and drawbacks of the swarm geometries are provided in Table \ref{tab:swarm_comp}.

\begin{table}[htb]
    \caption{A summary of the benefits and drawbacks of the two presented swarm geometries.}
    \label{tab:swarm_comp}
    \centering
    \begin{tabular}{p{1.7cm} p{2.2cm} p{2.2cm} p{4cm} p{2.6cm}} 
    \noalign{\hrule height 2pt}
    Swarm & Compensation for $\delta \mathbf{e}$ rotation & Safety  & Imaging Considerations & Collision Risk \\
    \noalign{\hrule height 2pt}
    E-I Vector &  yes & passive & Limited imaging angles & RN plane\\
    \hline
     Condensed & no & active control required & Varying imaging angles in subsequent orbits & RT plane \\
     \noalign{\hrule height 2pt}
\end{tabular}
\end{table}

\subsection{Control}

In derivation of the following state-space control law, impulsive maneuvers are assumed.  This follows from the fact that low-thrust actuators provide a thrust on the same order of magnitude of the two-body gravitational acceleration of the asteroid.  For example, the acceleration due to Keplerian gravity for a satellite 70 km from the center of mass of Eros is 91 $\mu N$, and the cubesat low-thrust actuator produced by Busek \cite{Busek_thrust, Lemmer} can produce $5-100 \mu N$ of instantaneous thrust.  Furthermore, the small $\Delta v's$ required result in no propellant consumption payoff from using low-thrust propulsion.  Finally, impulsive maneuvers ensure reconfigurations are quick to avoid interference with gravity recovery.

The following control algorithm and safety constraints are defined such that implementation is distributed.  Therefore the individual swarm spacecraft control their own position in ROE space with respect to the reference and require no information about the other swarm satellites.

The control law description is broken into three sections.  First, the in-plane description of the control approach includes four relevant ROE, $a\delta a, a\delta\lambda,  a\delta e_x, $ and $a\delta e_y$, and two maneuvering directions, radial and along-track.  This is followed by a discussion of the out-of-plane control with the relevant ROE $a\delta i_x$ and $a\delta i_y$ and out-of-plane maneuver direction.  Next, these planar control laws are combined into a single control law.  Throughout the following discussion, the utilized ROE are mean.

\subsubsection{In-plane.}
The in-plane control law has three objectives: 1) control $\delta\mathbf{e}$ to maintain formation geometry, 2) ensure small $\delta a$ for passively bounded motion and safety, and 3) bound $\delta \lambda$ for evaporation prevention and safety.  All three objectives are achieved using efficient maneuvers with one in-plane thruster to consider limited resources on cubesats. 

Based on the dynamics of the perturbed relative motion presented previously, drifts in $\delta \mathbf{e}$ are dominant compared to the other ROE drifts.  Furthermore, the control matrix in Equation~\eqref{eq:gamma} indicates along-track maneuvers are more efficient than radial maneuvers in changing the relative eccentricity vector.\cite{Chernick_sol}  However, any vector adjustment in $\delta \mathbf{e}$ created by along-track maneuvers result in a change of $\delta a$  with the same magnitude. This can create larger-than-desired changes in the relative semimajor axis when using a single along-track maneuver.  Alternatively, a pair of along-track maneuvers can be implemented that are separated by a half-orbit period in order to achieve the desired change in the relative eccentricity vector and contain the change in the relative semimajor axis.\cite{Chernick_sol}  The pair of maneuvers extends reconfiguration time by at least a half orbit and requires multiple separate maneuvers, which interferes with science operations. Furthermore, maneuvering around asteroids is inexpensive given the relatively small gravitational force which enables the thrust acceleration to produce large changes in ROE.  Thus, the extra delta-v from using radial maneuvers is not a concern.  As a result, a combination of radial and along-track maneuvers is considered preferable, but still the along-track component will be maximized to keep fuel costs low.  

\begin{figure}[htb]
	\centering\includegraphics[width=2in]{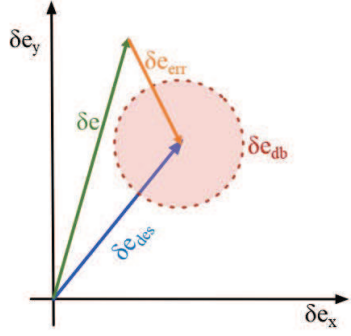}
	\caption{Graphical representation of the definition of the error vector $\delta \mathbf{e}_{err}$, allowable error $\delta \mathbf{e}_{db}$, nominal vector $\delta \mathbf{e}$, and desired vector $\delta\mathbf{e}_{des}$ in relative eccentricity space.}
	\label{fig:de_err}
\end{figure}

To begin, the desired relative eccentricity vector is referred to as $\delta \mathbf{e}_{des}$.  The user-specified, allowable error in the relative eccentricity vector is referred to as $\delta \mathbf{e}_{db}$, where $db$ stands for deadband.  The vector error between the current relative eccentricity vector $\delta \mathbf{e}$ and the desired vector is referred to as $\delta \mathbf{e}_{err}$.  It is defined in a Cartesian and polar reference as
\begin{equation}
\begin{aligned}
	\label{eq:de_err}
    \delta\mathbf{e}_{err} = ||\delta\mathbf{e}_{err}|| \begin{bmatrix} \cos{\nu} \\
    \sin{\nu}
    \end{bmatrix} = \begin{bmatrix} \delta\mathbf{e}_{x,des}-\delta\mathbf{e}_x \\
    \delta\mathbf{e}_{y,des}-\delta\mathbf{e}_y
    \end{bmatrix}
\end{aligned}
\end{equation}
where $\nu$ represents the phase of the error vector as measured counter-clockwise from the positive x-axis. It should be noted that the error is defined with respect to the center of the deadband, as shown in Figure \ref{fig:de_err}.  By defining the error to the center of the deadband, the control law is robust to uncertainty in the knowledge of the gravity coefficients, as the distance to all edges of the deadband is maximized and no predictions from the dynamics are used. The objective is to contain $\delta \mathbf{e}$ within this defined deadband, which requires constant maneuvers due to perturbations from SRP described previously.

To determine the largest available along-track maneuver to correct $\delta\mathbf{e}_{err}$, $\delta a$ and $\delta\lambda$ parameters must be considered.  Under SRP and gravitational perturbations, $\delta a$ is unchanged.  Comparatively, $\delta\lambda$ experiences a drift dominated by Keplerian effects.  The Keplerian drift $\delta \dot{\lambda}_{kep}$ is given by  \cite{Koenig}
\begin{equation}
\begin{aligned}
	\label{eq:kepler}
	\delta \dot{\lambda}_{kep} = -\frac{3}{2}n\delta a
\end{aligned}
\end{equation}
According to Equation~\eqref{eq:kepler}, a nonzero $\delta a$ induces a constant secular drift rate in $\delta \lambda$ with a direction opposite the sign $\delta a$. By requiring a user-specified time between maneuvers, $T_{bm}$, the change in $\delta\lambda$ over this time can be determined analytically.  By centering this change around $a\delta\lambda = 0$, the swarm remains co-located, which encourages the satellites to view the same features to aid in stereoscopic imaging. However, the approach can easily be extended for non-co-located swarms .  The limits in $\delta\lambda$ using the allowed time $T_{bm}$ are therefore defined as 
\begin{equation}
\begin{aligned}
	\label{eq:dlambda}
	\delta\lambda_{lim} =\pm \frac{T_{bm}}{2}\delta\dot{\lambda}_{kep} = \pm \frac{3T_{bm}}{4}n\delta a
\end{aligned}
\end{equation}

\begin{figure}[htb]
	\centering\includegraphics[width=1.8in]{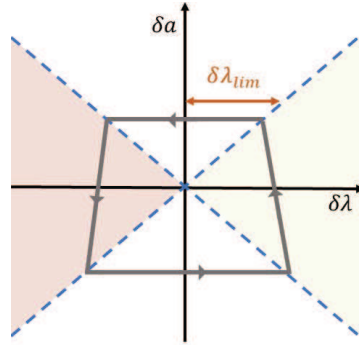}
	\caption{Graphical representation of the control deadband designed in the mean $\delta a-\delta\lambda$ plane.  The red and green regions represent ROE combinations requiring an anti-along-track and along-track direction maneuver, respectively.  The gray line represents a sample trajectory in the plane, and the red $\delta\lambda_{lim}$ represents the parameter defining the width of the drift window.}
	\label{fig:da_dl}
\end{figure}

Once $\delta \lambda$ exceeds the positive or negative limit in Equation \eqref{eq:dlambda}, the control algorithm begins to look for a maneuver.  The maneuver must produce an $\delta a$ that negates  $\delta \lambda$ through a Keplerian drift.  As such, a positive $\delta a$ is required when $\delta \lambda$ is positive, meaning a maneuver in the +T direction is necessary and vice versa.  This results in a deterministic limit cycle in the $\delta  a-\delta \lambda$ plane.  The limit cycle is illustrated in  Figure~\ref{fig:da_dl}.  The white area represents periods of free drift, and the red and green signify conditions for maneuvers in the positive and negative along-track direction, respectively. A sample trajectory induced by the limit is also included in gray with directional motion indicated. 
 
However, an adjustment is required for safety.  Safety is ensured by enforcing an upper limit on $\delta \lambda$ and $\delta a$, which were defined as $\delta a_{s,ei}$ and $\delta \lambda_{s,cond}$ in the previous section. Note that the E-I vector separation swarm requires a safety limit on $\delta a$ and the condensed swarm requires a safety limit on $\delta \lambda$.  However, due to the coupling in the $\delta a$-$\delta \lambda$ plane from the deadband, there is a corresponding constraint on $\delta a$ for the condensed swarm denoted $\delta a_{s,cond}$ and a corresponding constraint on $\delta \lambda$ denoted $\delta \lambda_{s,ei}$ for the E-I vector separation swarm.  However, to guarantee that the $\delta \lambda$ limits $\delta \lambda_{s(\cdot)}$ are not exceeded, the maximum time to find a maneuver once the $\delta a-\delta \lambda$ condition is violated must be considered because only a maneuver that decreases errors in all in-plane ROE will be allowed.  Since any desired directional change in $\delta \bm{e}$ by an in-plane reconfiguration maneuver can be achieved within one orbit, $T_{orb}$, the coupling between $\delta \lambda_{s(\cdot)}$ and $\delta a_{s(\cdot)}$ can be calculated by considering the drift in $\delta \lambda$ over half of the allowed drift time (since the change in $\delta \lambda$ is centered at $\delta \lambda = 0$) plus an additional orbit period $T_{orb}$+$T_{bm}/2$ and is described by
\begin{equation}
\begin{aligned}
	\label{eq:safety_da}
	\delta a_{s(\cdot)} = \frac{2}{3n}\frac{\delta\lambda_{s(\cdot)}}{T_{bm}/2+T_{orb}}
\end{aligned}
\end{equation}

 The control algorithm searches for a maneuver once $\delta \lambda$ violates the limit $\delta \lambda_{s(\cdot)}$ described in Equation~\eqref{eq:safety_da} to ensure safety conditions are met for the condensed swarm and discourage $\delta a$ from exceeding safety limits for the E-I vector swarm due to coupling with $\delta \lambda$.

When the deadband is violated, the desired relative semimajor axis is chosen to maximize the along-track component of the maneuver subject to a constraint on the magnitude of $\delta a$.  This is achieved by equating  $\delta a_{des}$ to the value on the opposite side of the $\delta a- \delta \lambda$ limit but with the same relative mean longitude. By combining Equations~\eqref{eq:dlambda} and~\eqref{eq:kepler}, $\delta a_{des}$ is described by
\begin{equation}
\begin{aligned}
	\label{eq:da_err}
	\delta a_{des} = \frac{2}{3n}\bigg(\frac{2\delta\lambda}{T_{bm}}\bigg)
\end{aligned}
\end{equation}
 
However, this $\delta a_{des}$ must be limited to a maximum $\delta a_{s(\cdot)}$ defined previously in order to ensure safety.

A parameter $\delta a_{err}$ is defined as the difference between the current relative semimajor axis and the desired relative semimajor axis, $\delta a_{des}$, as determined by 
\begin{equation}
\begin{aligned}
	\label{eq:da_err}
	\delta a_{err} = \delta a_{des} - \delta a
\end{aligned}
\end{equation}

By isolating the portion of the control matrix relevant to $\delta a$ and $\delta\mathbf{e}$, the $\Delta v$ is determined from
\begin{equation}
\begin{aligned}
	\label{eq:inplane_err}
	\begin{bmatrix}
	\delta a_{err} \\
	\delta \mathbf{e}_{err} \\
	\end{bmatrix} = \begin{bmatrix}
	\delta a_{err} \\
	||\delta \mathbf{e_{err}}|| \cos{\nu} \\
	||\delta \mathbf{e_{err}}|| \sin{\nu} \\
	\end{bmatrix} = \begin{bmatrix}
	0 & 2 \\
	\sin{u_c} & 2\cos{u_c} \\
	-\cos{u_c} & \sin{u_c} \\
	\end{bmatrix} \begin{bmatrix}
	\Delta v_R \\
	\Delta v_T \\
	\end{bmatrix}
\end{aligned}
\end{equation}
By solving for the optimal $[\Delta v_R$    $\Delta v_T]^T$ and normalizing the vector, the optimal in-plane maneuver direction $\hat{\delta \mathbf{p}}_{ip}$ and location $u_{opt,ip}$ are computed from
\begin{equation}
\begin{aligned}
	\label{eq:inplane_calc}
	\hat{\delta \mathbf{p}}_{ip} = \frac{1}{\sqrt{\Delta v_R^2 +\Delta v_T^2}}\begin{bmatrix}
	\Delta v_R \\
	\Delta v_T \\
	\end{bmatrix} \\
	\Delta v_T = \frac{\delta a_{err}}{2}\\
	u_{opt,ip} = -\arccos{\bigg(\frac{\delta a_{err}}{||\delta \mathbf{e_{err}}||}\bigg)}+\nu = \pm\theta +\nu\\
	\Delta v_R = ||\delta \mathbf{e_{err}}||\sin{(u_{opt}-\nu)}= ||\delta \mathbf{e_{err}}||\sin{(\pm\theta)}\\
\end{aligned}
\end{equation} 

Notably, the calculation in expression \eqref{eq:inplane_calc} is only necessary if a radial $\Delta v$ is required.  For completeness, in the case that the error in $\delta a$ is greater than the error in $\delta \mathbf{e}$, only an along track maneuver is necessary at $u_{opt} = \nu$. If in addition the error in $\delta \mathbf{e}$ is small, less than 15\% of $\delta e_{db}$, then any along-track maneuver that produces the desired $\delta a$ is allowed. This prevents stalling in control due to noise. 

If a radial component is necessary, there are two solutions for the optimal location, denoted $\pm \theta + \nu$.  It should be noted that the radial maneuvers are in opposite directions for the two candidates due to the symmetry of the sine function in Equation \eqref{eq:inplane_calc}.  

\begin{figure}[htb]
	\centering\includegraphics[width=1.8in]{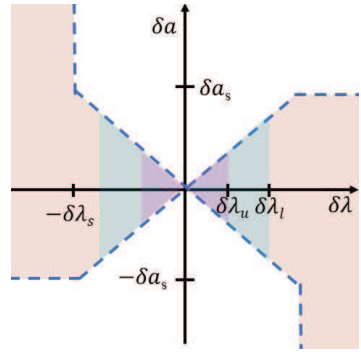}
	\caption{Graphical representation of the limits in the $\delta a - \delta \lambda$ plane.  The red region represents conditions where radial maneuvers are considered that decrease the magnitude of $\delta \lambda$.  The purple region represents conditions where radial maneuvers are considered that increase the magnitude of $\delta \lambda$.  Finally, the blue region indicates conditions where any radial maneuver is allowed.}
	\label{fig:dl_adj}
\end{figure}

The divisions of  $\delta \lambda$ into regions determines which of the maneuvers to implement.  At large magnitudes of $\delta \lambda$, a reduction of $\delta \lambda$ results in a reduction of $\delta a_{des}$ due to coupling in the deadband from Equation \eqref{eq:dlambda}.  This stabilizes the control law and prevents uncontained growth in the swarm size.  Therefore, when the magnitude of $\delta \lambda$ is greater than the user-specified $\delta \lambda_{u}$ limit, only maneuvers that decrease the magnitude of $\delta \lambda$ are permitted.  However, at small $\delta \lambda$ magnitudes, noise in the estimated mean state can cause oscillation across both deadband limits, which creates issues with maneuvering logic.  Therefore, at $\delta \lambda$ magnitudes below the user-specified limit $\delta \lambda_{l}$, maneuvers are implemented that only increase $\delta \lambda$. To smooth the transition between the regions delineated by $\delta \lambda_{u}$ and $\delta \lambda_{l}$ , a third region is added between them that allows either maneuver.  These limits are shown in Figure \ref{fig:dl_adj}.

\subsubsection{Out-of-plane.}
The out-of-plane control problem is solved using a similar approach to the in-plane problem and again assumes one thruster. The desired relative inclination vector is referred to as $\delta \mathbf{i}_{des}$. The user-specified, allowable error in the relative inclination vector is referred to as $\delta \mathbf{i}_{db}$. The objective is to maintain the error within the deadband.  However, unlike the relative eccentricity vector, the drifts in $\delta i$ are small.  Therefore, a maneuver may not need to be considered every time $\delta a$-$\delta \lambda $ violations occur.  Instead, a maneuver is only considered when the error magnitude is greater than half the deadband radius.  In this case, the vector error between the current relative inclination vector $\delta \mathbf{i}$ and the desired vector is referred to as $\delta \mathbf{i}_{err}$.  This is defined in a Cartesian and polar reference as
\begin{equation}
\begin{aligned}
	\label{eq:di_err}
    \delta\mathbf{i}_{err} = ||\delta\mathbf{i}_{err}|| \begin{bmatrix} \cos{\gamma} \\
    \sin{\gamma}
    \end{bmatrix} = \begin{bmatrix} \delta\mathbf{i}_{x,des}-\delta\mathbf{i}_x \\
    \delta\mathbf{i}_{y,des}-\delta\mathbf{i}_y
    \end{bmatrix}
\end{aligned}
\end{equation}
where $\gamma$ represents the phase of $\delta \mathbf{i}_{err}$ as measured counter-clockwise from the positive x-axis.    The relationship between these parameters can be described by a portion of the control matrix as
\begin{equation}
\begin{aligned}
	\label{eq:outplane_err}
	\begin{bmatrix}
	\delta \mathbf{i}_{err} \\
	\end{bmatrix} = \begin{bmatrix}
	||\delta \mathbf{i_{err}}|| \cos{\gamma} \\
	||\delta \mathbf{i_{err}}|| \sin{\gamma} \\
	\end{bmatrix} = \begin{bmatrix}
	\cos{u} \\
	\sin{u} \\
	\end{bmatrix} 
	\Delta v_N \\
\end{aligned}
\end{equation}

Equation~\eqref{eq:outplane_err} has two solutions separated by half an orbit in location given by
\begin{subequations}
\begin{eqnarray}
	\label{eq:outplane_sol1}
	u_{opt,oop} = \gamma, \Delta v_n = ||\delta \mathbf{i_{err}}|| \\ \label{eq:outplane_sol2}
	u_{opt,oop} = \gamma+\pi, \Delta v_n = -||\delta \mathbf{i_{err}}|| 
\end{eqnarray}
\end{subequations}

Both maneuvers are equivalent for implementation.  However, the next section demonstrates that one of these maneuvers is more desirable depending on the 6-D ROE error state.

\subsubsection{Combining Planes.}
In the previous sections, the optimal maneuver directions and locations for the planar problems were obtained.  However, a maneuver that decreases both $\delta \mathbf{e}$ and $\delta \mathbf{i}$ error is more desirable to avoid disruption of data acquisition.  

\begin{figure}[htb]
	\centering\includegraphics[width=2in]{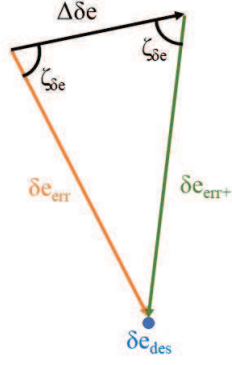}
	\caption{Graphical representation of the angular change between a maneuver and the error vector to maintain the same error magnitude.}
	\label{fig:zeta}
\end{figure}

As stated in recent work by Koenig \cite{Koenig_swarm}, any maneuver that produces a change within 90\degree \ of the error direction reduces state error for a continuous thrust maneuver.   For an impulsive maneuver, the angle is slightly less than 90\degree.  Let $\zeta_{\delta e}$ denote the angle between $\delta \mathbf{e}_{err}$ and a change in $\delta\mathbf{e}$ produced by a maneuver.  Consider the relative eccentricity problem defined in Figure \ref{fig:zeta}, where $\delta \mathbf{e}_{err}$ and $\delta \mathbf{e}_{err+}$ are the errors in the relative eccentricity vector before and after a maneuver, respectively, and $\Delta \delta \mathbf{e}$ indicates the change in the relative eccentricity vector from a maneuver.  The angle $\zeta_{\delta e}$ required to ensure the magnitude of $\delta \mathbf{e}_{err}$ and $\delta \mathbf{e}_{err+}$ are equivalent is found through simple trigonometry of an isosceles triangle to be

\begin{equation}
\begin{aligned}
	\label{eq:zeta}
	\zeta_{\delta e} = \arccos \bigg ( \frac{\Delta \delta \mathbf{e}/2}{||\delta \mathbf{e}_{err}||}\bigg )
\end{aligned}
\end{equation}

Any angle smaller than $ \zeta_{\delta e} $ will result in $||\delta \mathbf{e}_{err+}|| \leq ||\delta \mathbf{e}_{err}||$, which is a decrease in the error magnitude. However, $\zeta_{\delta e}$ remains large for large errors because the argument of the arccosine function approaches zero, so the arccosine function approaches 90\degree. For the thruster and orbit simulated here, the angle $\zeta_{\delta e}$ remains greater than 80\degree for any $||a\delta\mathbf{e}_{err}||$ greater than 17 m and greather than 60\degree for any errors greater than 3 m. 

The same equations and angular limitations apply to $\zeta_{\delta i}$, which denotes the angle between $\delta \mathbf{i}_{err}$ and a change in $\delta\mathbf{i}$ produced by a maneuver.

\begin{figure}[htb]
	\centering
	\begin{subfigure}[htb]{0.3\textwidth}
	\centering
	\caption{}
	\includegraphics[width=1.5in]{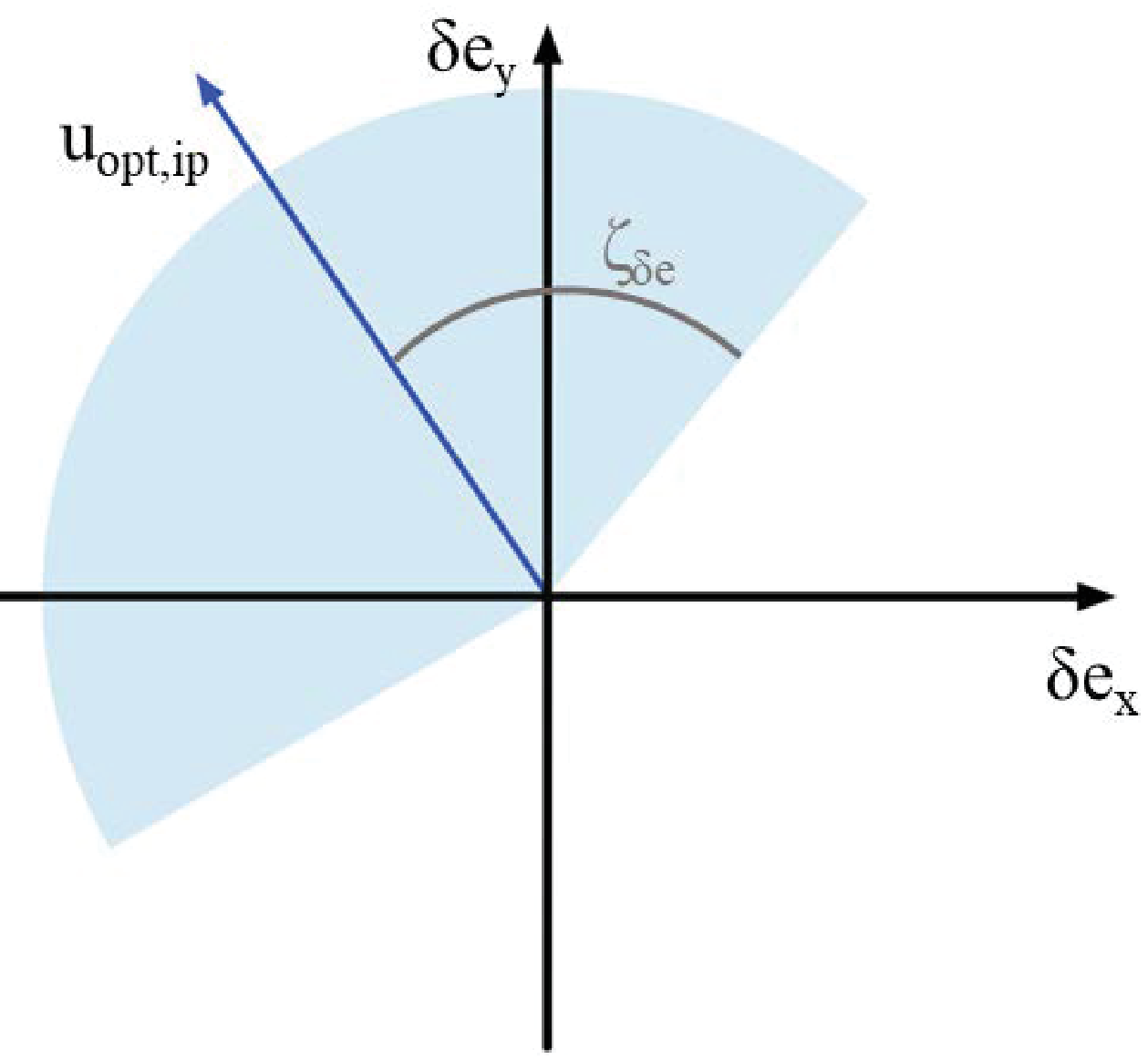}
	\end{subfigure}
	\begin{subfigure}[htb]{0.3\textwidth}
	\centering
	\caption{}
	\includegraphics[width=1.5in]{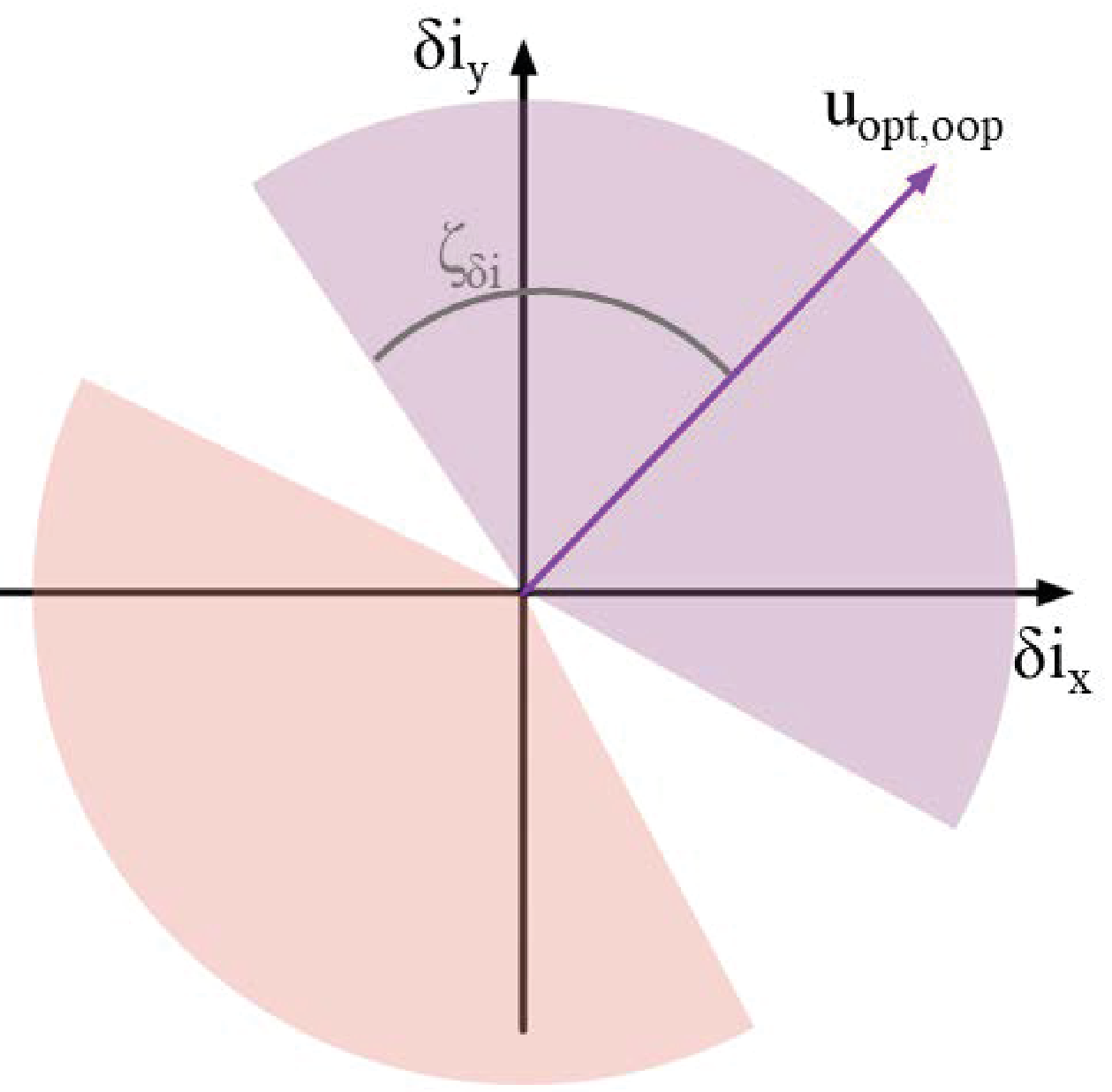}
	\end{subfigure}
	\begin{subfigure}[htb]{0.3\textwidth}
	\centering
	\caption{}
	\includegraphics[width=1.5in]{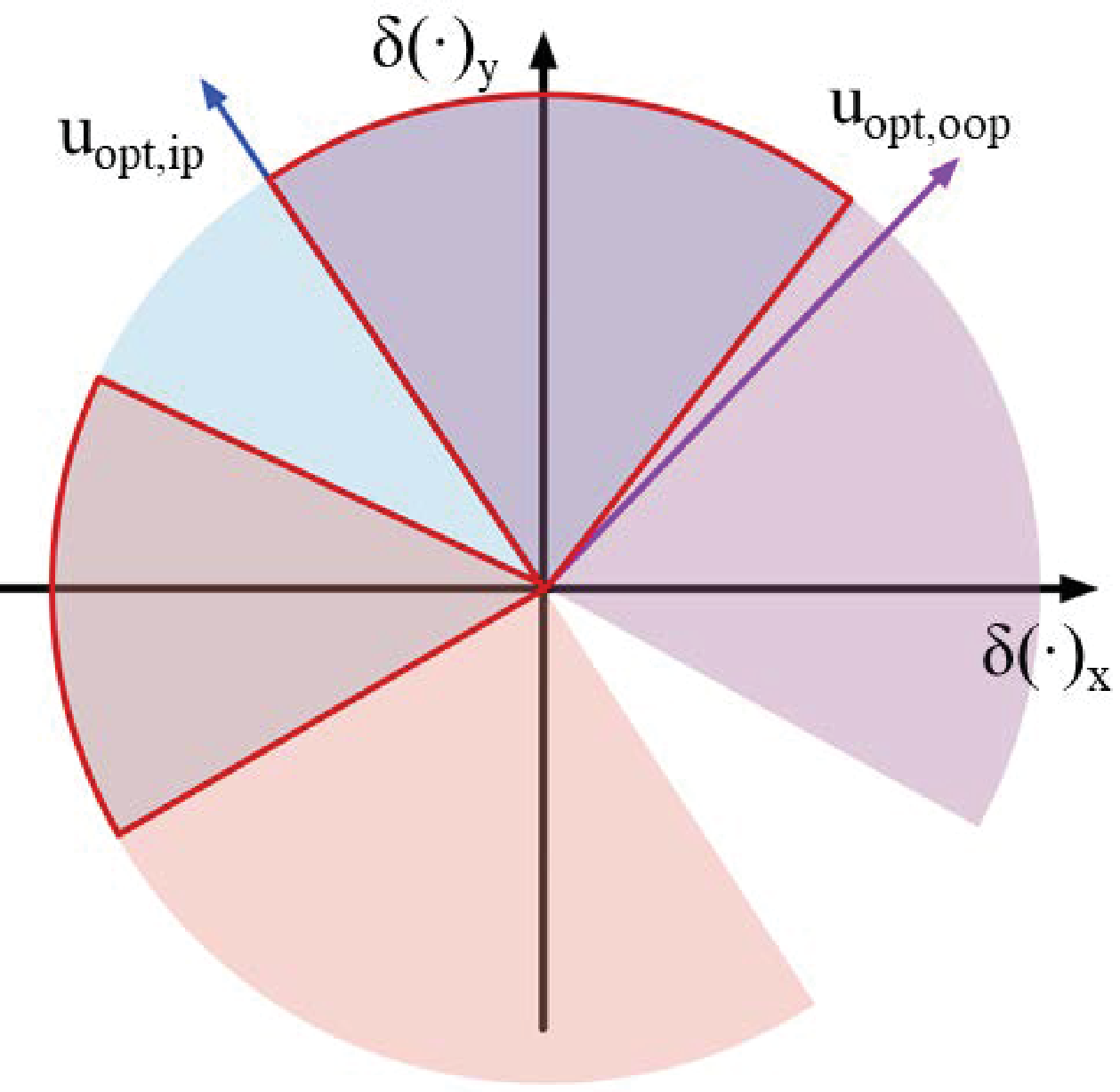}
	\end{subfigure}
	\caption{Subfigure a represents the desired change in mean $\delta \mathbf{e}$ defined by the maneuver location $u_{opt,ip}$ with a blue line, and the light blue region demonstrates a sub-optimal maneuver that reduces the $\delta \mathbf{e}$ tracking error. Subfigure b represents the optimal desired change in mean $\delta \mathbf{i}$ defined by a maneuver in the positive out-of-plane direction at the location $u_{opt,oop}$ with a solid line and a maneuver along the negative out-of-plane direction at 180\degree from $u_{opt,oop}$ with a dashed line.  The regions defined by purple defines sub-optimal maneuvers that still reduces the $\delta \mathbf{i}$ tracking error. Subfigure c illustrates the regions when maneuvers for both $\delta \mathbf{e}$  and $\delta \mathbf{i}$  tracking errors decrease.}
	\label{fig:overlap}
\end{figure}

The allowable optimal maneuvers are shown in Figure~\ref{fig:overlap}. The purple and blue line denote an optimal out-of-plane and in-plane maneuver, respectively.  These maneuvers correspond to the solutions of Equations~\eqref{eq:outplane_sol1} or ~\eqref{eq:outplane_sol2} and~\eqref{eq:inplane_calc}, respectively.  The corresponding color-coded areas represent the allowable changes $\zeta_{\delta e}$ in blue and $\zeta_{\delta i}$ in purple that reduce the control tracking error, where the control tracking error refers to the instantaneous error in $\delta \mathbf{e}$ and $\delta \mathbf{i}$.  

Figure \ref{fig:overlap} demonstrates that multiple maneuver locations exist such that the errors in $\delta \mathbf{e}$ and $\delta \mathbf{i}$ decrease, as represented by regions outlined in red.  Minimum overlap in feasible regions for out-of-plane and in-plane maneuvers occurs when the desired maneuver locations for the in-plane and out-of-plane maneuvers are separated by 90\degree in mean argument of latitude.  Therefore, to guarantee a possible maneuver exists, the following condition must hold 
\begin{equation}
\begin{aligned}
	\label{eq:zeta}
	\zeta_{\delta e}+\zeta_{\delta i} \geq 90\degree
\end{aligned}
\end{equation}

Using these allowable regions, a maneuver with RTN-direction components calculated from Equations~\eqref{eq:inplane_calc} and \eqref{eq:outplane_sol1} or ~\eqref{eq:outplane_sol2} is implemented at the current time if the following angular conditions are met
\begin{subequations}
\begin{eqnarray}
	\label{eq:inplane_ang}
	\frac{\hat{\delta \mathbf{e}_i}}{||\hat{\delta \mathbf{e}_i}||} \cdot \frac{\delta \mathbf{e}_{err}}{||\delta \mathbf{e}_{err}||} \geq \cos{\zeta_{\delta e}}
	\\ \label{eq:outplane_ang}
	\frac{\hat{\delta \mathbf{i}_i}}{||\hat{\delta \mathbf{i}_i}||} \cdot \frac{\delta \mathbf{i}_{err}}{||\delta \mathbf{i}_{err}||} \geq \cos{\zeta_{\delta i}} 
\end{eqnarray}
\end{subequations}
where $\hat{\delta \mathbf{e}_i}$ and $\hat{\delta \mathbf{i}_i}$ correspond to the changes in $\delta \mathbf{e}$  and $\delta \mathbf{i}$ created by the maneuver.  Note that only the condition in Equation \eqref{eq:inplane_ang} must be met when an out-of-plane maneuver is not necessary.  In this case, the window for the in-plane maneuver is expanded to the entire range 2$\zeta_{\delta e}$ in mean argument of latitude.  The algorithm enacts a specified pulse if conditions are met.

A summary of the control law is provided in Algorithm I where $\mathbf{q}_{(\cdot)}$ refers to potential maneuver candidates calculated by Equations~\eqref{eq:inplane_calc}, \eqref{eq:outplane_sol1}, and~\eqref{eq:outplane_sol2} at the current time step, and D is the scaling of this maneuver candidate by the thruster acceleration.  The inputs to the algorithm are the user defined parpameters ($\delta \bm{e}_{db}$, $\delta \bm{i}_{db}$, $\zeta_{\delta \bm{e}}$, $\zeta_{\delta \bm{i}}$), thruster-produced acceleration $D$ given the satellite mass $m$, the current ROE, and the desired ROE.  The output is the commanded thruster acceleration.

The algorithm only considers the current time horizon in executing a pulse, simplifying computation by avoiding predictive calculations and reducing reliance on the dynamic propagation.  Furthermore, the algorithm reduces state error in all components using a specified window for maneuvering for full state, decoupled plane control and ensures periods of time free from maneuvers, which could not be guaranteed with decoupled planar control.  Future work will extend this algorithm to eccentric orbits and investigate consideration of future time horizons to pre-plan impulsive maneuvers.  

\begin{algorithm}
\label{alg}
    Calculate $\delta\lambda_{lim}$ \\
    \eIf{$|\delta\lambda|$ $>$ $|\delta\lambda_{lim}|$}
      {
        Calculate $\delta a_{err}$, $\delta \mathbf{e_{err}}$, and $\delta \mathbf{i_{err}}$ \\
        Calculate $\hat{\delta \mathbf{p}}_{ip}$ \\
        \eIf{ $||\delta \mathbf{i_{err}}||$ $>$ $\delta i_{db}/2$}
        {
            $\delta \mathbf{q}_1$ = D$\begin{pmatrix} \hat{\delta \mathbf{p}}_{ip} \\ 1  \end{pmatrix}$ \\
            $\delta \mathbf{q}_2$ = D$\begin{pmatrix} \hat{\delta \mathbf{p}}_{ip} \\ -1  \end{pmatrix}$ \\
            Calculate $\hat{\delta\alpha}_i = \mathbf{\Gamma}(\alpha(t))\delta\mathbf{q_i}$ \\
            \eIf{$\frac{\hat{\delta \mathbf{e}_i}}{||\hat{\delta \mathbf{e}_i}||}\cdot\frac{\delta\mathbf{e}_{err}}{||\delta\mathbf{e}_{err}||} \geq \cos{\zeta_{\delta e}}$ and $\frac{\hat{\delta \mathbf{i}_i}}{||\hat{\delta \mathbf{i}_i}||}\cdot\frac{\delta\mathbf{i}_{err}}{||\delta\mathbf{i}_{err}||} \geq \cos{\zeta_{\delta i}}$}
            {
            $\delta\mathbf{p} = \delta\mathbf{q}_i$
            }
            {
            $\delta\mathbf{p} = 0$
            }
        }
            {
            $\delta \mathbf{q}$ = $\begin{pmatrix} \hat{\delta \mathbf{p}}_{ip} \\ 0  \end{pmatrix}$ \\
            \eIf{$\frac{\hat{\delta \mathbf{e}_i}}{||\hat{\delta \mathbf{e}_i}||}\cdot\frac{\delta\mathbf{e}_{err}}{||\delta\mathbf{e}_{err}||} \geq \cos{\zeta_{\delta e}}$}
                        {
            $\delta\mathbf{p} = \delta\mathbf{q}$
            }
            {
            $\delta\mathbf{p} = 0$
            }
        }
      }
      {
      $\delta\mathbf{p} = 0$
      }
    \caption{Autonomous formation-keeping algorithm for asteroid-orbiting swarms}
\end{algorithm}

\section{Validation}
The previously described advancements in the dynamics model, formation keeping, and osculating-to-mean conversions are validated through simulation of an example mission about the Eros-like asteroid presented in the Orbit Propagation section.  The parameters for the ground truth generation from numerical integration are identical to the propagator specifications described in Table~\ref{tab:dynamic_param}.  Notably, the ground-truth dynamics differs from the dynamics used in the guidance and control section.  Therefore, process noise exists which the EKF and control must overcome.

In this simulation, the deputy spacecraft are modeled with a cross-sectional area of 0.09 m\textsuperscript{2} and a mass of 5 kg to represent a 3U cubesat with small deployable solar panels.  For the solar panels, an SRP coefficient of 1.2 is used. A differential ballistic coefficient with respect to the reference used in simulation is 2\%.  This represents a few degrees of attitude error from a nominal orientation or a difference from a representative larger satellite that could be the mothership.   

Table~\ref{tab:mother} shows the initial condition of the chief satellite for both the condensed and E-I vector swarm simulations.   The reference orbit was chosen to be within the lower half of the altitudes NEAR-Shoemaker visited\cite{NEAR} and to have passive stability over 100 orbits, hence the highly retrograde inclination.  The passive stability over 100 orbits is desired to ensure all errors originate from the relative motion EKF and formation-keeping algorithm and not from absolute orbit maintenance.  Notice that the orbit of E-I vector swarm is simulated at a higher altitude.  This results from the fact that the rotation rate of $\delta \mathbf{e}$ due to $J_2$ at a semimajor axis of 70 km becomes too fast for close spacecraft operation without requiring maneuvers every orbit or expanding the swarm size to ensure safety under these large drifts.  Note that one of the assumptions of ANS is that the $J_2$ parameter is known reasonably well due to light-curve data and previous mission phases.  Therefore, the altitude for reasonable E-I vector separation swarm operation is known before the asteroid is visited. The control approach is robust to errors ($< 10$\%) in this estimate, but to provide uninterrupted imaging on the order of orbit periods, some low orbit altitudes are not realistic. Due to this fact, the E-I vector swarm is smaller and consists only of three spacecraft with 2 deputies and one chief, as larger swarms require larger relative eccentricity vectors, which exacerbates the perturbation from $J_2$. Comparatively, the condensed swarm consists of 5 satellites with 4 deputies and a chief.  The condensed swarm is unaffected by the rotation rate of $J_2$ because it does not compensate for this rotation. Table~\ref{tab:swarm} shows the geometry of the swarm spacecraft relative orbits upon initialization. The swarm geometries are presented as magnitudes and phases of the relative inclination and relative eccentricity vector, where the phase of  $\delta \mathbf{e}$ and $\delta \mathbf{i}$ are represented by $\phi$ and $\psi$, respectively.

\begin{table}[htb]
    \caption{Initial orbit parameters for the mothership used in the realistic mission simulation.}
    \label{tab:mother}
    \centering
    \begin{tabular}{c c c c c c }
     \noalign{\hrule height 2pt}
    Swarm &a (km) & e(-) & i(\degree) & $\Omega$(\degree) & $\omega$(\degree)  \\ [0.5ex] 
    \hline\hline
     E-I vector & 90 & 0.0001 & 160 & 0 & 136 \\ 
     \hline
     Condensed & 70 & 0.0001 & 160 & 0 & 136 \\ 
    \noalign{\hrule height 2pt}
    \end{tabular}
\end{table}

\begin{table}[htb]
    \caption{Initial mean relative orbit parameters for swarm geometries used in the realistic mission simulation.}
    \label{tab:swarm}
    \centering
    \begin{tabular}{c c c c c c} 
    \noalign{\hrule height 2pt}
    Swarm  &  Deputy  & $a\delta \mathbf{e}$ (m) & $\phi$(\degree) & $a\delta \mathbf{i}$ (m) & $\psi$(\degree)  \\ [0.5ex] 
     \hline\hline
     \multirow{2}{*}{E-I Vector} & 1 & 230 & 90 & 230 & 90 \\ 
     & 2 & 230 & -90 & 230 & -90 \\
     \hline
     \multirow{4}{*}{Condensed} & 1 & 600 & 0 & 600 & -90 \\ 
      & 2 & 600 & 90 & 600 & 90 \\
      & 3 & 600 & 180 & 600 & 90 \\
      & 4 & 600 & -90 & 600 & -90 \\
     \noalign{\hrule height 2pt}
\end{tabular}
\end{table}
Table~\ref{tab:control} provides relevant control and noise parameters.  The parameters are chosen based on the estimated drift rates due to SRP and gravitational potential terms $J_2$ through $J_4$ given the swarm spacecraft and chief orbit. Furthermore, the thruster acceleration is chosen to match the throttled Busek BET-100 \cite{Busek_thrust,Lemmer} actuator because it is representative of a commercially available, low-thrust propulsion system designed for cubesats and can supply a $\Delta v$ up to 36 m/s for a 5 kg cubesat.  Additionally, Gaussian white noise is added to the osculating absolute and relative orbital elements to represent realistic navigation measurements. \cite{Stacey} Notably, the two $T_{bm}$ values are different.  This ensures that the drifts can be maintained to within specified windows, as the E-I vector swarm must counter a larger drift because the rotation of $\delta \mathbf{e}$ is also compensated for.  Using these parameters, a simulation is conducted for 100 orbits, which is equivalent to 200 Earth days for the condensed swarm and 293 Earth days for the E-I vector swarm. 

\begin{table}[htb!]
    \caption{Control and noise parameters implemented in the realistic mission simulation.}
    \label{tab:control}
    \centering
    \begin{tabular}{p{3cm} p{3cm}| p{3cm} p{3cm} } 
    \noalign{\hrule height 2pt}
    Parameter & Value & Parameter & Value \\
    \noalign{\hrule height 2pt}
    $\zeta_{de}$ & 50\degree & $a\delta \lambda_{u}$ E-I vector & 190 m  \\ 
     \hline
     $\zeta_{di}$ & 60\degree  &  $\epsilon$ & 100 m  \\ 
     \hline
     $\delta e_{db}$  & 100 m &  thrust & 50 $\mu$N   \\ 
     \hline
      $\delta i_{db}$ & 60 m & $\sigma^2_{\delta \alpha}$ & 5 m$^2$ \\ 
     \hline
      $T_{bm}$ E-I vector &  2 Orbits   & $\sigma^2_a$ & $1.7$ m$^2$ \\
     \hline
      $T_{bm}$ Condensed & 3 orbits & $\sigma^2_u$   & $2.5 e{-10}$ rad$^2$ \\
     \hline
      $a\delta \lambda_{l}$ condensed & 210 m  & $\sigma^2_{e_x}$ & $5.0 e{-11}$   \\
     \hline
      $a\delta \lambda_{u}$ condensed & 280 m   & $\sigma^2_{e_y}$  & $4.4 e{-11}$\\ \hline
      $a\delta \lambda_{l}$ E-I vector & 140 m  & & \\
     \noalign{\hrule height 2pt}
\end{tabular}
\end{table}

The results from the formation-keeping implementation are shown in Table \ref{tab:keep}.  They demonstrate that the minimum separation of 100 m is met for both swarms.  Notably, the separation for the condensed swarm is much larger.  This is due to the fact that the minimum separation is based on a worst case scenario where $\delta \lambda$ is at the safety limit and $\delta \mathbf{e}$ is at opposing sides of the deadband for two spacecraft.  The likelihood that this occurs simultaneously is small, given the fact that the SRP causes $\delta \mathbf{e}$ to drift dominantly in the same direction.  Even so, the minimum separation is smaller than that predicted by the instantaneous ROE in simulation.  This results from the fact that the osculating absolute orbit eccentricity has risen to 0.17 from SRP and the gravity potential perturbations, causing deterioration of the near-circular mapping.  Future implementation of absolute orbit control will rectify this issue.

\begin{table}[htb!]
    \caption{Formation-Keeping results from sample asteroid mission simulation.}
    \label{tab:keep}
    \centering
    \begin{tabular}{p{9cm} p{2.5cm} p{2.5cm}} 
    \noalign{\hrule height 2pt}
       Evaluation Metric & Condensed & E-I Vector \\
       \noalign{\hrule height 2pt}
       Minimum Separation (m) & 206 & 107 \\
       \hline
       Deputy 1 $\Delta v$ (cm/s)  & 10.5 & 7.8 \\
       \hline
       Deputy 2 $\Delta v$ (cm/s)  & 6.7 & 6.5\\
       \hline
       Deputy 3 $\Delta v$ (cm/s)  & 7.7 & -- \\
       \hline
       Deputy 4 $\Delta v$ (cm/s)  & 11.8 & --  \\
       \hline
       Average Time Between Maneuver Windows (Orbits) & 2.92 & 1.92 \\
       \noalign{\hrule height 2pt}
\end{tabular}
\end{table}

In terms of fuel usage, all spacecraft require less than 12 cm/s  and 8 cm/s of $\Delta v$ over 100 orbit periods for the condensed and E-I vector swarm, respectively.  This fuel usage is less than 1\% of the available.  Generally, fuel usage varies by satellite in both swarms due to various, unmodelled perturbations and long-period effects that depend on the specific ROE.  One benefit of the condensed swarm is that these effects will likely balance over longer periods as the $\delta \mathbf{e}$ vectors rotate over time.  

\begin{figure}[htb]
	\centering
	\begin{subfigure}[htb]{0.3\textwidth}
	\centering
	\caption{}
	\includegraphics[width=1.5in]{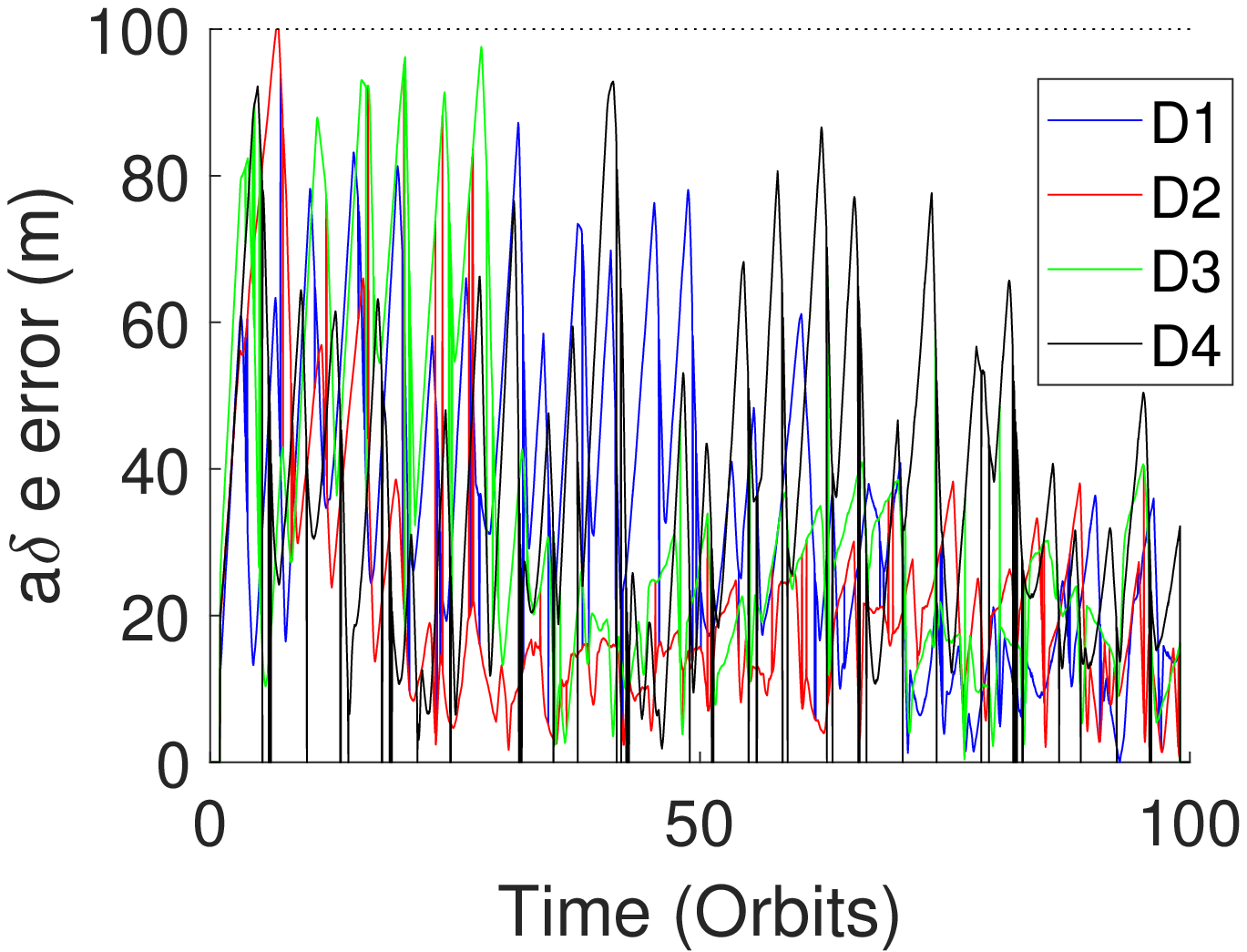}
	\end{subfigure}
	\begin{subfigure}[htb]{0.3\textwidth}
	\centering
	\caption{}
	\includegraphics[width=1.5in]{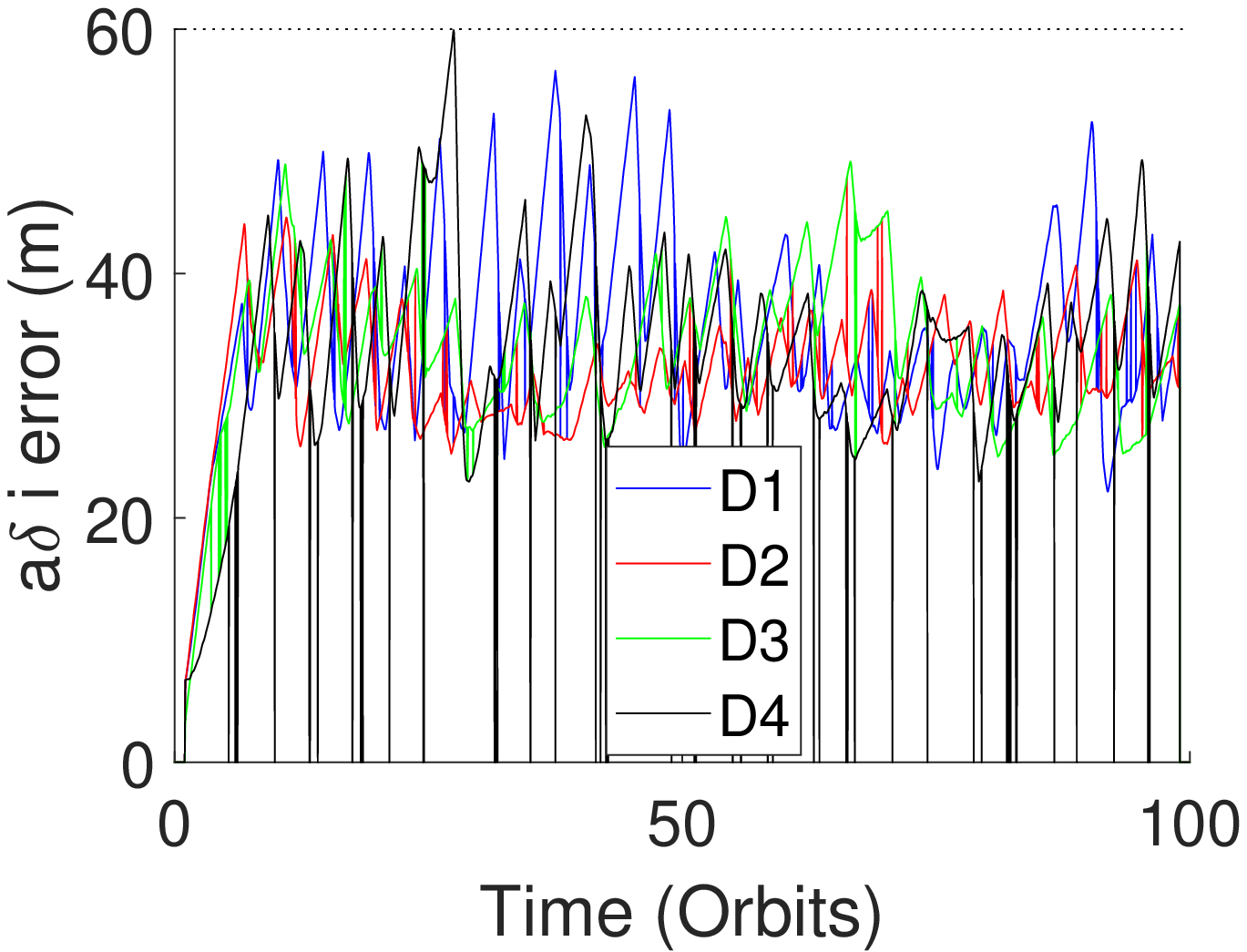}
	\end{subfigure}\\
	\centering
	\begin{subfigure}[htb]{0.3\textwidth}
	\centering
	\caption{}
	\includegraphics[width=1.5in]{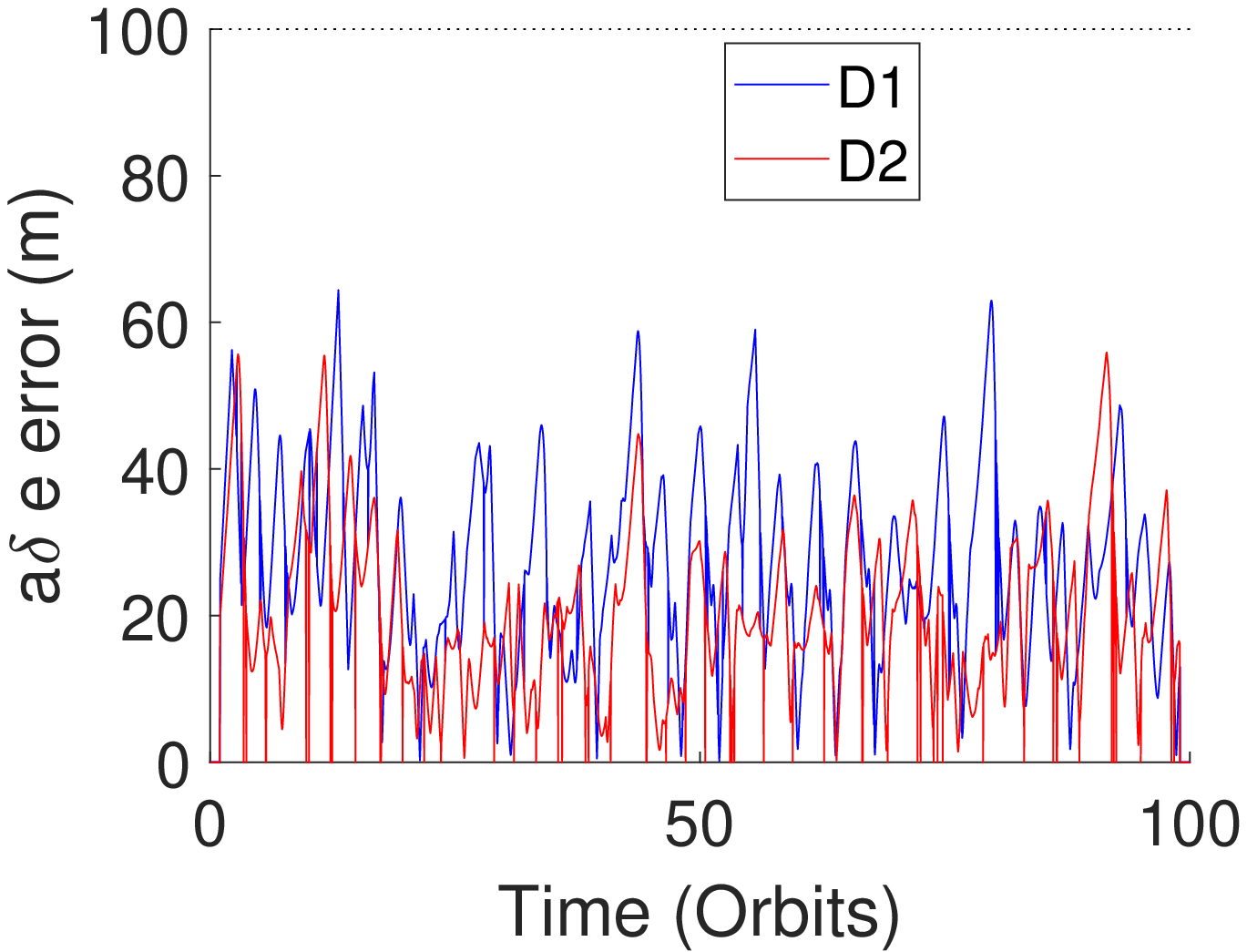}
	\end{subfigure}
	\begin{subfigure}[htb]{0.3\textwidth}
	\centering
	\caption{}
	\includegraphics[width=1.5in]{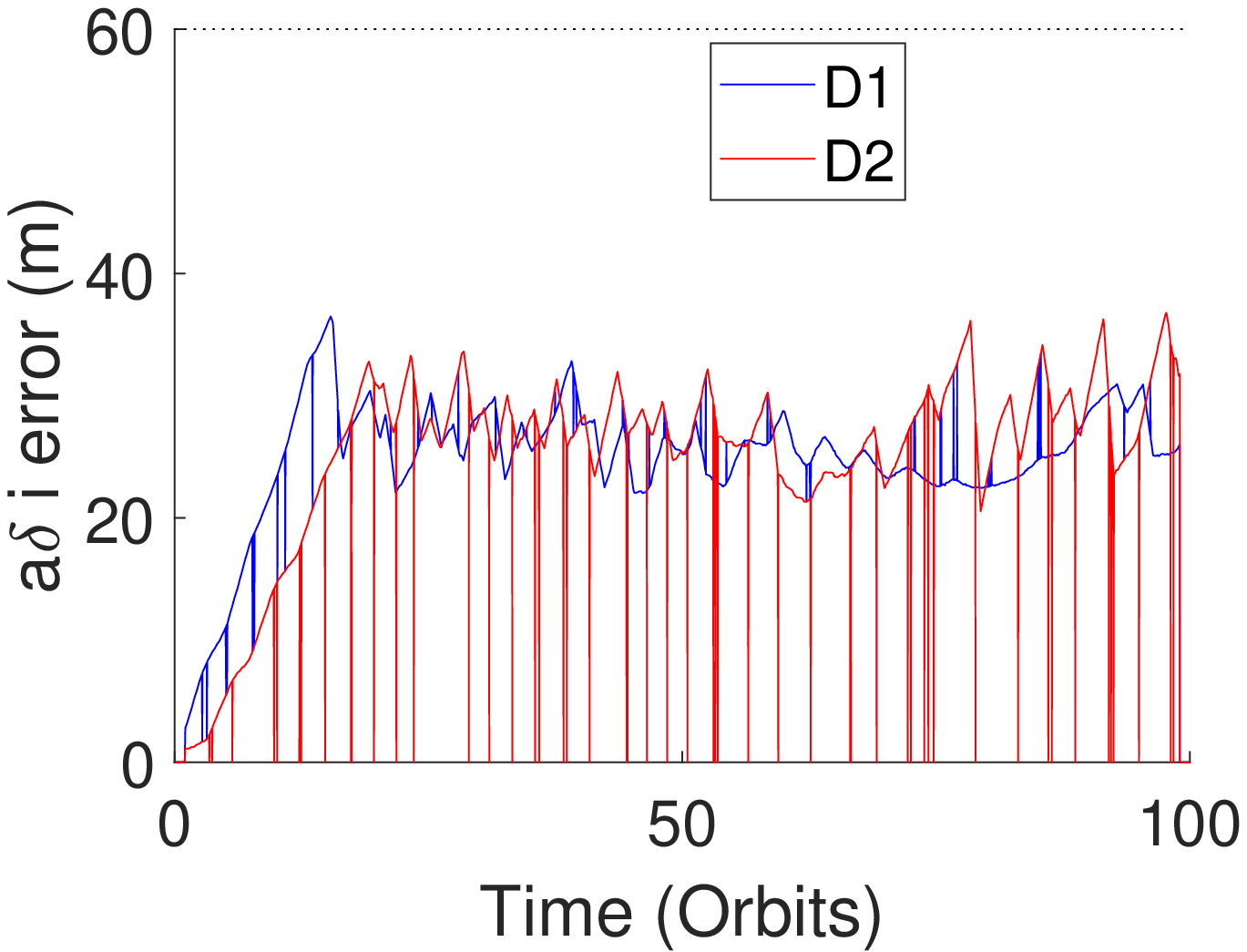}
	\end{subfigure}\\
	\caption{Error in the mean relative eccentricity and inclination vectors plotted over simulation time.  Subfigures a and b present the data for the condensed swarm design, and subfigures c and d present the data for the E-I vector swarm design.  The deadband limits are shown using the dashed black line.  The D refers to the different deputies, where the number matches the assignment in Table \ref{tab:swarm}.}
	\label{fig:fk_res}
\end{figure}

Figure  \ref{fig:fk_res} shows the error achieved between the desired and ground-truth mean relative eccentricity and inclination vectors.  Note that the deadband limits are shown using a dashed black line.  The results for the condensed swarm are presented in subfigures a and b.  The results for the E-I vector swarm are presented in subfigures c and d.  As demonstrated, the control law continually corrects for the drifts that create deviation over the mission lifetime and reduces error.  Notably, the error accumulated between maneuvers varies significantly due to unmodeled dynamics and the fact that the nominal vector is not always returned to the exact center of the deadband, which can actually increase the time the error stays in the deadband.  The relative eccentricity and relative inclination vector deadbands are not violated with the exception of one occurrence.  In this occurance, $\delta \mathbf{e}_{db}$ is exceeded by less than 2 m when the ground-truth mean ROE are poorly defined.  The ground-truth mean ROE are poorly defined within a half orbit period of a maneuver because the averaging period includes an instantaneous jump in the ROE.  Therefore, the instantaneous jump is smoothed over one orbit period of the mean ROE.

\section{Conclusion} 
To enable autonomous operations about asteroids, the Autonomous Nanosatellite Swarming (ANS) architecture was proposed to reduce reliance on the Deep Space Network and improve asteroid characterization accuracy through autonomy.  The ANS architecture imposes three guidance and control requirements: an accurate but computationally efficient dynamics model, an asteroid agnostic osculating-to-mean conversion, and a formation-keeping strategy.  All of these requirements have been addressed in this paper.

For one, a dynamics model is developed using averaging-theory models, which are adapted and combined with an available solar radiation pressure perturbation model to produce a propagator for absolute and relative motion. The resulting semi-analytical propagator is computationally efficient for autonomous control purposes and demonstrates accuracy for relative motion prediction within 30 m in terms of relative orbital elements (ROE) over 5 orbit periods in a worst-case asteroid gravity environment.  

Second, an asteroid-agnostic osculating-to-mean state estimation is presented for relative motion. This is accomplished by leveraging the previously developed dynamics model and the slowly varying mean orbital element state in an extended Kalman filter. The extended Kalman filter demonstrated meter level precision.

Thirdly, an understanding of the dynamics informed the development of a formation-keeping guidance profile, which was defined for two swarm geometries: the condensed and E-I vector swarm.  The guidance profile ensures swarm safety but also enables stereoscopic imaging and radiometric cross-link measurements for science operations by guaranteeing maneuver-free operation windows.  

Simulations of a reference mission demonstrate the benefits of the three presented developments.  The resulting formation keeping profile was demonstrated to use little delta-v with satellites using less than 12 cm/s  and 8 cm/s over 100 orbit periods for the condensed and E-I vector swarm, respectively.  Thus, the use of both radial and along-track maneuvers are justified.  Furthermore, the swarm did not evaporate, and the minimum guaranteed separation of 100 m was reached for both swarms. 

However, a limitation of the presented work is the absence of absolute orbit control.  Currently, it is assumed that the spacecraft are controlled to a quasi-stable reference that is allowed to evolve over time. Maintenance to a desired reference has not been addressed.  Furthermore, improvement of the osculating-to-mean conversion can be achieved using adaptive filtering processes to make the filter more robust to varying separations between spacecraft, as the oscillations in the osculating orbital elements increase with inter-spacecraft separation.  Also, future work should address thruster failure in order to guarantee safety in such cases.  Given these limitations, future work should focus on absolute orbit maintenance, advancement of the osculating-to-mean extended Kalman Filter, and contingency planning.  

This paper advances the state-of-the-art in autonomous guidance and control of asteroid swarming missions.  For one, an asteroid dynamics model and osculating-to-mean conversion are developed for arbitrary asteroids.  Secondly, this paper addresses, for the first time, ROE-based control of a swarm in asteroid orbits. These developments aim to enable swarming about arbitrary bodies for future science missions and even commercial mining.

\section{Acknowledgment}
This material is based upon work supported by the National Science Foundation Graduate Research Fellowship Program under Grant No. DGE-1656518. Any opinions, findings, and conclusions or recommendations expressed in this material are those of the authors and do not necessarily reflect the views of the National Science Foundation. This research is also part of the Autonomous Nanosatellite Swarming (ANS) Using Radio-Frequency and Optical Navigation project supported by the NASA Small Spacecraft Technology Program cooperative agreement number 80NSSC18M0058.




\newpage
\section{Notation}

\begin{table}[htb]
    \caption{Symbols}
    \label{tab:not1}
    \centering
    \begin{tabular}{p{1cm} p{5.7cm} p{1cm} p{5.7cm} } 
    \noalign{\hrule height 2pt}
    $t$ & continuous time &  $\phi$ & phase of relative eccentricity vector     \\
     $a$ &  semimajor axis &  $\delta \mathbf{p}$ & vector of thruster acceleration \\
    $e$ &  eccentricity magnitude & $\mathbf{\Gamma}$ & ROE near-circular orbit control matrix   \\
    $i$ &  inclination & $\nu$ & phase of $\delta\mathbf{e}$ errors\\
    $\Omega$ &  right ascension of the ascending node & $T$ & specified window of time  \\
    $\omega$ &  argument of perigee & $\Delta v$ & instantaneous change velocity \\
     $e_x$ &  eccentricity vector x-component & $\theta$ & arccosine of in-plane errors  \\
      $e_y$ &  eccentricity vector y-component & $\hat{\delta \mathbf{p}_{ip}}$ & in-plane maneuver direction  \\
    $u$ &  mean argument of latitude & $\gamma$ & phase of the relative inclination vector  \\
    $n$ & mean motion  & $\zeta$ & range of allowable maneuvers \\
    $R$ & disturbing function &  $D$ & thrust magnitude \\
    $\delta \alpha$ & set of relative orbital elements & $\mathbf{q}$ &  maneuver candidate \\ 
    $\delta a$ & relative semimajor axis &  $\epsilon$ & user-specified safe separation \\
    $\delta \lambda$ & relative mean longitude &  $\rho$ & in-plane separation parameter \\
    $\delta \mathbf{e}$ & relative eccentricity vector & $\xi$ & in-plane separation parameter\\
    $\delta \mathbf{i}$ & relative inclination vector & $\tau$ & in-plane safety parameter \\
    $\delta \mathbf{r}$ & relative position  &  $\bm{\beta}$ & Jacobian \\
    $\delta \mathbf{v}$ & relative velocity &  $\bm{\kappa}$ & diagonal matrix of maneuver uncertainty \\
    $J_{(\cdot)}$ & normalized zonal coefficient & $\xi$ & maneuver uncertainty parameter \\
     $B$ & SRP ballistic coefficient &   $\mathbf{P}$ & current covairiance matrix   \\
      $\Delta B$ & differential ballistic coefficient &  $\mathbf{Q}$ & process noise matrix\\
   $C_{SRP}$ & reflectivity coefficient & $\mathbf{K}$ & Kalman Gain \\
    $A$ &  cross-sectional area &  $\mathbf{H}$ & measurement sensitivity matrix  \\
    $m$ & mass & $\mathbf{S}$ & measurement covariance\\
    $\mu$ & gravitational parameter of central body  & & \\
    \noalign{\hrule height 2pt}
\end{tabular}
\end{table}

\begin{table}[htb!]
    \caption{Subscript}
    \label{tab:not2}
    \centering
    \begin{tabular}{p{1cm} p{5cm} p{1cm} p{5cm} } 
    \noalign{\hrule height 2pt}
   $x$ & x-component of vector &  $opt$ & optimal  \\
   $y$ & y-component of vector & $ip$ & in-plane \\
   $c$ & chief satellite & $oop$ & out-of-plane \\
   $d$ & deputy satellite &  $\delta e$ & parameter relevant to $\delta e$ plane \\
   $r$ & radial & $\delta i$ & parameter relevant to $\delta i$ plane \\
   $t$ & along-track &  $s$ & safety limit \\
   $n$ & out-of-plane & $orb$ & orbit \\
   $des$ & desired vector & $bias$ & constant bias \\
   $db$ & deadband &   $k$ & current time step \\
   $err$ & error & $u$ & upper limit\\
   $kep$ & kepler effects & $l$ & lower limit\\
   $bm$ & between maneuver & & \\
   \noalign{\hrule height 2pt}
\end{tabular}
\end{table}

\appendix
\section{Geopotential Effects}
\subsection{$J_2$}
\begin{equation}
\begin{split}
\frac{da}{dt} = 0 \\
\\
\frac{du}{dt} = \frac{3}{4} n J_2 \bigg(\frac{R_E}{a(1-(e_x^2+e_y^2))}\bigg)^2 \bigg ( \sqrt{1-(e_x^2+e_y^2)} (3 \cos^2 i -1) + (5 \cos^2 i -1) \bigg )\\
\\
\frac{de_x}{dt} = -\frac{3}{4} n J_2\bigg(\frac{R_E}{a(1-(e_x^2+e_y^2))}\bigg)^2 e_y(5 \cos^2 i -1)  \\
    \\
\frac{de_y}{dt} = \frac{3}{4} n J_2\bigg(\frac{R_E}{a(1-(e_x^2+e_y^2))}\bigg)^2 e_x(5 \cos^2 i -1)  \\
    \\
\frac{di}{dt} = 0 \\
\\
\frac{d\Omega}{dt} = -\frac{3}{2} n J_2^2 \bigg(\frac{R_E}{a(1-(e_x^2+e_y^2))}\bigg)^2 \cos i\\
\end{split}
\end{equation}

\subsection{$J_2^2$}

\begin{equation}
\begin{split}
\frac{da}{dt} = 0 \\
\\
\frac{du}{dt} = \frac{3}{8} n J_2^2  \bigg(\frac{R_E}{a(1-(e_x^2+e_y^2))}\bigg)^4 \frac{1}{\sqrt{1-(e_x^2+e_y^2)}} \bigg[ 3 \bigg( 3-\frac{15}{2} \sin^2i + \frac{47}{8} \sin^4i+\bigg(\frac{3}{2}-5 \sin^2i+\\
\frac{117}{16} \sin^4i\bigg) (e_x^2+e_y^2) 
-\frac{1}{8} (1+5 \sin^2i-\frac{101}{8} sin^4i) (e_x^2+e_y^2)^2 \bigg)  +\\
\frac{(e_x^2-e_y^2)}{8} \sin^2i (70-123 \sin^2i+(56-66 \sin^2i) (e_x^2+e_y^2))
    +\frac{27}{128}  \sin^4i \ ((e_x^2-e_y^2)^2-4e_y^2e_x^2) \\
    +\frac{1}{2} \bigg(48-103 \sin^2i+ \frac{215}{4} \sin^4i+(7-\frac{9}{2} \sin^2i-\frac{45}{8} \sin^4i) \ (e_x^2+e_y^2) \\
    +6 (1-\frac{3}{2} \sin^2i) (4-5 \sin^2i) \sqrt{1-(e_x^2+e_y^2)} -\frac{1}{4} (2 (14-15 \sin^2i) \sin^2i\\
    -(28-158 \sin^2(i)+135 \sin^4i) (e_x^2-e_y^2) \bigg)\bigg] \\
\\
\frac{de_x}{dt} = -\frac{3}{32} n J_2^2 \bigg(\frac{R_E}{a(1-(e_x^2+e_y^2))}\bigg)^4 \bigg [\sin^2i (14-15 \sin^2i)  (1-(e_x^2+e_y^2)) \frac{2e_ye_x^2}{(e_x^2+e_y^2)} \\
+ 2 e_y \bigg(48-103 \sin^2i+ \frac{215}{4} \sin^4i+(7-\frac{9}{2} \sin^2i-\frac{45}{8} \sin^4i) \ (e_x^2+e_y^2) \\
    +6 (1-\frac{3}{2} \sin^2i) (4-5 \sin^2i) \sqrt{1-(e_x^2+e_y^2)} -\frac{1}{4} (2 (14-15 \sin^2i) \sin^2i\\
    -(28-158 \sin^2(i)+135 \sin^4i) (e_x^2-e_y^2) \bigg) \bigg ]\\
\\
\frac{de_y}{dt} =-\frac{3}{32} n J_2^2 \bigg(\frac{R_E}{a(1-(e_x^2+e_y^2))}\bigg)^4 \bigg[ \sin^2i (14-15 \sin^2i)  (1-(e_x^2+e_y^2)) \frac{2e_y^2 e_x}{e_x^2+e_y^2} \\
-2 e_x \bigg(48-103 \sin^2i+ \frac{215}{4} \sin^4i+(7-\frac{9}{2} \sin^2i-\frac{45}{8} \sin^4i) \ (e_x^2+e_y^2) \\
    +6 (1-\frac{3}{2} \sin^2i) (4-5 \sin^2i) \sqrt{1-(e_x^2+e_y^2)} -\frac{1}{4} (2 (14-15 \sin^2i) \sin^2i\\
    -(28-158 \sin^2(i)+135 \sin^4i) (e_x^2-e_y^2) \bigg) \bigg ]\\
\\
\frac{di}{dt} = \frac{3}{64} n J_2^2 \bigg(\frac{R_E}{a(1-(e_x^2+e_y^2))}\bigg)^4 \sin2i \ (14-15 \sin^2i) 2e_xe_y \\
\\
\frac{d\Omega}{dt} = -\frac{3}{2} n J_2^2  \bigg(\frac{R_E}{a(1-(e_x^2+e_y^2))}\bigg)^4 \cos i \bigg(\frac{9}{4}+\frac{3}{2} \sqrt{1-(e_x^2+e_y^2)}-\sin^2i \bigg(\frac{5}{2}+\frac{9}{4} \sqrt{1-(e_x^2+e_y^2)}\bigg) \\ 
+\frac{(e_x^2+e_y^2)}{4} (1+\frac{5}{4} \sin^2i)+\frac{(e_x^2-e_y^2)}{8} (7-15 \sin^2i) \bigg) \\
\\
\end{split}
\end{equation}

\subsection{$J_3$}
\begin{equation}
\begin{split}
\frac{da}{dt} = 0 \\
\\
\frac{du}{dt} = \frac{3}{8}nJ_3\bigg(\frac{R_E}{a(1-(e_x^2+e_y^2))}\bigg)^3\bigg[ \bigg((4-5\sin^2i)\bigg(\frac{\sin^2i -(e_x^2+e_y^2)\cos^2i}{\sqrt{e_x^2+e_y^2} \sin i}\bigg) \\
+2\sin i (13-15 \sin^2 i)\sqrt{e_x^2+e_y^2}\bigg)\frac{e_y}{\sqrt{e_x^2+e_y^2}} \\
-\sin i(4-5\sin^2i)\frac{(1-4(e_x^2+e_y^2))}{e_x^2+e_y^2} e_y \sqrt{1 - (e_x^2+e_y^2)}\bigg ]
\\
\frac{de_x}{dt} = -\frac{3}{8}nJ_3\bigg(\frac{R_E}{a(1-(e_x^2+e_y^2))}\bigg)^3 \bigg [ \sin i(4-5\sin^2i)(1-(e_x^2+e_y^2))\frac{e_x^2}{e_x^2+e_y^2} \\
+\bigg((4-5\sin^2i)\bigg(\frac{\sin^2i -(e_x^2+e_y^2)\cos^2i}{\sqrt{e_x^2+e_y^2} \sin i}\bigg) \\
+2\sin i (13-15 \sin^2 i)\sqrt{e_x^2+e_y^2}\bigg)\frac{e_y^2}{\sqrt{e_x^2+e_y^2}} \bigg] \\
\\
\frac{de_y}{dt} =  -\frac{3}{8}nJ_3\bigg(\frac{R_E}{a(1-(e_x^2+e_y^2))}\bigg)^3 \bigg [ \sin i(4-5\sin^2i)(1-(e_x^2+e_y^2))\frac{e_xe_y}{e_x^2+e_y^2}\\
-\bigg((4-5\sin^2i)(\frac{\sin^2i -(e_x^2+e_y^2)\cos^2i}{\sqrt{e_x^2+e_y^2} \sin i}) \\
+2\sin i (13-15 \sin^2 i)\sqrt{e_x^2+e_y^2}\bigg)\frac{e_xe_y}{\sqrt{e_x^2+e_y^2}} \bigg ] \\
\\
\frac{di}{dt} = \frac{3}{8}nJ_3\bigg(\frac{R_E}{a(1-(e_x^2+e_y^2))}\bigg)^3\cos i (4-5\sin^2i )e_x \\
\\
\frac{d\Omega}{dt} = -\frac{3}{8}nJ_3\bigg(\frac{R_E}{a(1-(e_x^2+e_y^2))}\bigg)^3(15\sin^2i-4)e_y\cot(i) \\
\end{split}
\end{equation}

\subsection{$J_4$}
\begin{equation}
\begin{split}
\frac{da}{dt} = 0 \\
\\
\frac{du}{dt} = -\frac{45}{128} n J_4 \bigg(\frac{R_E}{a(1-(e_x^2+e_y^2))}\bigg)^4 \bigg [ (8-40 \sin^2i+35 \sin^4i) (e_x^2+e_y^2) \sqrt{1-(e_x^2+e_y^2)} \\
-\frac{2}{3}  \sin^2i (6-7 \sin^2(i)) (2-5 (e_x^2+e_y^2)) \sqrt{1-(e_x^2+e_y^2)} \frac{e_x^2-e_y^2}{e_x^2+e_y^2} \\
+ \frac{4}{3} \bigg(16-62 \sin^2i+49 \sin^4i+ \frac{3}{4} (24-84 \sin^2(i) 
+63 \sin^4i) (e_x^2+e_y^2) \\ +(\sin^2i (6-7 \sin^2i)-\frac{1}{2} (12-70 \sin^2i 
+63 \sin^4i) (e_x^2+e_y^2)) \frac{e_x^2-e_y^2}{e_x^2+e_y^2}\bigg) \bigg]\\
\\
\frac{de_x}{dt} = -\frac{15}{32} n J_4   \bigg(\frac{R_E}{a(1-(e_x^2+e_y^2))}\bigg)^4 \bigg[ \sin^2i (6-7 \sin^2i) (1-(e_x^2+e_y^2)) \frac{2e_ye_x^2}{e_x^2+e_y^2} \\
 - \bigg(16-62 \sin^2i+49 \sin^4i+ \frac{3}{4} (24-84 \sin^2(i) 
+63 \sin^4i) (e_x^2+e_y^2) \\ +(\sin^2i (6-7 \sin^2i)-\frac{1}{2} (12-70 \sin^2i 
+63 \sin^4i) (e_x^2+e_y^2)) \frac{e_x^2-e_y^2}{e_x^2+e_y^2}\bigg)e_y \bigg ] \\
\\
\frac{de_y}{dt} =  -\frac{15}{32} n J_4 \bigg(\frac{R_E}{a(1-(e_x^2+e_y^2))}\bigg)^4 \bigg [ \sin^2i (6-7 \sin^2i) (1-(e_x^2+e_y^2)) \frac{2e_y^2e_x}{e_x^2+e_y^2} \\
+ \bigg(16-62 \sin^2i+49 \sin^4i+ \frac{3}{4} (24-84 \sin^2(i) 
+63 \sin^4i) (e_x^2+e_y^2) \\ 
+(\sin^2i (6-7 \sin^2i)-\frac{1}{2} (12-70 \sin^2i
+63 \sin^4i) (e_x^2+e_y^2)) \frac{e_x^2-e_y^2}{e_x^2+e_y^2}\bigg)e_x \bigg ]\\
\\
\frac{di}{dt}  = \frac{15}{64} n J_4 \bigg(\frac{R_E}{a(1-(e_x^2+e_y^2))}\bigg)^4 \sin2 i (6-7 \sin^2i) 2e_xe_y \\
\\
\frac{d\Omega}{dt}  = \frac{15}{16} n J_4 \bigg(\frac{R_E}{a(1-(e_x^2+e_y^2))}\bigg)^4 \cos i \bigg((4-7 \sin^2i) (1+\frac{3}{2} (e_x^2+e_y^2))-\\
(3-7 \sin^2i) (e_x^2-e_y^2) \bigg) \\
\\
\end{split}
\end{equation}
\section{SRP}
\begin{equation}\label{Eq7}
\begin{split}
\frac{da}{dt} = 0 &\\
\\
\frac{de}{dt}= \frac{3 \sqrt{1-e^2}}{2 n a} T^p_{srp} &\\
\\
\frac{di}{dt} = -\frac{3 e \cos{\omega}}{2 n a \sqrt{1-e^2}} N_{srp} &\\
\\
\frac{d\Omega}{dt} = -\frac{3 e \sin{\omega}}{2 n a \sqrt{1-e^2} \sin{i}} N_{srp} &\\
\\
\frac{d\omega}{dt} = - \frac{3 \sqrt{1-e^2}}{2 n a e} R^p_{srp} - \ \frac{d\Omega}{dt}  \cos{i} &\\
\\
\frac{dM}{dt} = \frac{9 e}{2 n a} R^p_{srp} - \sqrt{1-e^2} \Big(\frac{d\omega}{dt} + \frac{d\Omega}{dt} \cos{i}\Big) &
\end{split}
\end{equation}

with 

\begin{equation}
\begin{split}
& R^p_{srp} = - F_{srp} (\mathcal{A} \cos{\omega} + \mathcal{B} \sin{\omega})\\
\\
& T^p_{srp} = - F_{srp} (-\mathcal{A} \sin{\omega} + \mathcal{B} \cos{\omega})\\
\\
& N_{srp} = - F_{srp} \mathcal{C}
\end{split}
\end{equation}

where 

\begin{equation}
\begin{split}
\mathcal{A} &= \cos(\Omega - \Omega_\odot)\cos{u_\odot} + \cos{i_\odot} \sin{u_\odot} \sin(\Omega-\Omega_\odot)\\
\\
\mathcal{B} &= \cos{i} [-\sin(\Omega-\Omega_\odot)\cos{u_\odot} + \cos{i_\odot}\sin{u_\odot}\cos(\Omega-\Omega_\odot)]+\sin{i}\sin{i_\odot}\sin{u_\odot}\\
\\
\mathcal{C} &= \sin{i}[\sin(\Omega-\Omega_\odot)\cos{u_\odot}-\cos{i_\odot}\sin{u_\odot}\cos(\Omega-\Omega_\odot)] + \cos{i}\sin{i_\odot}\sin{u_\odot}
\end{split}
\end{equation}

and 

\begin{equation}\label{Eq5}
F_{srp} = B \frac{\Phi}{c} \Big(\frac{1 AU}{r_\odot}\Big)^2 \end{equation}
where $\Phi = 1367 W m^{-2}$ is the solar flux at $1AU$ from the Sun, $c$ is the light speed, $r_\odot$ is the distance of the Sun from the asteroid and $B = \frac{C_r A}{m}$ is the spacecraft ballistic coefficient, with $C_r$ reflectivity coefficient, $A$ illuminated area, $m$ spacecraft mass.

\bibliographystyle{AAS_publication}   
\bibliography{IWSCFF19_template}   

\end{document}